\documentclass[11pt]{article}
\usepackage{amssymb,amsmath,graphicx,color}
\usepackage{amssymb} 
\usepackage{euscript}
\def\be{\begin{eqnarray}}
\def\ee{\end{eqnarray}}
\def\al{\alpha}
\def\vt{\vartheta}
\def\vtb{{\bar\vartheta}}
\def\xib{{\bar\xi}}
\def\dal{\dot\alpha}
\def\dbeta{\dot\beta}
 \def\0{\nonumber}
 
\def\sfM{{\sf M}}
\def\mfs{{\mathfrak s}}
\def\mfsb{\bar{\mathfrak s}}

\def\d{\partial}
\def\EA{\EuScript{A}}
\def\EC{\EuScript{C}}
\def\ED{\EuScript{D}}

\def\EF{\EuScript{F}}
\def\EW{\EuScript{W}}

\def\bfh{{\bf h}}

\def\bfG{{\bf G}}
\def\sb{{\bar s}}
\def\sfd{{\sf d}}

\begin{document}
\begin{flushright}
SISSA 17/2021/FISI 
\end{flushright}
\vskip 2cm
\begin{center}

{\LARGE {\bf BRST and superfield formalism. A review}}
 \vskip 2cm

{\large  L.~Bonora${}^{a}$, R.~P.~Malik${}^{b}$,
\\\textit{${}^{a}$ International School for Advanced Studies (SISSA),\\Via
Bonomea 265, 34136 Trieste, Italy \\
\textit{${}^{b}$ Physics Department, Center of Advance Studies, Institute of Science, Banaras Hindu University, Varanasi - 221 005, (U.P.), India
 } }}

\end{center}

\vskip 2cm {\bf Abstract}. {\sf This article, which is a review with substantial original material, is meant  to offer a comprehensive description of the superfield representations of the BRST and anti-BRST algebras and their applications to some field-theoretic topics. After a review of the superfield formalism for gauge theories we present the same formalism for gerbes and diffeomorphism invariant theories. The application to diffeomorphisms leads, in particular, to a horizontal Riemannian geometry in the superspace. We then illustrate the application to the description of consistent gauge anomalies and Wess-Zumino terms, for which the formalism seems to be particularly tailor-made. The next subject covered is the higher spin YM-like theories and their anomalies. Finally, we show that the BRST superfield formalism applies as well to the N=1 super-YM theories formulated in the supersymmetric superspace, for the two formalisms go along with each other very well. }

\vskip 2cm
 
\section{Introduction}

The discovery of the BRST symmetry in gauge field theories, \cite{BRS,T}, has been a fundamental achievement in quantum field theory. This symmetry is not only the building block of the renormalization  programs, but it has opened the way to an incredible number of applications. Beside gauge field theories, all theories with a local symmetry are characterized by a BRST symmetry: theories of gerbes, sigma models, topological field theories, string and superstring theories, to name the most important ones. Whenever  a classical theory is invariant under local gauge transformations, its quantum counterpart has a BRST-type symmetry that governs its quantum behavior. Two main properties characterize the BRST symmetry. The first is its group theoretical nature: {performing two (different) gauge transformations one after the other and then reversing the order of them does not lead to the same result (unless the original symmetry is Abelian), but the two different results are related by a group theoretical rule. This is contained in the nilpotency of the BRST transformations.}

The second important property of the BRST transformations is their nilpotency itself. It is inherited, via the Faddeev-Popov quantization procedure, from the anticommuting nature of the ghost and anti-ghost fields. This implies that, applying twice the same transformation, we get 0. The two properties together give rise to the Wess-Zumino consistency conditions, a fundamental tool in the study of anomalies. It must be noted that while the first property is classical, the second is entirely quantum. In other words, the BRST symmetry is a quantum property. 

It was evident from the very beginning that the roots of BRST symmetry are geometrical. The relevant geometry was linked at the beginning to the geometry of the principal fiber bundles, \cite{TM,Quiros} (see also \cite{Stora0}). Although it is deeply rooted in it,  the BRST symmetry is rather connected to the geometry of infinite dimensional bundles and groups, in particular the Lie group of gauge transformations, \cite{BCR,BCRS}. One may wonder where this infinite dimensional geometry is stored in a perturbative quantum gauge theory. The answer is: in the anticommutativity of the ghosts and in the nilpotency of the BRST transformations themselves, together with their group theoretical nature. There is a unique way to synthesize these quantum properties, and this is the superfield formalism. The BRST symmetry calls for the introduction of the superfield formulation of quantum field theories. One might even dare say that the superfield formalism is the genuine language of a quantum gauge theory.  This is the subject of the present article, which is both a review of old results and a collection of new ones, with the aim of highlighting the flexibility of the superfield approach to BRST symmetry (it is natural to extend it to include also the anti-BRST symmetry, \cite{CF,KO}). Here the main focus is on the algebraic aspects and on the ample realm of applications, leaving the more physical aspects (functional integral and renormalization) for another occasion. We will meet general features - we can call them universal - which appear in any application and for any symmetry.  One is the so-called horizontality conditions, i.e. the vanishing of the components along the anticommuting directions, which certain quantities must satisfy. Another is the  so-called Curci-Ferrari conditions, \cite{CF}, which always appear when  both (non-Abelian) BRST and anti-BRST symmetries are present.

Before passing to a description of how the present review is organized, let us comment on the status of anti-BRST. It is an algebraic structure that  comes up naturally as a companion of the BRST one, but it is not necessarily a symmetry of any gauge-fixed action. It holds, for instance, for linear gauge fixing, and some implications have been studied to some extent in \cite{KO} and also in \cite{BT}.  But it is fair to say that no fundamental role for this symmetry has been uncovered so far, although it is also fair to say that the research in this field has never overcome a preliminary stage \footnote{{But there seem to  exist models where the anti-BRST symmetry is of particular interest, for  instance the one of the vector supersymmetry (combining with BRST and anti-BRST symmetries)
in certain topological field theories like Chern-Simons theory \cite{Delduc1} for which this
symmetry is at the origin of the finiteness of models \cite{Delduc2}. More recently,  the fundamental role of the anti-BRST symmetry in the construction of Hodge-type theories was pointed out in \cite{Malik-Hodge}}} . In this review we will consider BRST and anti-BRST together in the superfield formalism, but, whenever, it is more convenient and expedient to use only the BRST symmetry we will focus only on it.

We start in section 2 with a review of the well-known superfield formulation of BRST and anti-BRST of non-Abelian gauge theories, which is obtained by enlarging the spacetime with two anticommuting coordinates $\vt,\vtb$. Section 3 is devoted to gerbe theories, which are close to ordinary gauge theories. After a short introduction, we show that it is simple and natural to reproduce the BRST and anti-BRST symmetries with the superfield formalism. As always, when both BRST and anti-BRST are involved, we come across specific CF conditions. The next two sections are devoted to diffeomorphisms. Diffeomorphisms are a different kind of local transformations, therefore it is interesting to see first of all if the superfield formalism works. In fact in section 4 we find horizontality and CF conditions 
for which BRST and anti-BRST transformations are reproduced by the superfield formalism. We show, however, that the super-metric, i.e. the metric with components in the anticommuting directions, is not invertible. So a super-Riemannian geometry is not possible in the superspace, but, in exchange, we can define a horizontal super-geometry, with Riemann and Ricci tensors defined on the full superspace. In section 5 we deal with frame superfields and define fermions in superspace. In summary, there are no obstructions to formulate quantum gravitational theories in the superspace.

The second part of the paper concerns applications of the superfield method to some practical problems, notably to anomalies. Consistent anomalies are a perfect playground 
for the superfield method, as we show in section 6. We show that not only all the formulas concerning anomalies in any even dimension are easily reproduced but actually the superfield formalism seems to be tailor-made for them. A particularly sleek result is the way one can extract Wess-Zumino terms from it. In section 7 we apply the superfield formalism to HS-YM-like theories. After a rather detailed introduction to such novel models we show that the superfield method fits perfectly well and is instrumental in deriving the form of anomalies, which would be otherwise of hard access. Section 8 is devoted to the extension of the superfield method in still another direction, that of supersymmetry. We show, as an example, that the supersymmetric superspace formulation of $N=1$ SYM theory in 4d can be easily enlarged by extending the superspace with the addition of $\vt,\vtb$, while respecting the supersymmetric geometry (constraints) . In section 9, we make some concluding remarks and comments on some salient features of our present work.

The Appendices contain auxiliary materials, except the first (Appendix A) which might seem a bit off-topic with respect to the rest of the paper. We deem it useful to report it  in order to clarify the issue of the classical geometric description of the BRST symmetry. As mentioned above this description is possible. However one must formulate this problem in the framework of the geometry of the infinite dimensional groups of gauge transformations (which are in turn rooted in the geometry of principal fiber bundles). The  appropriate mathematical tool is the evaluation map. One can easily see how the superfield method formulation parallels the geometrical description. 

Finally let us add that this review covers only a part of the applications of the superfield approach that have appeared in the literature. We must mention \cite{Lavrov,Reshetnyak, Boldo} and \cite{Malik,malikdiff} for further extensions of the method and additional topics not presented here.
{A missing subject in this paper, as well as,  to the best of our knowledge, in the present literature, is the exploration of the possibility to extend the superfield method to the  Batalin-Vilkovisky approach to field theories with local symmetries.}

\vskip 1cm
{
 {\bf Notations and Conventions} The superspace will be represented by super-coordinates $X^M=\{x^\mu,\vt,\vtb\}$, where $x^\mu$ $(\mu=0,1,\ldots, d-1)$ are ordinary commuting coordinates, while $\vt$ and $\vtb$ are anticommuting: $\vt^2=\vtb^2=\vt\vtb+\vtb\vt=0$, but commute with $x^\mu$. We will make use of a generalized differential geometric notation: the exterior differential $d=\frac{\partial}{\partial x^\mu} dx^\mu$ will be generalized to $\tilde d= d+\frac {\partial}{\partial \vt}d\vt + \frac {\partial}{\partial \vtb}d\vtb$. Correspondingly, mimicking the ordinary differential geometry,  we will introduce super-forms, for instance
$\widetilde \omega= \omega_\mu (x) dx^\mu+ \omega_\vt(x) d\vt +\omega_\vtb(x) d\vtb$, where $\omega_\mu$ are ordinary commuting intrinsic components, while $\omega_\vt, \omega_\vtb$ anticommute with each other and commute with $\omega_\mu$. In the same tune we will introduce also super-tensors, such as the super-metric, see section 4.4. As far as commutativity properties (gradings) are concerned, the intrinsic components of forms and tensors on one side and the symbols $d^\mu, d\vt,d\vtb$, on the other, constitute separate mutually commuting sets. When $\tilde d$ acts on a super-function $\widetilde F(X)$, it is understood that the derivatives act on it from the left to form the components of a 1-super-form: 
\be
\tilde d \widetilde F(X)= \frac {\partial}{\partial x^\mu} \widetilde F(X) dx^\mu+  \frac {\partial}{\partial \vt} \widetilde F(X)d\vt + \frac {\partial}{\partial \vtb} \widetilde F(X)d\vtb
\ee
When it acts on a 1-super-form it is understood that the derivatives act on the intrinsic components from the left and the accompanying symbol $dx^\mu, d\vt,d\vtb$ gets juxtaposed to the analogous symbols of the super-form from the left, to form the combinations $dx^\mu\wedge dx^\nu, dx^\mu\wedge d\vt, d\vt\wedge d\vt, d\vt \wedge d\vtb, \ldots$, with the usual rule for the spacetime symbols, and  $dx^\mu \wedge d\vt=-d\vt\wedge dx^\mu, dx^\mu \wedge d\vtb=-d\vtb\wedge dx^\mu$, but $d\vt\wedge d\vtb=d\vtb\wedge d\vt$, and $d\vt\wedge d\vt$ and $d\vtb\wedge d\vtb$ are non-vanishing symbols. In a similar way one proceeds with higher degree super-forms. More specific notations will be introduced later when necessary.
}

\section{The superfield formalism in gauge field theories}

The superfield formulation of the BRST symmetry in gauge field theories was proposed in \cite{BT}, for an earlier version see \cite{FPSF}. Here we limit ourselves to a summary. Let us consider a generic gauge theory in $\sfd$ dimensional {Minkowski spacetime} $\sfM$, with connection $A_\mu ^a T^a$ ($\mu=0,1,\ldots,\sfd-1$), valued in a
Lie algebra ${\mathfrak g}$ with anti-hermitean generators $T^a$ , such that 
$[T^a , T^b ] = f^{abc} T^c$ .
In the following it is convenient to use the more compact form notation and
represent the connection as a one-form $A = A_\mu ^a T^a dx^\mu$. The curvature and gauge transformation are
\be
F = d A+\frac 12[A,A]\quad\quad{\rm and}\quad\quad\delta_\lambda A = d \lambda +[A, \lambda],\label{lambdaetransf}
\ee
with $\lambda(x) = \lambda^a(x) T^a$ and $d= dx^\mu \frac {\partial}{\partial x^\mu}$.
{The infinite dimensional Lie algebra of gauge transformations and its cohomology can be formulated in a simpler and more effective way} if we
promote the gauge parameter $\lambda$ to an anticommuting ghost field $c = c^a T^a$
and define the BRST transform as\footnote{ The symbol $[\cdot,\cdot]$ denotes an ordinary commutator when both entries are non-anticommuting, and an anticommutator when both entries are anticommuting.}
\be
\mfs A\equiv dc+[A,c], \quad\quad \mfs \,c=-\frac 12 [c,c]. \label{BRST1}
\ee
As a consequence of this, we have
\be
{\boldsymbol {\cal F}}\equiv  (d - \mfs)(A +c)+\frac 12[A+c ,A+c]= F,\label{russian}
\ee
{which is sometime referred to as {\it Russian formula}, \cite{BT,BauTh,Stora1,Stora2} }. Eq.\eqref{russian} is true 
provided we assume that
\be
[A,c]=[c,A].\label{AccA}
\ee
i.e. if we assume that $c$ behaves like a one-form in the commutator with ordinary forms and with itself. It can be in fact related to the Maurer-Cartan form in $\sf G$. This explains its anticommutativity.

A very simple way to reproduce the above formulas and properties is by enlarging the space to a superspace with coordinates $(x^\mu, \vt)$, where $\vt$ is anticommuting, and promoting the connection $A$ to a  one-form superconnection $\widetilde A= \phi (x,\vt)+ \phi_{\vt}(x,\vt)d\vt$ with expansions
\be
\phi (x,\vartheta)={\sf A}(x)+ \vt \Gamma(x), \quad\quad \phi_\vt (x,\vt) ={\sf c}(x) +\vt G(x),\label{Aphi}
\ee
and two-form supercurvature
\be
\widetilde F= \widetilde d \widetilde A +\frac 12 [\widetilde A,\widetilde A],\quad\quad \widetilde F = \Phi(x,\vt) + \Phi_\vt(x,\vt)d\vt+\Phi_{\vt\vt}(x,\vt) d\vt \wedge d\vt ,\label{curvatureF}
\ee
with $\Phi(x,\vt)={\sf F}(x)+\vt \Lambda(x)$ and
$\widetilde d=d+\frac {\partial }{\partial \vt} d\vt$. Notice that since $\vt^2=0$,  $ d\vt \wedge d\vt\neq 0$, while $dx^\mu \wedge d\vt=-d\vt \wedge dx^\mu$. Then we impose the `horizontality' condition
\be
\widetilde F = \Phi(x,\vt), \quad\quad {\rm i.e.} \quad\quad  \Phi_\vt(x,\vt) =0= \Phi_{\vt\vt}(x,\vt) . \label{widetildeF}
\ee
The last two conditions imply
\be
\Gamma(x) = d c(x) +[{\sf A}(x),c(x)], \quad\quad G(x)=-\frac 12 [{\sf c}(x),{\sf c}(x)] .\0
\ee
Moreover $\Lambda(x)= [{\sf F}(x), {\sf c}(x)]$. 

This means that we can identify ${\sf c}(x)\equiv c(x)$, $ {\sf A}\equiv A$, ${\sf F}\equiv F$,  and the $\vt$ translation with the BRST transformation $\mfs$, i.e. $\mfs\equiv \frac {\partial}{\partial\vt}$. In this way all the previous transformations, including the, at first sight strange looking, eq.\eqref{AccA}, are naturally explained. It is also possible to push further the use of the superfield formalism by noting that, after imposing the horizontality condition, we have
\be
\widetilde A = e^{-\vt c} A \,e^{\vt c} + e^{-\vt c} \widetilde d e^{\vt c},\quad\quad
\widetilde F = e^{-\vt c} F \,e^{\vt c}. \label{expgaugetransf}
\ee

A comment is in order concerning the horizontality condition (HC). This condition is suggested by the analogy with the principal fiber bundle geometry. In the total space of a principal fiber bundle one can define horizontal (or basic) forms. These are forms with no components in the vertical direction: for instance, given a connection its curvature is horizontal. In our superfield approach the $\vt$ coordinate mimics the vertical direction: for the curvature $\widetilde F$ does not have components in that direction. This horizontality principle can be extended also to other quantities, for instance to covariant derivatives of matter fields and, in general, to all quantities which are invariant under local gauge transformations.

\subsection{Extension to anti-BRST transformations}

The superfield representation of the BRST symmetry with one single anticommuting variable is, in general, not sufficient for ordinary Yang-Mills theories, because gauge fixing requires in general other fields beside $A_\mu$ and  $c$. For instance, in the Lorenz gauge the Lagrangian density {takes}  the form 
\be
{\cal L}_{YM} = -{\rm t}r\left(\frac 1{4g^2} F_{\mu\nu}F^{\mu\nu}+ A_\mu \partial^\mu B -\partial^\mu \bar c D_\mu c+\frac {\alpha}2 B^2\right),\label{calLYM}
\ee
where two new fields have been introduced, the antighost field $\bar c(x)$ and the Nakanishi-Lautrup field $B(x)$. It is necessary to enlarge the algebra \eqref{BRST1} as follows
\be
\mfs\, \bar c= B,\quad\quad \mfs B=0, \label{BRST2}
\ee
in order to get a symmetry of \eqref{calLYM}. But, at this point, ${\cal L}_{YM}$ is invariant under a larger symmetry, whose transformations, beside \eqref{BRST1} and \eqref{BRST2},  are the anti-BRST ones
\be
\bar \mfs\, A= d\bar c +[A,\bar c],\quad\quad \bar \mfs\, \bar c= -\frac 12 [\bar c,\bar c],
\quad\quad \bar \mfs\, c = \bar B, \quad\quad \bar \mfs\, \bar B=0,\label{antiBRST}
\ee
provided
\be
B+\bar B+[c,\bar c]=0.\label{curciferrari}
\ee
This is the Curci-Ferrari condition, \cite{CF}. 

The BRST and anti-BRST transformation are nilpotent and anticommute:
\be
\mfs^2=0,\quad\quad \bar \mfs^2=0,\quad\quad \mfs\, \bar \mfs + \bar \mfs\, \mfs =0.\label{milpotence}
\ee

The superfield formalism applies well to this enlarged symmetry provided we introduce another anticommuting coordinate $\vtb$: $\vtb^2=0, \vt \vtb+\vtb \vt=0$. Here we do not repeat the full derivation as in the previous case, but simply introduce the supergauge transformation, \cite{BT,malikU},
\be
U(x,\vt,\vtb) = \exp {[\vt \bar c(x)+\vtb c(x) + \vt\vtb (B(x)+[c(x),\bar c (x)])]}, \label{Ucttb}
\ee
and generate the superconnection
\be
\widetilde A(x,\vt,\vtb) =U(x,\vt,\vtb)^\dagger\left( \widetilde d + A(x)\right) U(x,\vt,\vtb) , \label{superconnectionA}
\ee
where $\widetilde d= d + d\vt \frac {\partial}{\partial\vt}+ d\vtb \frac {\partial}{\partial\vtb}$ and the hermitean operation is defined as follows
\be
\vt^\dagger =\vt, \quad \vtb^\dagger =-\vtb,\quad\quad (c^a)^\dagger=c^a, \quad\quad  (\bar c^a)^\dagger=-\bar c^a,\0
\ee
while the $B^a(x),\bar B^a(x)$ are real.  
Then the superconnection is
\be
\widetilde A(x,\vt,\vtb)= \Phi(x,\vt,\vtb) + \eta(x,\vt,\vtb) d\vtb+ \bar \eta(x,\vt,\vtb) d\vt.\label{suoeconnA1}
\ee
The one-form $\Phi$ is
\be
 \Phi(x,\vt,\vtb) =A(x) + \vt D\bar c(x) +\vtb D c(x)+ \vt \vtb (DB(x) +[Dc(x),\bar c(x)]),\label{Phitvtb}
\ee
where $D$ denotes the covariant differential: $Dc= dc +[A,c]$, etc., and the anticommuting functions $\eta, \bar \eta $ are
\be
\eta(x,\vt,\vtb)& =& c(x)+\vt\, \bar B(x) -\frac 12 \vtb \,[c(x),c(x)] + \vt\vtb\, [\bar B(x), c(x)], \label{etavtvtb}\\
\bar\eta(x,\vt,\vtb)& =&\bar c(x)-\frac 12 \vt\, [\bar c(x),\bar c(x)] +\vtb\, B(x) + \vt\vtb\, [\bar c(x), B(x)], \label{baretavtvtb}
\ee
together with the condition \eqref{curciferrari}. One can verify that the supercurvature $\widetilde F$ satisfies the horizontality condition
\be
\widetilde F(x,\vt,\vtb) = d\Phi(x,\vt,\vtb)+\frac 12[\Phi(x,\vt,\vtb),\Phi(x,\vt,\vtb)].\label{horizontalF}
\ee
The BRST transformation correspond to $\vtb$ translations and the anti-BRST to $\vt$ ones:
\be
\mfs =\left.\frac {\partial}{\partial \vtb}\right\vert_{\vt=0}, \quad\quad \bar\mfs =\left.\frac {\partial}{\partial \vt}\right\vert_{\vtb=0}.
\ee

At the end this short review it is important to highlight an important fact. As anticipated above the Lagrangian density \eqref{calLYM} is invariant under {\it both} the BRST and anti-BRST transformations (\ref{BRST1},\ref{BRST2}) and \eqref{antiBRST}, provided \eqref{curciferrari} is satisfied. However, while {
the Lagrangian density contains a specific gauge-fixing, the BRST and anti-BRST {\it algebra} (when it holds) is independent of any gauge fixing condition. We can change the gauge-fixing but the BRST  and anti-BRSTalgebra (when it is present), as well as their superfield representation, are always the same.} This algebra can be considered the {\it quantum version} of the original classical gauge algebra. A classical geometrical approach based on fiber bundle geometry was originally proposed in \cite{TM} and \cite{Quiros}. Subsequently the nature of the BRST transformations was clarified in \cite{BCR} and in \cite{BCRS}. In fact it is possible to uncover the BRST algebra in the geometry of principal fiber bundles, in particular in terms of the evaluation map, as shown in Appendix A. However, while classical geometry is certainly the base of classical gauge theories, it becomes very cumbersome and actually intractable for perturbative quantum gauge theories. On the other hand, in dealing with the latter, anticommuting ghost and antighost fields and the (graded) BRST algebra seems to be the natural tools. Therefore, a as noted previously, one may wonder whether the natural language for a quantum gauge field theory is in fact the superfield formalism. We leave this idea for future developments.

\vskip 1cm 
Here ends our short introduction of the superfield formalism in gauge field theories, which was historically the first application. Later on, we shall see a few of its applications.  But now we would like to explore the possibility to apply this formalism to other local symmetries. The first example, and probably the closest to the one presented in this section, is a theory of gerbes. A gerbe is a mathematical construct which, in a sense, generalizes the idea of gauge theory. From the field theory point of view, the main difference with the {\it latter} is that it is not based on a single connection, but, beside one-forms, it contains also other forms. Here we  consider the simplest case, an Abelian 1-gerbe, see \cite{bonora-malik}.

\section{1- gerbes}

Let us recall a few basic definitions. A 1--gerbe, \cite{gerbes}, is a mathematical object that can be described with
a triple $(B,A,f)$, formed by the 2-form $B$, 1-form $A$ and 0-form $f$, respectively. 
These are related in the following way. Given a covering $\{U_i\}$ of 
the manifold $\sf M$, we associate to each $U_i$ a two--form $B_i$. On a double intersection $U_i
\cap U_j$, we have $B_i-B_j = dA_{ij}$. On the triple
intersections $U_i \cap U_j\cap U_k$, we must have $A_{ij}+
A_{jk}+A_{ki}= d f_{ijk}$ ($B_i$ denotes $B$ in $U_i$, $A_{ij}$denotes $A$ in $U_i \cap U_j$, etc. ). Finally, on the quadruple intersections
$U_i \cap U_j\cap U_k\cap U_l$, the following integral cocycle
condition must be satisfied by $f$:
\be
f_{ijl}-f_{ijk}+f_{jkl}-f_{ikl}= 2\;\pi \;n, 
\qquad n = 0, 1, 2, 3.........\label{cocyclefijk} 
\ee
This integrality condition will not concern us in our Lagrangian
formulation but it has to be imposed as an external condition.

Two triples, represented by $(B,A,f)$ and $(B',A',f')$,
respectively, are said to be  gauge equivalent if they satisfy the 
following relations
\be
&&B_i'=B_i+ dC_i \quad\quad {\rm on} \quad U_i, \label{gaugeeq1}\\
&&A_{ij}' = A_{ij} + C_i-C_j + d\lambda_{ij} \quad\quad {\rm on}
\quad U_i\cap U_j, \label{gaugeeq2}\\ && f_{ijk}' = f_{ijk} +
\lambda_{ij}+\lambda_{ki}+ \lambda_{jk} \quad\quad {\rm on} \quad
U_i\cap U_j\cap U_k, \label{gaugeeq3} 
\ee 
for the one--forms $C$ and the zero--forms $\lambda$.

Let us now define the BRST and anti--BRST transformations corresponding to these
geometrical transformations. 
It should be recalled that, while the above geometric transformations are defined on
(multiple) neighborhood overlaps, the BRST and anti--BRST transformations, in quantum field theory,  
are defined on a single local coordinate patch. These (local, field-dependent) transformations are the means QFT uses to record the underlying geometry.

The appropriate BRST and anti--BRST transformations are:
\begin{eqnarray}\begin{matrix}
 \mfs\, B = d C, & \mfs\, A= C + d \lambda,&\mfs\, f=\lambda+\mu,\cr
\mfs \,C= -d h,  &    \mfs\,\lambda = h, 
& \mfs\,\mu=-h, \cr
 \mfs \,\bar C = - K, &\mfs\,\bar K = d \rho, &\mfs\,\bar\mu=-g, \cr
 \mfs \,\bar\beta = - \bar \rho,  & \mfs\,\bar \lambda= g,  &
\mfs\, \bar g= \rho, \end{matrix}\label{BRST3}
\end{eqnarray}
together with $\mfs \; [\rho, \bar\rho, g, K_\mu, \beta] = 0$, and
\begin{eqnarray}
&&\mfsb\, B = d\bar C, 
\qquad\mfsb\, A = \bar C + d \bar \lambda,\qquad  \mfsb\, f=\bar\lambda+\bar \mu,\nonumber\\
&&  \mfsb\,\bar C = + d\bar  h, \qquad
\mfsb \, \bar \lambda  = - \bar  h, 
\qquad \mfsb\, \bar \mu=-\bar h, \nonumber\\
&&\mfsb\, C = + \bar  K,
\qquad \mfsb \, K = - d \bar \rho , \qquad \mfsb\,\bar \mu =\bar g,  \nonumber\\
&&\mfsb\,\beta = + \rho,    \qquad \mfsb\,\lambda= - \bar g, \qquad
\mfsb\, g = - \bar \rho, \label{antiBRST3}
\end{eqnarray}
while $\mfsb \;[\bar\beta, \bar g, \bar K,\mu, \rho, \bar\rho]= 0$.

In these formulas $C, \bar C$ are anticommuting 1-forms, $K, \bar K$ are commuting 1-forms. The remaining fields are scalars, which are commuting if denoted by latin letters, anticommuting if denoted by greek letters.

It can be easily verified that $(\mfs+\mfsb)^2=0$ if the following constraint
is satisfied:
\begin{equation}
\bar K- K=d\, \bar g -d\, g.\label{CF2}
\end{equation}
This condition is both BRST and anti--BRST invariant.
It is the analogue of the Curci-Ferrari condition in non--Abelian 1-form 
gauge theories and we will refer to it with the same name.

Before we proceed to the superfield method, we would like to note that the 
above realization of the BRST and anti--BRST algebra is not the only 
possibility. In general, it may be possible to augment it by the 
additng a sub-algebra of elements which are all in the kernel of both $s$ and $\bar s$ or,
if it contains such a sub-algebra, the latter could be moded out. 
For instance, in equations (\ref{BRST3},\ref{antiBRST3}) $\rho$ and $\bar \rho$
form an example of this type of subalgebra. It is easy to see that $\rho$ and 
$\bar \rho$ can be consistently set equal to 0.
 
\subsection{The superfield approach to gerbes}

We introduce superfields whose lowest components are $B,A$ and $f$.
\be
\widetilde {\cal B}=\widetilde {\cal B}_{MN}(X)dX^M\wedge dX^N&=& {\cal B}_{\mu\nu}(X)dx^\mu\wedge dx^\nu + {\cal B}_{\mu\vt} (X)dx^\mu\wedge  d\vt+ {\cal B}_{\mu\vtb} (X)dx^\mu\wedge  d\vtb,\0\\
&& + {\cal B}_{\vt\vt} (X) d\vt\wedge  d\vt+ {\cal B}_{\vtb\vtb} (X)d\vtb\wedge  d\vtb +{\cal B}_{\vt\vtb} (X)d\vt \wedge \vtb, \label{widetildecalB}\\
\widetilde{\cal A}=\widetilde {\cal A}_M(X)dX^M &=& {\cal A}_\mu(X) dx^\mu + {\cal A}_\vt (X) d\vt +{\cal A}_{\vtb} (X) d\vtb,\label{widetildecalA}\\
\widetilde{\sf f}(X)&=&f(x)+ \vt \bar \phi(x)+\vtb \phi(x) +\vt\vtb F(x).\label{widetildecalf}
\ee
{where $X$ denotes the superspace point and $X^M=(x^\mu,\vt,\vtb)$ the superspace coordinates.}
All the intrinsic components are to be expanded like \eqref{widetildecalf}. Then we impose the horizontality conditions. There are two:
\be
\tilde d \, \widetilde {\cal B}&=& d{\cal B},\quad\quad \quad\quad\quad\quad{\cal B}={\cal B}_{\mu\nu}(X) dx^\mu\wedge dx^\nu,\label{firsthor}\\
\widetilde{\cal B} - \tilde d \, \widetilde{\cal A } &=& {\cal B}- d {\cal A},\quad\quad \quad\quad {\cal A}={\cal A}_\mu(X) dx^\mu.\label{secondhor}
\ee
The first is suggested by the invariance of  $H=dB$ under $B\to B+d\Lambda$, where $\Lambda$ is a 1-form. The second by the invariance of $B-dA$ due to the transformations $B\to B+d\Sigma, A\to A+\Sigma$, where $\Sigma$ is also a 1-form.

Using the second we can eliminate many components of $\widetilde {\cal B}$ in favor of the components of $\widetilde{\cal A}$:
\be
 {\cal A}_\mu(X)&=& A_\mu(x) +\vt\, \bar \alpha_\mu(x) +\vtb\, \alpha_\mu(x) +\vt\vtb\, \EA_\mu(x),\label{calAmu}\\
 {\cal A}_\vt (X)&=& \gamma(x)+\vt\, \bar e(x) +\vtb\, e(x) +\vt
\vtb\, \Gamma(x),\label{calAtheta}\\
{\cal A}_{\vtb} (X)&=& \bar\gamma(x)+\vt\, \bar a(x) +\vtb\, a(x) +\vt\vtb \,\bar\Gamma(x).\label{calAthetab}
\ee
Imposing \eqref{secondhor} $\widetilde {\cal B}$ takes the form
\be
 {\cal B}_{\mu\nu}(X)&=& B_{\mu\nu}(x) + \vt\, \bar \beta_{\mu\nu}(x) +\vtb \,\beta_{\mu\nu}(x) +\vt\vtb\, M_{\mu\nu}(x),\label{calBmunu}\\
{\cal B}_{\mu\vt}(X) &=& \bigl(-\bar\alpha_\mu(x) +\partial_\mu \gamma(x)\bigr)+\vt \,\partial_\mu \bar e(x)+\vtb \, \bigl( \partial_\mu e(x) - \EA_\mu(x)\bigr) +\vt\vtb\, \partial_\mu \Gamma(x), \label{calBmutheta}\\
{\cal B}_{\mu\vtb}(X) &=& \bigl(-\alpha_\mu(x) +\partial_\mu \bar\gamma(x)\bigr)+\vt \,\partial_\mu\bigl(\bar a(x)+\EA_\mu(x)\bigr)+\vtb \,  \partial_\mu a(x)+\vt\vtb\, \partial_\mu \bar\Gamma(x), \label{calBmutheta}\\
{\cal B}_{\vt\vt}(X) &=& \bar e(x) +\vtb \,\Gamma(x),\label{calBvtvt}\\
{\cal B}_{\vtb\vtb}(X) &=& a(x) -\vt \,\bar\Gamma(x),\label{calBvtbvtb}\\
{\cal B}_{\vt\vtb}(X) &=&(e(x)+ \bar a(x)) +\vtb \,\bar\Gamma(x)-\vt \Gamma(x),\label{calBvtvtb}
\ee
where all the component fields on the RHS's are so far unrestricted. If we now impose \eqref{firsthor} we get further restrictions
\be
\bar\beta_{\mu\nu}(x)&=&-(d\beta)_{\mu\nu}(x)=  (d\bar \alpha)_{\mu\nu}(x),\label{betamunu}\\
\beta_{\mu\nu}(x)&=&- (d\bar \beta)_{\mu\nu}(x)= (d\alpha)_{\mu\nu}(x ),\label{barbetamunu}\\
M_{\mu\nu}(x)&=& (d\EA)_{\mu\nu}(x) ,\label{Mmunu}
\ee
where $\beta, \bar \beta, \EA$ denote one-forms with components $\beta_\mu(x),\bar\beta_\mu(x) ,\EA_\mu(x)$, respectively.

We will also consider, instead of $\widetilde {\cal A}$ the superfield $ \widetilde {\cal A}- \tilde d \,\widetilde {\sf f}$, and, in particular, we will replace $A$ with $A'=A-df$.

From the previous equations we can read off the BRST trabsformations of the independent component fields. Dropping the argument $(x)$ and using the form notation for the BRST transformations we have
\be\begin{matrix}
\mfs B=d\bar\alpha, \quad \quad&\mfs A'=\bar \alpha-d\bar\phi , \quad \quad&\mfs f=\bar\phi,\\
\mfs \alpha =- d\EA,\quad\quad&\mfs\gamma=\bar e,\quad\quad & \mfs e=-\Gamma,\\
\mfs \bar \gamma=\bar a,\quad \quad& \mfs a= -\bar \Gamma,\quad\quad&\mfs\phi = -F,
\end{matrix}\label{sB}
\ee
all the other $\mfs$ transformations being {\it trivial}.  For the anti-BRST transformations we have
\be\begin{matrix}
\mfsb B =d\alpha ,\quad\quad& \mfsb A'= \alpha - d \phi,\quad\quad &\mfsb f =\phi,\\
\mfsb \bar\alpha = \EA,\quad \quad&\mfsb \bar\gamma= a ,\quad\quad & \mfsb\bar a = \bar \Gamma,\\
\mfsb \gamma= e,\quad\quad & \mfsb \bar e =\bar \Gamma,\quad\quad &\mfsb \bar \phi =F,
\end{matrix}\label{sBbar}
\ee
All the other anti-BRST transformations are {\it trivial}.

The system (\ref{sB},\ref{sBbar}) differs from (\ref{BRST3},\ref{antiBRST3}) only by field redefinitions. For let us set
\be
&&C=\bar \alpha+d \gamma,\quad \quad \lambda = -\gamma-\bar \phi,\label{redef1}\\
&& \overline C= \alpha-d \bar\gamma,\quad\quad \bar\lambda =\bar\gamma- \phi .\label{redef2}
\ee
Then  the first equation of \eqref{sB} and the first of \eqref{sBbar} become
\be
\begin{matrix}
\mfs B = dC,\quad\quad& \mfs A'= C+d\lambda,\quad\quad &\mfs f= \lambda+\gamma, \label{firstsB}\\
\mfsb\bar B = d\overline C,\quad \quad& \mfsb A'= \overline C + d \bar \lambda,\quad\quad &\mfsb f= \bar \lambda -\bar \gamma.\end{matrix}\label{firstsBbar}
\ee
Next we define
\be
K_\mu = \EA_\mu +\partial_\mu \bar a, \quad\quad \overline K_\mu= \EA_\mu +\partial_\mu e.\label{KmuKmubar}
\ee
The remaining $s$ and $\sb$ transformations become
\be\begin{matrix}
\mfs C= d\bar e, \quad\quad&\mfs \lambda =- \bar e,\quad\quad&\mfs \gamma= \bar e,\\
\mfs \overline C=- K,\quad\quad &\mfs\overline K =-d\Gamma, \quad\quad&\mfs\bar \gamma=\bar a,\\
\mfs a= -\overline\Gamma,\quad\quad &\mfs\bar \lambda =\bar a +F,\quad\quad & \mfs a= -\bar \Gamma,
\end{matrix}\label{sBext}
\ee
and
\be\begin{matrix}
\mfsb \overline C= -d a, \quad\quad&\mfsb \bar\lambda =a,\quad\quad&\mfsb\bar \gamma= a,\\
\mfsb C=\overline K,\quad\quad & \mfsb K =-d\overline \Gamma, \quad\quad&\mfsb \gamma= e,\\
\mfsb \bar e= \Gamma,\quad\quad &\mfsb \lambda =-e-F,\quad\quad & \mfsb \bar a= \bar \Gamma.
\end{matrix}\label{sBbarext}
\ee
Moreover we have the CF-like condition
\be
\overline K-K = d(e-\bar a).\label{CFlike}
\ee
These relations coincide with those of the 1-gerbe provided we make the following replacements:
$\gamma \to \mu,\,  \bar \gamma \to- \bar \mu,\, a\to -\bar \beta, \,\bar a \to g, \,e \to \bar g,\, \bar e \to -\beta$ and $\Gamma\to -\rho,\, \overline \Gamma\to \bar \rho$.

There is only one difference: the presence of $F$ in two cases in last lines of both \eqref{sBext} and \eqref{sBbarext}. This is an irrelevant term as it belongs to the kernel of both $s$ and $\sb$.

{\bf Remark.} One can also impose the horizontality condition $\widetilde {\cal A} -\tilde d\,\tilde {\sf f}= {\cal A} -d\,  {\sf f}$, but this doesn't change much the final result: in fact, the resulting 1-gerbe algebra is the same.

\section{Diffeomorphisms and the superfield formalism}
 
After the successful extension of the superfield formalism to gerbes, we wish to deal with an entirely different type of symmetry, the diffeomorphisms. Our aim is 
to answer a few questions:
\begin{itemize}
\item is the superfield formalism applicable to diffeomorphisms?
\item what are the horizontality conditions for the latter?
\item what are the CF conditions?
\item can we generalize the Riemannian geometry to the superspace?
\end{itemize}
In the sequel we will answer all these questions. The answer to the last question will be partly negative, because an inverse supermetric does not exist. But it is nevertheless possible to develop a superfield formalism in the horizontal (commuting) directions.

A first proposal of a superfield formalism for diffeomorphisms was made by 
\cite{Delbourgo}. Here we present another approach, started in \cite{bonora}, closer in spirit to the standard (commutative) geometrical approach. 

Diffeomorphisms, or general coordinate transformations, are given in terms 
generic (smooth) functions of $x^\mu$, 
\be
x^\mu \to x^{'\mu}=f^\mu(x).\0
\ee
An infinitesimal diffeomorphism is defined by means of a local parameter 
$\xi^\mu(x)$: $f^\mu(x)= x^\mu- \xi^\mu(x)$. In a quantized theory this is 
promoted to an anticommuting field and the BRST transformations for a scalar field, a vector field, the metric and $\xi$, respectively, are
\be 
&&\delta_\xi \varphi= \xi^\lambda \partial_\lambda\varphi,\label{diffphi}\\
&& \delta_\xi A_\mu = \xi^\lambda \partial_\lambda A_\mu + \partial_\mu 
\xi^\lambda A_\lambda,\label{diffAmu}\\
&& \delta_\xi g_{\mu\nu}= \xi^\lambda \partial_\lambda g_{\mu\nu} + 
\partial_{\mu} \xi^\lambda g_{\lambda\nu}+
\partial_{\nu} \xi^\lambda g_{\mu\lambda},\label{dsiffg}\\
&& \delta_\xi \xi^\mu = \xi^\lambda \partial_\lambda \xi^\mu,\label{diffxi}
\ee
It is easy to see that these transformations are nilpotent. We wish now to define the analogs of anti-BRST transformations. To this end 
we introduce another anticommuting field, $\xib$, and a $\delta_{\xib}$ 
transformation, which transforms a scalar, vector, the metric and $\bar\xi$ 
in just the same way as $\delta_\xi$ (these transformations will not be rewritten here). In addition we have the cross-transformations
\be 
&&\delta_\xi {\xib}^\mu=  b^\mu,  \quad\quad  \delta_{\xib} \xi^\mu= \bar 
b^\mu,\label{xixib}\\
&&\delta_\xi b^\mu =0,\quad\quad \delta_\xi \bar b^\mu= -\bar b \!\cdot\! 
\partial \xi^\mu+ \xi  \!\cdot\! \partial \bar b^\mu,\label{xib}\\
&&\delta_{\xib} \bar b^\mu =0,\quad\quad \delta_{\xib} b^\mu= - b \!\cdot\! 
\partial {\xib}^\mu+ \xib  \!\cdot\! \partial b^\mu,\label{xibb}
\ee

It follows that the overall transformation $\delta_\xi +\delta_{\xib}$ is nilpotent:
\be
(\delta_\xi +\delta_{\xib})^2=0.\0
\ee

\subsection{The superfield formalism}

Our aim now is to reproduce the above transformations by means of the superfield formalism.
The superspace coordinates are $X^M=(x^\mu,\vt,\vtb)$, where $\vartheta,\vtb$ are 
the same anticommuting variables as above.
A (super)diffeomorphism is represented by a superspace transformation 
$X^M=(x^\mu,\vartheta,\vtb)\rightarrow \tilde X^M=(F^\mu (X^M), 
\vartheta,\vtb)$, where \footnote{There are also more general superdiffeomorphisms, which 
we ignore here.},
\be
F^\mu(X^M) =f^\mu(x)-\vt \,\xib^\mu -\vtb\, \xi^\mu(x) + \vt\vtb\, 
h^\mu(x)\label{FxM}.
\ee   
Here $f^\mu(x)$ is an ordinary diffeomorphism, $\xi,\xib$ are the generic 
anticommuting functions introduced before and $h^\mu$
a generic commuting one.

The horizontality condition is formulated by selecting appropriate {\it 
invariant geometric expressions} in ordinary
spacetime and identifying them with the same expressions extended to the 
superspace. To start with we work out explicitly the case of a scalar field.
 
\subsection{The scalar}

The diffeomorphism transformation properties of an ordinary scalar field are 
\be
\tilde\varphi(f^\mu(x))=\varphi(x^\mu).\label{difftransf}
\ee

Now we embed the scalar field $\varphi$ in a superfield\footnote{Whenever possible we use Greek letters for anticommuting auxiliary fields and Latin letters for commuting ones.}
\be
\Phi (X)= \varphi(x) + \vartheta\,\bar \beta(x)+\vtb\, \beta(x)+\vt\vtb 
\,C(x),\label{PhiF}
\ee
The BRST interpretation is $\delta_\xi = \frac {\partial}{\partial \vtb}\Big{\vert}_{\vt=0},
\delta_{\xib} = \frac {\partial}{\partial \vt}\Big{\vert}_{\vtb=0}$.
The horizontality condition, suggested by \eqref{difftransf}, is
\be
\Phi (F(X))= \varphi(x).\label{horizonscalar}
\ee
Using (\ref{FxM}) with $f(x^\mu)=x^\mu$,
this becomes 
\be
\Phi(F(X))&=& \varphi(x) - (\vartheta\, {\xib}(x)+
\vtb\,\xi(x)-\vt\vtb\, h(x) )\!\cdot\!\partial\varphi(x) 
+\vt \left( \bar \beta(x) -\vtb \xi(x)\!\cdot\! \partial\bar \beta(x) 
\right)\label{scalar1}\\ 
&&
+ \bar\vartheta\left( \beta(x) - \vartheta\bar\xi \!\cdot\!\partial \beta(x)\right)+ 
\vt\vtb\left( C(x)- {\xib}^\mu \xi^\nu \partial_\mu \partial_\nu 
\varphi(x)\right)\0\\
&=&\varphi(x) +\vartheta\left( \bar \beta(x) 
-{\xib}\!\cdot\!\partial\varphi(x)\right)+ \bar \vartheta\left(\beta(x)-
\xi \!\cdot\!\partial\varphi(x)\right) \0\\
&&+ \vt\vtb \left(C(x) - \xi \!\cdot\!\partial\bar \beta(x)+ {\xib} 
\!\cdot\!\partial \beta(x)+h(x) \!\cdot\! \partial \varphi(x)- {\xib}^\mu \xi^\nu 
\partial_\mu \partial_\nu \varphi(x) \right),\0
\ee
where $\cdot$ denotes index contraction. Then \eqref{horizonscalar} implies
\be 
\beta(x)&= &\xi \!\cdot\!\partial \varphi(x),\quad\quad \bar\beta(x)= {\xib}\!\cdot\! 
\partial \varphi(x),\0\\
C(x) &=&\xi\!\cdot\!\partial\bar \beta(x) - {\xib}\cdot\!\partial \beta(x) -\xi 
{\xib}\partial^2 \varphi(x)-h(x) \!\cdot\! \partial \varphi(x),\label{scalar2}
\ee
where $ \xi {\xib}\partial^2 \varphi(x)= \xi^\mu {\xib}^\nu 
\partial_\mu\partial_\nu \varphi(x)$.
Now the BRST interpretation implies
\be
\delta_{\xib} \varphi(x) = \bar \beta(x)= {\xib}\!\cdot\! \partial 
\varphi(x),\quad\quad  \delta_{\xi} \varphi(x) = \beta(x)=\xi \!\cdot\!\partial 
\varphi(x), \label{nilpotentscalar1}
\ee 
and $\delta_\xi \bar \beta(x)= C(x), \delta_{\xib} 
\beta(x)=-C(x)$.
Inserting $\beta$ and $\bar \beta$ into $C$ in (\ref{scalar2})  we obtain
\be
\delta_\xi \delta_{\xib} \varphi =b \!\cdot\!  \partial \varphi -\xib \!\cdot\! 
\partial \xi \!\cdot\! \partial\varphi-
\xib \xi \partial^2 \varphi.\label{dxidixibfi}
\ee
This coincides with the expression of $C$, \eqref{scalar2}, if 
\be
h^\mu = -b^\mu +\xi\!\cdot\partial{\xib}^\mu.\label{h1}
\ee
Likewise 
\be
\delta_{\xib} \delta_{\xi} \varphi =\bar b \!\cdot\!  \partial \varphi -\xi 
\!\cdot\! \partial \xib \!\cdot\! \partial\varphi-
\xi \xib \partial^2 \varphi,\label{dxibdixifi}
\ee
which coincides with the expression of  $-C$, \eqref{scalar2}, if 
\be
h^\mu = \bar b^\mu -\xib\!\cdot\partial{\xi}^\mu.\label{h2}
\ee
Equating \eqref{h1} with \eqref{h2} we get
\be
h^\mu(x) =-b^\mu(x) + \xi(x)\!\cdot\! \partial {\xib}^\mu(x)= \bar b^\mu(x) - 
\bar\xi(x)\!\cdot\! \partial {\xi}^\mu(x),
\label{nilpotentscalar}
\ee
which is possible if and only iff the CF condition
\be
b^\mu+\bar b^\mu= \xi^\lambda \partial_\lambda \bar \xi^\mu +  \bar\xi^\lambda 
\partial_\lambda \xi^\mu,\0
\ee 
is satisfied.  This condition is consistent, for applying $\delta_\xi$ and $\delta_{\xib}$ 
to both sides we obtain the same result. As we shall see, this condition is, so to speak, universal: it appears whenever BRST and anti-BRST diffeomorphisms are involved and it is the only required condition.

\subsection{The vector}

We now extend the previous approach to a vector field. In order to apply the horizontality 
condition we must first identify the appropriate expression. This is a 1-superform:
\be 
{\mathbb A}\equiv {\cal A}_M(X) dX^M = A_\mu (X) dx^\mu + A_\vartheta(X) d\vartheta+A_{\vtb}(X) 
d\vtb, \label{vector1}
\ee
where
\be
{\cal A}_\mu(X) &=&A_\mu(x)+ \vartheta\,\bar \phi_\mu (x)+ \vtb \phi_\mu(x) 
+\vt\vtb B_\mu(x),\label{Amu1}\\
{\cal A}_\vartheta (X)&=& \chi(x)+ \vartheta \bar C(x) +\vtb C(x) +\vt\vtb 
\psi(x),  \label{Atheta}\\
{\cal A}_{\vtb} (X)&=& \omega(x) + \vartheta \bar D(x) +\vtb D(x)+ \vt\vtb 
\rho(x).  \label{Athetabar}
\ee

According to our prescription horizontality means that
\be
{\cal A}_M(\tilde X) \tilde d\tilde X^M = A_\mu (x) dx^\mu, 
\label{horizonvector}
\ee
where $\tilde d = \frac {\partial} {\partial x^\mu} dx^\mu+ 
\frac{\partial}{\partial \vartheta} d\vartheta+ 
\frac{\partial}{\partial \bar\vartheta} 
d\bar\vartheta$. Thus, we obtain the following:
\be
\tilde d\tilde X^M &=&\Big{(} dx^\mu -\vt\, \partial_\lambda {\xib}^\mu 
dx^\lambda - 
\vtb \,\partial_\lambda \xi^\mu dx^\lambda+ \vt\vtb \partial_\lambda h^\mu 
dx^\lambda \0\\
&&\quad\quad  -({\xib}^\mu-\vtb h^\mu) d\vt 
- ({\xi}^\mu+\vt h^\mu) d\vtb,\, d\vt,\, d\vtb\Big{)}.\label{dXM}
\ee
It remains for us to expand the LHS of \eqref{horizonvector}. The explicit expression can be found in Appendix B. The commutation prescriptions are: $x^\mu , \vartheta,\vtb, \xi^\mu$ commute with $dx^\mu$ and $d\vartheta, d\vtb$; $\xi^\mu,\bar\xi^\mu$ anticommute with $\vartheta,\vtb$. 
From (\ref{vector3}) we obtain the following identifications:
\be
\phi_\mu &=&  \xi\!\cdot\!\partial A_\mu + \partial_\mu \xi^\lambda 
A_\lambda,\label{vectorphi1}\\
 \bar\phi_\mu &=&  \bar\xi\!\cdot\!\partial A_\mu + \partial_\mu \bar\xi^\lambda 
A_\lambda,\label{vectorphi2}\\
B_\mu &=&  \xi\!\cdot\!\partial \bar\phi_\mu- \bar\xi\!\cdot\!\partial \phi_\mu  
-\xi{\xib}\!\cdot\! \partial^2 A_\mu 
+\partial_\mu\bar\xi^\lambda  \xi\!\cdot\!\partial A_\lambda- 
\partial_\mu\xi^\lambda  {\xib}\!\cdot\!\partial A_\lambda\0\\
&&- \partial_\mu\bar\xi^\lambda \phi_\lambda + \partial_\mu\xi^\lambda \bar 
\phi_\lambda-h\!\cdot\!\partial A_\mu-\partial_\mu h\!\cdot\! A 
,\label{vectorphi3}
\ee
\be
\chi &=&A_\mu{\xib}^\mu, \label{vectorchi1}\\
C&=&  -\xi\!\cdot\!\partial A_\mu{\xib}^\mu +\phi_\mu {\xib}^\mu + 
\xi\!\cdot\!\partial\chi-h\!\cdot\!  A,\label{vectorchi2}\\
\bar C &=&  -{\xib}\!\cdot\!\partial A_\mu{\xib}^\mu +\bar\phi_\mu {\xib}^\mu + 
{\xib}\!\cdot\!\partial\chi,\label{vectorchi3}\\
\psi &=& \xi{\xib}\!\cdot\! \partial^2 A_\mu{\xib}^\mu - \xi\!\cdot\!\partial 
\bar\phi_\mu{\xib}^\mu+{\xib}\!\cdot\!\partial \phi_\mu {\xib}^\mu 
+B_\mu {\xib}^\mu-\xi{\xib}\!\cdot\! \partial^2\chi+ \xi\!\cdot\!\partial \bar C 
-{\xib}\!\cdot\!\partial C\label{vectorchi4}\\
&& -\bar\xi\!\cdot\!\partial A\!\cdot\! h+ \bar\phi\!\cdot\! h - 
h\!\cdot\!\partial  \chi+h\!\cdot\!\partial A_\mu \bar\xi^\mu,\0
\ee
and
\be
\omega &=&A_\mu\xi^\mu,\label{vectorchib1}\\
D&=&  -\xi\!\cdot\!\partial A_\mu\xi^\mu +\phi_\mu \xi^\mu + 
\xi\!\cdot\!\partial\omega,\label{vectorchib2}\\
\bar D &=&  -{\xib}\!\cdot\!\partial A_\mu\xi^\mu +\bar\phi_\mu \xi^\mu + 
{\xib}\!\cdot\!\partial\omega+h\!\cdot\! A  ,\label{vectorchib3}\\
\rho &=& \xi{\xib}\!\cdot\! \partial^2 A_\mu\xi^\mu - \xi\!\cdot\!\partial 
\bar\phi_\mu\xi^\mu+{\xib}\!\cdot\!\partial \phi_\mu \xi^\mu 
+B_\mu \xi^\mu-\xi{\xib}\!\cdot\! \partial^2\omega+ \xi\!\cdot\!\partial \bar D 
-{\xib}\!\cdot\!\partial D.\label{vectorchib4}\\
&& -\xi\!\cdot\!\partial A\!\cdot\! h+ \phi\!\cdot\! h - h\!\cdot\!\partial  
\omega+h\!\cdot\! \partial A_\mu \xi^\mu.\0
\ee

One can see that
\be
\phi_\mu= \delta_\xi A_\mu,\quad\quad \bar \phi_\mu = \delta_{\xib} A_\mu, 
\quad\quad B_\mu = \delta_{\xib} \phi_\mu=-\delta_\xi \bar\phi_\mu.
\label{vectorfinal1}
\ee
\be
D=\delta_\xi\omega, \quad\quad \bar D=\delta_{\xib}\omega, \quad\quad \rho= 
-\delta_{\xib} D = \delta_\xi \bar D,\label{vectorfinal2}
\ee
and
\be
C=\delta_\xi\chi, \quad\quad \bar C=\delta_{\xib}\chi, \quad\quad \psi= 
-\delta_{\xib} C = \delta_\xi \bar C, \label{vectorfinal4}
\ee
provided 
\be
h^\mu(x) =-b^\mu(x) + \xi(x)\!\cdot\! \partial {\xib}^\mu(x)= \bar b^\mu(x) - 
\bar\xi(x)\!\cdot\! \partial {\xi}^\mu(x),
\label{nilpotentscalar3}
\ee
which is possible if and only iff the CF condition
\be
b^\mu+\bar b^\mu= \xi^\lambda \partial_\lambda \bar \xi^\mu +  \bar\xi^\lambda 
\partial_\lambda \xi^\mu,\0
\ee 
is satisfied. In particular $\rho$ can be rewritten as
\be 
\rho&=& \xi\!\cdot\!\partial {\xib}\!\cdot\!\partial A_\mu \xi^\mu - 
{\xib}\!\cdot\!\partial \xi\!\cdot\!\partial A_\mu \xi^\mu+  
{\xib}\!\cdot\!\partial\xi^\mu \,\xi\!\cdot\!\partial A_\mu \xi^\mu\0\\
&&-  
\xi\!\cdot\!\partial{\xib}^\mu \,\xi\!\cdot\!\partial A_\mu \xi^\mu+ 
\xi{\xib}\!\cdot\! \partial^2 A_\mu \xi^\mu -\xi\!\cdot\!\partial A\!\cdot\! h+ 
\phi\!\cdot\! h - h\!\cdot\!\partial  \omega\0.\label{vectorfinal3}
\ee

\subsection{The metric}

A most important field for theories invariant under diffeomorphisms is the metric $g_{\mu\nu}(x)$.
To represent its BRST transformation properties in the superfield formalism, we embed it in a 
supermetric $ G_{MN}(X)$
and form the symmetric 2-superdifferential
\be
{\mathbb G}= G_{MN}(X) \tilde d X^M \vee \tilde dX^N, \label{supermetric}
\ee
where $\vee$ denotes the symmetric tensor product and
\be
G_{\mu\nu}(X) &=& g_{\mu\nu}(x) +\vartheta\, \bar\Gamma_{\mu\nu}(x)+\vtb 
\Gamma_{\mu\nu}(x)+ \vt\vtb V_{\mu\nu}(x),\0\\
G_{\mu\vartheta}(X)&=& \gamma_\mu(x)+\vartheta \, \bar g_\mu(x)+ \vtb g_\mu(x)+ 
\vt\vtb \Gamma_\mu(x)= G_{\vartheta\mu}(X),\0\\
G_{\mu\bar\vartheta}(X)&=&\bar\gamma_\mu(x)+\vartheta \, \bar f_\mu(x)+ \vtb 
f_\mu(x)+ \vt\vtb \bar\Gamma_\mu(x) = G_{\bar\vartheta\mu}(X),\label{superG}\\
\frac 12 G_{\vt\vtb}(X)&=& g(x) +\vartheta \, \bar \gamma(x)+ \vtb \gamma(x)+ \vt\vtb 
G(x) = -\frac 12 G_{\vtb\vt}(X),
\ee
while $G_{\vt\vt}(X)=0=G_{\vtb\vtb}(X)$, because the symmetric tensor product 
becomes antisymmetric for anticommuting variables: 
$d\vt \vee d\vt=0=d\vtb\vee d\vtb, d\vt\vee d\vtb=-d\vtb\vee d\vt$.

The horizontality condition is obtained by requiring:
\be
\widetilde G_{MN}(\tilde X) \tilde d\tilde X^M \vee \tilde d\tilde X^N = g_{\mu\nu}(x) 
dx^\mu \vee dx^\nu.\label{horizontalmetric}
\ee
The explicit expression of the LHS of this equation can be found again in Appendix B. From which the following identification follow 
\be 
\Gamma_{\mu\nu} &=& \xi\! \cdot \! \partial g_{\mu\nu} + \partial_\mu 
\xi^\lambda g_{\lambda\nu}+
\partial_\nu \xi^\lambda g_{\mu\lambda}=\delta_\xi g_{\mu\nu},\label{metric1}\\
\bar\Gamma_{\mu\nu} &=& \bar\xi\! \cdot \! \partial g_{\mu\nu} + \partial_\mu 
\bar\xi^\lambda g_{\lambda\nu}+
\partial_\nu \bar\xi^\lambda g_{\mu\lambda}=\delta_{\xib} 
g_{\mu\nu},\label{metric2}\\
V_{\mu\nu} &=& -\xi{\xib}\!\cdot\! \partial^2 g_{\mu\nu} +  \xi\! \cdot \! 
\partial \bar\Gamma_{\mu\nu}
+ \partial_\mu \xi^\lambda \bar\Gamma_{\lambda\nu}+\partial_\nu \xi^\lambda \bar 
\Gamma_{\mu\lambda}
-  \xib\! \cdot \! \partial \Gamma_{\mu\nu}-  \partial_\mu {\xib}^\lambda 
\Gamma_{\lambda\nu}
-\partial_\nu {\xib}^\lambda \Gamma_{\mu\lambda}\label{metric3}\\
&&+ \partial_\mu {\xib}^\lambda \xi\! \cdot \! \partial g_{\lambda\nu} + 
\partial_\nu {\xib}^\lambda \xi\! \cdot \! \partial g_{\mu\lambda}
- \partial_\mu \xi^\lambda \bar\xi\! \cdot \! \partial g_{\lambda\nu} - 
\partial_\nu \xi^\lambda \bar\xi\! \cdot \! \partial g_{\mu\lambda}+\partial_\mu {\xib}^\lambda \partial_\nu \xi^\rho g_{\lambda\rho}\0\\
&& +\partial_\nu {\xib}^\lambda \partial_\mu \xi^\rho 
g_{\lambda\rho}-h\!\cdot\!\partial g_{\mu\nu}-\partial_\mu h^\lambda 
g_{\lambda\nu}- \partial_\nu h^\lambda g_{\mu\lambda}\0\\
&=& \delta_\xi \bar\Gamma_{\mu\nu} =- \delta_{\xib} \Gamma_{\mu\nu}.\0 
\ee
Moreover,
\be
\gamma_\mu &=& g_{\mu\nu} \bar\xi^\nu,\label{metric4}\\
g_\mu &=& \partial_\mu \xi^\lambda g_{\lambda\nu}\bar\xi^\nu +\partial_\nu 
\xi^\lambda g_{\mu\lambda}\bar\xi^\nu 
+  \xi\! \cdot \! \partial g_{\mu\nu} \bar\xi^\nu+g_{\mu\nu} \xi\!\cdot\! 
\partial \bar\xi^\nu-g_{\mu\nu}h^\nu=\delta_{\xi} \gamma_\mu,\label{metric5}\\
\bar g_\mu& =& \xib\! \cdot \! \partial g_{\mu\nu} \bar\xi^\nu+ \partial_\mu 
\bar\xi^\lambda g_{\lambda\nu}\bar\xi^\nu =\delta_{\xib} 
\gamma_\mu,\label{metric6}\\
\Gamma_\mu &=& -\xi{\xib}\!\cdot\! \partial^2 \gamma_{\mu} +  \xi\! \cdot \! 
\partial \bar g_{\mu}
-\partial_\mu \bar\xi^\lambda g_{\lambda}- \xib\! \cdot \! \partial  
g_{\mu}+\partial_\mu \xi^\lambda \bar g_{\lambda}
 +\partial_\mu \bar\xi^\lambda \xi\! \cdot \! \partial \gamma_\lambda - 
\partial_\mu \xi^\lambda \xib\! \cdot \! \partial \gamma_\lambda \0\\
&&-\bar\xi\!\cdot\! \partial  h^\nu g_{\mu\nu}-\partial_\mu h^\lambda 
g_{\lambda\nu} \bar \xi^\nu+\bar \Gamma_{\mu\nu} h^\nu -h \!\cdot\!\partial\gamma_\mu - 
g_{\nu\lambda} h^\nu \partial_\mu\bar \xi^\lambda\0\\
&&  +\partial_\nu \xi^\lambda \overline \Gamma_{\mu\lambda} \bar \xi^\nu -\partial_\nu \bar\xi^\lambda \Gamma_{\mu\lambda} \bar \xi^\nu+\bar\xi \!\cdot\! \partial\bar\xi^\lambda\, \xi\!\cdot\! \partial g_{\mu\lambda}\0\\
&&-\bar\xi\! \cdot\! \partial\xi^\lambda\, \bar\xi\!\cdot\! \partial g_{\mu\lambda}-\bar\xi\! \cdot\! \partial\xi^\lambda\, g_{\lambda\rho} \partial_\mu \bar\xi^\rho+\bar\xi \!\cdot\! \partial\bar \xi^\lambda \,g_{\lambda\rho} \partial_\mu\xi^\rho
 =-\delta_{\xib}g_\mu\,\,=\,\, \delta_\xi \bar g_\mu,  \label{metric7}
\ee 
and
\be
\bar\gamma_\mu &=& g_{\mu\nu} \xi^\nu,\label{metric12}\\
\bar f_\mu &=& \partial_\mu \bar\xi^\lambda g_{\lambda\nu}\xi^\nu +\partial_\nu 
\bar\xi^\lambda g_{\mu\lambda}\xi^\nu 
+  \bar\xi\! \cdot \! \partial g_{\mu\nu} \xi^\nu+g_{\mu\nu} \bar\xi\!\cdot\! 
\partial \xi^\nu+g_{\mu\nu} h^\nu=\delta_{\xib} 
\bar\gamma_\mu,\label{metric13}\\
f_\mu& =& \xi\! \cdot \! \partial g_{\mu\nu} \xi^\nu+ \partial_\mu \xi^\lambda 
g_{\lambda\nu}\xi^\nu 
=\delta_{\xi} \bar\gamma_\mu,\label{metric14}\\
\bar\Gamma_\mu&=&\!\!-\xi\!\cdot\! \partial g_{\mu\nu} h^\nu 
-\partial_\mu h^\lambda g_{\lambda\nu} \xi^\nu+ \Gamma_{\mu\nu} h^\nu -h 
\!\cdot\! \partial \bar\gamma_\mu - \partial_\mu h^\lambda \bar\gamma_\lambda -\xi\!\cdot\! \partial h^\lambda g_{\lambda\mu}\0\\
&&+\partial_\mu \xi_\lambda \bar f^\lambda+\xi \!\cdot\! \partial \bar f_\mu-\bar \xi \!\cdot\! \partial  f_\mu-\xi^\lambda \xib^\rho \partial_\lambda\partial_\rho \bar\gamma_\mu-\partial_\mu \xib^\lambda \partial_\lambda \xi^\rho\bar\gamma_\rho\0\\
&&-\xib\!\cdot\! \partial\bar \gamma_\lambda \partial_\mu \xi^\lambda+ \partial_\mu \xib^\lambda \xi\!\cdot\! \partial\xi^\rho g_{\lambda\rho}-\xi\!\cdot\! \partial\xib^\rho \partial_\rho \xi^\lambda g_{\lambda\mu}+\xi \!\cdot\!\partial \xi^\rho \partial_\rho \xib^\lambda g_{\lambda\mu}\0\\
&=& \delta_\xi \bar f_\mu\,\,=\,\, -\delta_{\xib} f_\mu.\label{metric15}
\ee
Finally, we obtain 
\be 
g&=&g_{\mu\nu} \left({\xib}^\mu \xi^\nu-\xi^\mu{\xib}^\nu\right)= 2 g_{\mu\nu} 
{\xib}^\mu \xi^\nu,\label{metric16}\\
\gamma &=& 2 \xi\!\cdot\! \partial g_{\mu\nu} {\xib}^\mu \xi^\nu+2 g_{\mu\nu} 
\left(  \xi\!\cdot\! \partial\bar\xi^\mu- 
\bar\xi\!\cdot\! \partial\xi^\mu\right) \xi^\nu -2\bar\gamma_\mu 
h^\mu=\delta_\xi g,\label{metric17}\\
\bar\gamma &=& 2 \bar\xi\!\cdot\! \partial g_{\mu\nu} {\xib}^\mu \xi^\nu+2 
g_{\mu\nu} \left(  \xi\!\cdot\! \partial\bar\xi^\mu- 
\bar\xi\!\cdot\! \partial\xi^\mu\right) \bar\xi^\nu-2\gamma_\mu h^\mu 
=\delta_{\xib} g,\label{metric18}\\
G&=&\!\!2b^\mu \partial_\mu g +2 b^\mu g_\mu-2\xib \!\cdot\! \partial\gamma    -2\xib \!\cdot\! \partial \xib^\mu f_\mu -2b \!\cdot\! \partial \xib^\rho \bar \gamma_\rho-2 \xib \!\cdot\! \partial b^\rho \bar\gamma_\rho 
=\delta_\xi \bar \gamma\,\,=\,\,  
-\delta_{\xib} \gamma.\label{metric19}
\ee

This completes the verification of the horizontality condition. As expected, it leads to 
identifying the $\vtb$-and  $\vt$-superpartners of the metric as
BRST and anti-BRST transforms.  

\subsection{Inverse of $G_{\mu\nu}(X)$}

A fundamental ingredient of Riemannian geometry is the inverse metric. Therefore in order to see whether a super-Riemannian geometry can be introduced in the supermanifold we have to verify whether an inverse supermetric exists. We start by the inverse of $G_{\mu\nu}(X)$, which is defined  by first writing it as follows
\be
G_{\mu\nu}(X) &=& g_{\mu\lambda}(x)\left[\delta_\nu^\lambda  +g^{\lambda\rho} 
\left(\vartheta\, \bar\Gamma_{\rho\nu}(x)+\vtb \Gamma_{\rho\nu}(x)+ \vt\vtb 
V_{\rho\nu}(x)\right) \right]\equiv g_{\mu\lambda}(x) \left(1 
+X\right)^\lambda{}_\nu,\0
\ee
then in matrix terms
\be
\widehat G^{-1} = \left( 1-X+X^2\right)\hat g^{-1},\0
\ee
 where $\hat g^{-1}$ is the inverse of $g$, i.e.
\be
\widehat G^{\mu\nu} &=&  \left( 1-X+X^2\right)^\mu{}_\lambda \, 
\hat g^{\lambda\nu}\label{Ginverse}\\
&=& \left(\delta^\mu_\lambda - \hat g^{\mu\rho}\left(\vartheta\, 
\bar\Gamma_{\rho\lambda}(x)+\vtb \Gamma_{\rho\lambda}(x)+ \vt\vtb 
V_{\rho\lambda}(x)\right)+\vt\vtb\, \hat g^{\mu\rho}\left( \Gamma_{\rho\sigma} 
g^{\sigma\tau} \bar\Gamma_{\tau\lambda} - \bar \Gamma_{\rho\sigma} 
g^{\sigma\tau} \Gamma_{\tau\lambda}\right) \right)\hat g^{\lambda\nu}\0\\
 &\equiv&\hat g^{\mu\nu}(x) +\vartheta\, \widehat{\bar\Gamma}^{\mu\nu}(x)+\vtb 
\widehat{\Gamma}^{\mu\nu}(x)+ \vt\vtb \widehat{V}^{\mu\nu}(x),\0
\ee
and $\hat g^{\mu\nu}$ is the ordinary metric inverse. Moreover
\be
\widehat{\Gamma}^{\mu\nu}&=&-\hat g^{\mu\lambda} \Gamma_{\lambda\rho} \hat g^{\rho\nu},
\label{InverseGamma}\\
\widehat{\bar\Gamma}^{\mu\nu}&=&- \hat g^{\mu\lambda} \bar\Gamma_{\lambda\rho}\hat 
g^{\rho\nu},\label{InversebarGamma}\\
 \widehat{V}^{\mu\nu}&=&  \hat g^{\mu\lambda}\left( 
-V_{\lambda\rho}+\Gamma_{\lambda\sigma}\hat g^{\sigma\tau} \bar\Gamma_{\tau\rho} -
\bar \Gamma_{\lambda\sigma} \hat g^{\sigma\tau} \Gamma_{\tau\rho}  
\right)\hat g^{\rho\nu}.\label{InverseV}
\ee
This contains the correct BRST transformation properties. For instance, we have:
\be
\widehat\Gamma^{\mu\nu} &=&-  \hat g^{\mu\lambda}\left( \xi\! \cdot \! \partial 
g_{\lambda\rho} + \partial_\lambda \xi^\tau g_{\tau\rho}+
\partial_\rho \xi^\tau g_{\lambda\tau}\right)\hat g^{\rho\nu}\label{widehatGamma}\\
&=& \xi\! \cdot \! \partial \hat g^{\mu\nu} - \partial_\lambda \xi^\mu \hat g^{\lambda\nu} 
-\partial_\lambda \xi^\nu \hat g^{\nu\lambda}=\delta_{\xi}\hat g^{\mu\nu}.\0
\ee
Similarly, we obtain
\be
\widehat{\bar \Gamma}^{\mu\nu}&=&\delta_{\bar\xi}\hat g^{\mu\nu},\label{widehatbarGamma}\\
\widehat V^{\mu\nu}&=&\delta_{\xi}  \delta_{\bar\xi}\hat g^{\mu\nu}.\label{widehatV}
\ee
The simplest way to obtain \eqref{widehatV} is to proceed as follows 
\be
\delta_{\xi}  \delta_{\bar\xi} \hat g^{\mu\nu}&=& \delta_{\xi}  \delta_{\bar\xi}\left( \hat g^{\mu\lambda} g_{\lambda\rho} \hat g^{\rho\nu} \right)\label{deltaxixibargmunu}\\
&=& \delta_{\xi}\left( -  \hat g^{\mu\lambda} \delta_{\bar\xi} g_{\lambda\rho}\, \hat g^{\rho\nu}\right)\0\\
&=& -\hat  g^{\mu\lambda}\delta_\xi \delta_{\bar\xi} g_{\lambda\rho}\,\hat   g^{\rho\nu}
+ \hat  g^{\mu\lambda} \delta_{\xi} g_{\lambda\rho}\,  \hat g^{\rho\sigma}\delta_{\bar\xi} g_{\sigma\tau}\,\hat g^{\tau\nu}-\hat g^{\mu\lambda} \delta_{\bar\xi} g_{\lambda\rho}\,  \hat g^{\rho\sigma}\delta_{\xi} g_{\sigma\tau}\,\hat g^{\tau\nu}\0\\
&=& - \hat  g^{\mu\lambda}V_{\lambda\rho}\,  \hat g^{\rho\nu}
+\hat   g^{\mu\lambda} \Gamma_{\lambda\rho}\,\hat   g^{\rho\sigma}\bar \Gamma_{\sigma\tau}\,\hat g^{\tau\nu}-\hat g^{\mu\lambda} \bar \Gamma_{\lambda\rho}\, \hat  g^{\rho\sigma} \Gamma_{\sigma\tau}\,\hat g^{\tau\nu}.\0
\ee

\subsection{$\widehat G^{MN}$}

Now we are ready to tackle the problem of the supermetric inverse.
In ordinary Riemannian geometry the inverse $\hat g^{\mu\nu}$ of the metric is defined by:
$\hat g^{\mu\lambda}g_{\lambda\nu}= \delta^\mu_\nu$. But $\hat g^{\mu\nu}$ can also be considered as a bi-vector such that,
\be
\hat {\mathfrak g}=\hat g^{\mu\nu} \frac {\partial}{\partial x^\mu}\vee  \frac {\partial}{\partial x^\nu},\label{bivector}
\ee 
is invariant under diffeomorphisms. 

We can try to define the analog of \eqref{bivector} in the superspace, i.e.
\be
\widehat {\mathbb G}= \widehat G^{MN}(X) \frac{\tilde\partial}{\tilde\partial X^M} \vee \frac {\tilde\partial}{\tilde\partial X^N},\label{mathfrakG}
\ee
where $  \frac{\tilde\partial}{\tilde\partial X^M} = \left( \frac {\partial}{\partial x^\mu}, \frac {\partial}{\partial \vt},  \frac {\partial}{\partial \vtb}\right)$, and
\be
\widehat G^{\mu\nu}(X) &=&\hat g^{\mu\nu}(x) +\vartheta\, \widehat{ \bar\Gamma}^{\mu\nu}(x)+\vtb 
\widehat \Gamma^{\mu\nu}(x)+ \vt\vtb \widehat V^{\mu\nu}(x),\0\\
\widehat G^{\mu\vartheta}(X)&=&\hat \gamma^\mu(x)+\vartheta \, \hat{\bar g}^\mu(x)+ \vtb \hat g^\mu(x)+ 
\vt\vtb\widehat\Gamma^\mu(x)= \widehat G^{\vartheta\mu}(X),\0\\
\widehat G^{\mu\bar\vartheta}(X)&=&\hat{\bar\gamma}^\mu(x)+\vartheta \, \hat {\bar f}^\mu(x)+ \vtb 
\hat f^\mu(x)+ \vt\vtb \widehat{\bar\Gamma}^\mu(x) =\widehat G^{\bar\vartheta\mu}(X),\0\\
\frac 12 \widehat G^{\vt\vtb}(X)&=& \hat g(x) +\vartheta \,\hat{ \bar \gamma}(x)+ \vtb\hat \gamma(x)+ \vt\vtb \widehat
G(x) = -\frac 12\widehat G^{\vtb\vt}(X).\label{superGhat}
\ee

This suggests immediately the horizontality condition $\widehat {\mathfrak G}=\hat {\mathfrak g}$, i.e.
\be
\widetilde{\widehat  G}^{MN}(\tilde X) \frac{\partial}{\tilde \partial \tilde X^M} \vee \frac {\tilde\partial}{\tilde \partial \tilde X^N}= \hat g^{\mu\nu}(x) \frac {\partial}{\partial x^\mu}\vee  \frac {\partial}{\partial x^\nu}.\label{horhatGhatg}
\ee 

The partial derivative $\frac{\partial}{\tilde \partial \tilde X^M}$ can be derived from $\tilde d \tilde X$ by inverting the relation:
\be
\tilde d \tilde X^M = \left( \begin{matrix} \delta^\mu_\nu -\vt \partial_\nu {\xib}^\mu -\vtb \partial_\nu \xi^\mu+\vt\vtb \partial_\nu h^\mu & -{\xib}^\mu +\vtb h^\mu& -\xi^\mu -\vt h^\mu\\0&1&0 \\
0&0&1\end{matrix}\right) \left( \begin{matrix} dx^\nu\\ d\vt\\d\vtb\end{matrix}\right).\label{dXtilde=YdX}
\ee
 The matrix has the structure $\left(\begin{matrix} A&B\\C&D\end{matrix} \right)$, where $A,D$ are commuting square matrices, while $B,C$ are anticommuting rectangular ones (in this case $C=0$ and $D=1$). Its inverse is $ \left( \begin{matrix} A ^{-1} & -A^{-1} B\\
0 & 1\end{matrix} \right)$.
Therefore we have the following
\be
\frac {\tilde\partial}{\tilde \partial \tilde x^\mu}\!\!&=&\!\! \frac {\partial}{\partial x^\mu}+ \left(\vt \partial_\mu {\xib}^\nu +\vtb \partial_\mu \xi^\nu - \vt\vtb \left(\partial_\mu h^\nu + \partial_\mu\xib ^\lambda \partial_\lambda \xi^\nu -   \partial_\mu\xi ^\lambda \partial_\lambda \xib^\nu\right)\right) \frac {\partial}{\partial x^\nu},\label{ddxddtddtb}\\
\frac {\tilde\partial}{\tilde \partial \tilde \vt}\!\!& =&\!\! \frac {\partial}{ \partial  \vt} + \left( -\xib^\nu
+\vt \xib\!\cdot \! \partial\xib^\nu + \vtb \left( -h^\nu +\xib\!\cdot\!\partial \xi^\nu \right) +\vt\vtb \left(h\!\cdot\!\partial \xib^\nu -\xib \!\cdot\! \partial h^\nu - \xib \!\cdot\! \partial\xib\!\cdot\! \partial \xi^\nu 
+ \xib \!\cdot\! \partial\xi\!\cdot\! \partial \xib^\nu \right)\right) \frac {\partial}{\partial x^\nu},\0\\
\frac {\tilde\partial}{\tilde \partial \tilde \vtb}\!\!& =&\!\! \frac {\partial}{ \partial  \vtb} + \left( -\xi^\nu
+\vtb \xi\!\cdot \! \partial\xi^\nu + \vt \left( h^\nu +\xi\!\cdot\!\partial \xib^\nu \right) +\vt\vtb \left(h\!\cdot\!\partial \xi^\nu -\xi \!\cdot\! \partial h^\nu - \xi \!\cdot\! \partial\xib\!\cdot\! \partial \xi^\nu 
+ \xi \!\cdot\! \partial\xi\!\cdot\! \partial \xib^\nu \right)\right) \frac {\partial}{\partial x^\nu}.\0
\ee
The explicit form of the  RHS can be found on Appendix B,  see \eqref{horhatGhatg1}, from which  we can now proceed to identify the various fields in \eqref{superGhat}. 

From the $\frac {\partial}{ \partial  \vt}\vee\frac {\partial}{ \partial  \vtb}$ term we get the equation 
\be
  \hat g - \left(\vartheta {\xib}+\vtb \xi-\vt\vtb h\right)\!\cdot\! 
\partial \hat g +  
\vt \vtb\, \xi{\xib}\!\cdot\! \partial^2 \hat g+ \vt \left(\hat{ \bar \gamma} -\vtb \xi 
\!\cdot \! \partial\hat{ \bar \gamma}\right)+
\vtb \left(\hat \gamma -\vt \xi \!\cdot \!\hat \gamma \right)+\vt\vtb \widehat G   =0.\label{dtdtbr}
\ee
from which we deduce
\be
\hat g=0, \quad \hat\gamma=0,\quad \hat{\bar \gamma}=0,\quad \widehat G=0.\label{first0}
\ee
Similarly, from the $\frac {\partial }{\partial x^\mu}\vee \frac {\partial}{ \partial  \vt}$ term we deduce
\be
\hat \gamma^\mu=0,\quad \hat g^\mu=0,\quad \hat{\bar g}^\mu=0,\quad \widehat \Gamma^\mu=0,\label{second0}
\ee
and from $\frac {\partial }{\partial x^\mu}\vee \frac {\partial}{ \partial  \vtb}$ 
\be
\hat {\bar \gamma}^\mu=0,\quad \hat f^\mu=0,\quad \hat{\bar f}^\mu=0,\quad \widehat {\overline\Gamma}^\mu=0.\label{third0}
\ee

Therefore only the components of $\widehat G^{\mu\nu}(X) $  do not vanish. Eq.\eqref{horhatGhatg1} becomes
\be
&& \hat g^{\mu\nu}(x) \frac {\partial}{\partial x^\mu}\vee  \frac {\partial}{\partial x^\nu}=
\widetilde{\widehat  G}^{MN}(\tilde X) \frac{\partial}{\tilde \partial \tilde X^M} \vee \frac {\tilde\partial}{\tilde \partial \tilde X^N}\label{horhatGhatg2}\\
&=&\Big{(} \hat g^{\mu\nu} - \left(\vt {\xib}+\vtb \xi -\vt\vtb 
h\right)\!\cdot\! \partial \hat g^{\mu\nu} +  
\vt \vtb\, \xi{\xib}\!\cdot\! \partial^2 \hat g^{\mu\nu} + \vt \left(\widehat {\overline 
\Gamma}^{\mu\nu} -\vtb \xi \!\cdot \! \partial \widehat{\overline \Gamma}^{\mu\nu}\right) \0\\
&&
\quad\quad+\vtb \left( \widehat\Gamma^{\mu\nu} -\vt \xib \!\cdot \! 
\partial\widehat \Gamma^{\mu\nu}\right)+ \vt\vtb \,\widehat V^{\mu\nu}(x)
\Big{)}\0\\
&&\cdot\left(\frac {\partial}{\partial x^\mu}+ \left(\vt \partial_\mu {\xib}^\lambda +\vtb \partial_\mu \xi^\lambda - \vt\vtb \left(\partial_\mu h^\lambda + \partial_\mu\xib ^\sigma \partial_\sigma \xi^\lambda -   \partial_\mu\xi ^\sigma \partial_\sigma \xib^\lambda\right)\right) \frac {\partial}{\partial x^\lambda}\right)\0\\
&&\vee\left(\frac {\partial}{\partial x^\nu}+ \left(\vt \partial_\nu {\xib}^\rho +\vtb \partial_\nu \xi^\rho - \vt\vtb \left(\partial_\nu h^\rho + \partial_\nu\xib ^\tau \partial_\tau \xi^\rho -   \partial_\nu\xi ^\tau \partial_\tau\xib^\rho\right)\right) \frac {\partial}{\partial x^\rho}\right).\0
\ee
This implies 
\be
\widehat \Gamma^{\mu\nu} &=& \xi\!\cdot\! \partial \hat g^{\mu\nu}- \partial_\lambda \xi^\mu \hat g^{\lambda \nu} -\partial_\lambda \xi^\nu \hat g^{\mu\lambda} = \delta_\xi \hat g^{\mu\nu},\0\\
\widehat {\overline \Gamma}^{\mu\nu} &=& \xib\!\cdot\! \partial \hat g^{\mu\nu}- \partial_\lambda \xib^\mu \hat{  g}^{\lambda \nu} -\partial_\lambda \xib^\nu \hat{  g}^{\mu\lambda} = \delta_\xib \hat g^{\mu\nu},\0\\
\widehat V^{\mu\nu}& =&\delta_\xi\delta_\xib \hat g^{\mu\nu} .\label{GammaGammabarV}
\ee
If we impose $\hat g^{\mu\nu}(x)$ to be the inverse of $g_{\mu\nu}(x)$, these are identical to  eqs. \eqref{InverseGamma}, \eqref{InversebarGamma} and \eqref{InverseV}.

\subsection{Super-Christoffel symbols and super-Riemann tensor} 

From the previous results and from Appendix B.5, it is clear that we cannot define an inverse of $G_{MN}(X)$, therefore we must give up the idea of mimicking the Riemannian geometry in the superspace. However, no obstacles exist if we limit ourselves to $G_{\mu\nu}(X)$. We have seen that its inverse exists. Therefore we can introduce a horizontal Riemannian geometry in the superspace, that is a Riemannian geometry where the involved tensors are horizontal, i.e. they do not have components in the anticommuting directions. To start with, we can define the super-Christoffel symbol as follows
\be
{\mathbf \Gamma}_{\mu\nu}^\lambda &=& \frac 12 \widehat G^{\lambda\kappa} \left( 
\partial_\mu  
G_{\nu\kappa} +\partial_\nu G_{\mu\kappa} -\partial_\kappa  
G_{\mu\nu}\right)\label{superChris}\\
&=& \Gamma_{\mu\nu}^\lambda +\vartheta \overline K_{\mu\nu}^\lambda +\bar\vartheta  K_{\mu\nu}^\lambda+\vartheta\bar\vartheta  H_{\mu\nu}^\lambda,\label{superChris1}
\ee
where
\be
\overline K_{\mu\nu}^\lambda&= &\frac 12 \left(\widehat {\overline \Gamma}^{\lambda \rho} \left( \partial_\mu g_{\rho \nu} + \partial_\nu g_{\mu\rho} -\partial_\rho g_{\mu\nu} \right) + \hat g^{\lambda\rho} \left(  \partial_\mu\overline \Gamma_{\rho \nu} + \partial_\nu \overline\Gamma_{\mu\rho} -\partial_\rho\overline \Gamma_{\mu\nu} \right) \right)\label{barKmunul}\\
&=& \frac 12 \left(\delta_{\bar\xi}\hat g^{\lambda \rho} \left( \partial_\mu g_{\rho \nu} + \partial_\nu g_{\mu\rho} -\partial_\rho g_{\mu\nu} \right) +\hat  g^{\lambda\rho} \left(  \partial_\mu \delta_{\bar\xi}g_{\rho \nu} + \partial_\nu  \delta_{\bar\xi}g_{\mu\rho} -\partial_\rho   \delta_{\bar\xi}g_{\mu\nu} \right) \right)\0\\
&=& \delta_{\bar\xi}\Gamma_{\mu \nu}^\lambda.\0
\ee
Similarly, we note that
\be
K_{\mu\nu}^\lambda =  \delta_{\xi}\Gamma_{\mu \nu}^\lambda,\label{Kmunul}
\ee
and 
\be
H_{\mu\nu}^\lambda&=& \frac 12 \left( \widehat {V}^{\lambda \rho} \left( \partial_\mu g_{\rho \nu} + \partial_\nu g_{\mu\rho} -\partial_\rho g_{\mu\nu} \right) +\hat  g^{\lambda\rho} \left(  \partial_\mu V_{\rho \nu} + \partial_\nu V_{\mu\rho} -\partial_\rho V_{\mu\nu} \right)\right.\label{Hmunu}\\
&&+ \left. \widehat {\Gamma}^{\lambda \rho} \left( \partial_\mu \overline\Gamma_{\rho \nu} + \partial_\nu \overline \Gamma_{\mu\rho} -\partial_\rho  \overline\Gamma_{\mu\nu} \right) - \widehat{ \overline\Gamma}^{\lambda\rho} \left(  \partial_\mu \Gamma_{\rho \nu} + \partial_\nu \Gamma_{\mu\rho} -\partial_\rho \Gamma_{\mu\nu} \right)\right)\0\\
&=& \frac 12 \left( \delta_\xi \delta_{\xib} \hat g^{\lambda \rho} \left( \partial_\mu g_{\rho \nu} + \partial_\nu g_{\mu\rho} -\partial_\rho g_{\mu\nu} \right) +\hat  g^{\lambda\rho} \left(  \partial_\mu \delta_\xi \delta_{\xib} g_{\rho \nu} + \partial_\nu \delta_\xi \delta_{\xib} g_{\mu\rho} -\partial_\rho \delta_\xi \delta_{\xib} g_{\mu\nu} \right)\right.\0\\
&&+ \left. \delta_\xi\hat  g^{\lambda \rho} \left( \partial_\mu  \delta_{\xib} g_{\rho \nu} + \partial_\nu \delta_{\xib} g_{\mu\rho} -\partial_\rho \delta_{\xib} g_{\mu\nu} \right) - \delta_{\xib} \hat g^{\lambda\rho} \left(  \partial_\mu \delta_\xi g _{\rho \nu} + \partial_\nu  \delta_\xi g_{\mu\rho} -\partial_\rho  \delta_\xi g_{\mu\nu} \right)\right)\0\\
&=& \delta_\xi \delta_{\xib} \Gamma_{\mu\nu}^\lambda, \0
\ee
in agreement with  \eqref{superChris1}.

The super-Riemann curvature is
\be
{\bf R}_{\mu\nu\lambda}{}^\rho &=& - \partial_\mu {\mathbf 
\Gamma}_{\nu\lambda}^\rho 
+ \partial_\nu {\mathbf \Gamma}_{\mu\lambda}^\rho - {\mathbf 
\Gamma}_{\mu\sigma}^\rho
{\mathbf \Gamma}_{\nu\lambda}^\sigma+
 {\mathbf \Gamma}_{\nu\sigma}^\rho {\mathbf 
\Gamma}_{\mu\lambda}^\sigma\0\\
&=& R_{\mu\nu\lambda}{}^\rho + \vt \,\overline \Omega_{\mu\nu\lambda}{}^\rho  +\vtb\, \Omega _{\mu\nu\lambda}{}^\rho + \vt\vtb\, S_{\mu\nu\lambda}{}^\rho,\label{superRiemann}
\ee
where
\be
\overline \Omega_{\mu\nu\lambda}{}^\rho  &=& - \partial_\mu {\overline K}_{\nu\lambda}^\rho 
+ \partial_\nu {\overline K}_{\mu\lambda}^\rho - {
\Gamma}_{\mu\sigma}^\rho
{\overline K}_{\nu\lambda}^\sigma+
 { \Gamma}_{\nu\sigma}^\rho {\overline K}_{\mu\lambda}^\sigma- {
\overline K}_{\mu\sigma}^\rho
{\Gamma}_{\nu\lambda}^\sigma+
 { \overline K}_{\nu\sigma}^\rho {\Gamma }_{\mu\lambda}^\sigma\0\\
&=&  - \partial_\mu {\delta_{\xib} \Gamma}_{\nu\lambda}^\rho 
+ \partial_\nu {\delta_{\xib} \Gamma}_{\mu\lambda}^\rho - {
\Gamma}_{\mu\sigma}^\rho
{ \delta_{\xib} \Gamma}_{\nu\lambda}^\sigma+
 { \Gamma}_{\nu\sigma}^\rho { \delta_{\xib} \Gamma}_{\mu\lambda}^\sigma- {\delta_{\xib} \Gamma
 }_{\mu\sigma}^\rho
{\Gamma}_{\nu\lambda}^\sigma+
 { \delta_{\xib} \Gamma}_{\nu\sigma}^\rho {\Gamma }_{\mu\lambda}^\sigma\0\\
&=&  \delta_{\xib}\, R_{\mu\nu\lambda}{}^\rho.\label{deltaxibR}
\ee
Likewise
\be
\Omega_{\mu\nu\lambda}{}^\rho = \delta_\xi \,R_{\mu\nu\lambda}{}^\rho, \label{deltaxiR}
\ee
and
\be
S_{\mu\nu\lambda}{}^\rho &=&  \partial_\mu {H}_{\nu\lambda}^\rho 
+ \partial_\nu {H}_{\mu\lambda}^\rho - {H}_{\mu\sigma}^\rho
{\Gamma}_{\nu\lambda}^\sigma+
 {H}_{\nu\sigma}^\rho {\Gamma}_{\mu\lambda}^\sigma- {
\Gamma}_{\mu\sigma}^\rho
{H}_{\nu\lambda}^\sigma+
 { \Gamma}_{\nu\sigma}^\rho {H }_{\mu\lambda}^\sigma\0\\
&&  + {\overline K}_{\mu\sigma}^\rho
{ K}_{\nu\lambda}^\sigma-
 { K}_{\mu\sigma}^\rho {\overline K}_{\nu\lambda}^\sigma- {
\overline K}_{\nu\sigma}^\rho
{K}_{\mu\lambda}^\sigma+
 { K}_{\nu\sigma}^\rho {\overline K}_{\mu\lambda}^\sigma\0\\
&=&-  \partial_\mu {\delta_\xi \delta_{\xib} \Gamma}_{\nu\lambda}^\rho 
+ \partial_\nu {\delta_\xi \delta_{\xib} \Gamma}_{\mu\lambda}^\rho - {\delta_\xi \delta_{\xib} \Gamma}_{\mu\sigma}^\rho
{\Gamma}_{\nu\lambda}^\sigma+
 {\delta_\xi \delta_{\xib} \Gamma}_{\nu\sigma}^\rho {\Gamma}_{\mu\lambda}^\sigma- {
\Gamma}_{\mu\sigma}^\rho
{\delta_\xi \delta_{\xib} \Gamma}_{\nu\lambda}^\sigma+
 { \Gamma}_{\nu\sigma}^\rho {\delta_\xi \delta_{\xib} \Gamma }_{\mu\lambda}^\sigma\0\\
&&  + {\delta_{\xib} \Gamma}_{\mu\sigma}^\rho{ \delta_\xi \Gamma}_{\nu\lambda}^\sigma-
 { \delta_\xi \Gamma}_{\mu\sigma}^\rho {\delta_{\xib}  \Gamma}_{\nu\lambda}^\sigma- {
\delta_{\xib} \Gamma}_{\nu\sigma}^\rho
{ \delta_\xi \Gamma}_{\mu\lambda}^\sigma+
 {  \delta_\xi \Gamma}_{\nu\sigma}^\rho {\delta_{\xib} \Gamma}_{\mu\lambda}^\sigma\0\\
&=&\delta_\xi \delta_{\xib} \,R_{\mu\nu\lambda}{}^\rho . \label{deltaxideltaxibR}
\ee

This gives immediately the super-Ricci tensor
\be
{\bf R}_{\mu\lambda}\equiv {\bf R}_{\mu\nu\lambda}{}^\nu
&=& R_{\mu\lambda} + \vt \,\overline \Omega_{\mu\lambda} +\vtb\, \Omega _{\mu\lambda} + \vt\vtb\, S_{\mu\lambda},\label{superRicci}
\ee
with $\overline \Omega_{\mu \lambda} =\overline \Omega_{\mu\nu\lambda}{}^\nu,\,\,
\Omega_{\mu \lambda} = \Omega_{\mu\nu\lambda}{}^\nu$ and $S_{\mu \lambda} = S_{\mu\nu\lambda}{}^\nu$. Of course
\be
\overline \Omega_{\mu\lambda}= \delta_{\xib} R_{\mu\lambda}, \quad\quad\Omega_{\mu\lambda}= \delta_{\xi} R_{\mu\lambda}, \quad\quad {\rm and} \quad\quad  S_{\mu\lambda}= \delta_\xi\delta_{\xib} R_{\mu\lambda}.\label{deltaxiRicci}
\ee

The super-Ricci scalar is
\be
{\bf R}\equiv \widehat G^{\mu\nu} \, {\bf R}_{\mu\nu}= R +\vt \, \overline \Omega+ \vtb\, \Omega +\vt\vtb\, S .\label{bfR}
\ee
It's easy to show that
\be
\overline \Omega= \delta_{\xib}R, \quad\quad \Omega =\delta_\xi R,\quad\quad
S=\delta_\xi\delta_{\xib} R .\label{deltaxiR}
\ee

\section{The vielbein}

If we want to include fermions in a theory {in curved spacetime} we need frame fields.  This section is devoted to introducing vielbein in the superspace. We define the supervierbein as $d$-vector 1-form
\be
{\mathbb {E}}^a = E^a_M (X)\tilde d X^M, \label{E}
\ee
 where
\be
E_\mu ^a (X) &=& e^a_\mu(x) +\vt \bar \phi^a_\mu(x)+\vtb \phi^a_\mu(x) +\vt\vtb f^a_\mu(x),\0\\
E_\vt ^a (X) &=& \chi^a (x) + \vt \bar C^a(x) +\vtb C^a(x) + \vt\vtb \psi^a(x) ,\0\\
E_\vtb ^a (X) &=& \lambda^a (x) + \vt \bar D^a(x) +\vtb D^a(x) + \vt\vtb \rho^a(x).\label{E(X)}
\ee
The natural horizontality condition is
\be
\widetilde E^a_M (\widetilde X)\tilde d \widetilde X^M=  e^a_\mu(x) dx^\mu .\label{horizon}
\ee
This is the same condition as for a vector field. So we get immediately the results
\be
\phi^a_\mu &=&  \xi\!\cdot\!\partial e^a_\mu + \partial_\mu \xi^\lambda 
e^a_\lambda= \delta_\xi e^a_\mu,\label{fierbein1}\\
 \bar\phi^a_\mu &=&  \bar\xi\!\cdot\!\partial e^a_\mu + \partial_\mu \bar\xi^\lambda 
e^a_\lambda= \delta_\xib e^a_\mu,\label{vierbein2}\\
f^a_\mu &=&  \xi\!\cdot\!\partial \bar\phi^a_\mu- \bar\xi\!\cdot\!\partial \phi^a_\mu  
-\xi{\xib}\!\cdot\! \partial^2 e^a_\mu 
+\partial_\mu\bar\xi^\lambda  \xi\!\cdot\!\partial e^a_\lambda- 
\partial_\mu\xi^\lambda  {\xib}\!\cdot\!\partial e^a_\lambda\0\\
&&- \partial_\mu\bar\xi^\lambda \phi^a_\lambda + \partial_\mu\xi^\lambda \bar 
\phi^a_\lambda-h\!\cdot\!\partial e^a_\mu-\partial_\mu h\!\cdot\! e^a =-\delta_\xib \phi^a_\mu=  \delta_\xi \bar \phi^a_\mu,\label{vierbein3}
\ee
\be
\chi ^a&=&e^a_\mu{\xib}^\mu, \label{vierbeinchi1}\\
C^a&=&  -\xi\!\cdot\!\partial e^a_\mu{\xib}^\mu +\phi^a_\mu {\xib}^\mu + 
\xi\!\cdot\!\partial\chi^a-h\!\cdot\!  e^a=  \delta_\xi \chi^a,\label{vierbeinchi2}\\
\bar C^a &=&  -{\xib}\!\cdot\!\partial e^a_\mu{\xib}^\mu +\bar\phi^a_\mu {\xib}^\mu + 
{\xib}\!\cdot\!\partial\chi^a=  \delta_\xib \chi^a,\label{vierbeinchi3}\\
\psi^a &=& \xi{\xib}\!\cdot\! \partial^2 e^a_\mu\,{\xib}^\mu - \xi\!\cdot\!\partial 
\bar\phi^a_\mu{\xib}^\mu+{\xib}\!\cdot\!\partial \phi^a_\mu {\xib}^\mu 
+f^a_\mu {\xib}^\mu-\xi{\xib}\!\cdot\! \partial^2\chi^a+ \xi\!\cdot\!\partial \bar C^a 
-{\xib}\!\cdot\!\partial C^a\label{vierbeinchi4}\\
&& -\bar\xi\!\cdot\!\partial e^a\!\cdot\! h+ \bar\phi^a\!\cdot\! h - 
h\!\cdot\!\partial  \chi^a+h\!\cdot\! \partial e^a_\mu\,\bar \xi^\mu
 =   \delta_\xi \bar C^a= -\delta_\xib  C^a,\0
\ee
and
\be
\lambda^a &=&e^a _\mu\, \xi^\mu, \label{vierbeinchib1}\\
D^a&=&  -\xi\!\cdot\!\partial e^a_\mu\xi^\mu +\phi^a_\mu \xi^\mu + 
\xi\!\cdot\!\partial\lambda^a= \delta_\xi \lambda^a,\label{vierbeinchib2}\\
\bar D^a &=&  -{\xib}\!\cdot\!\partial e^a_\mu\xi^\mu +\bar\phi^a_\mu \xi^\mu + 
{\xib}\!\cdot\!\partial\lambda^a+h\!\cdot\! e^a = \delta_\xib \lambda^a ,\label{vierbeinchib3}\\
\rho^a &=& \xi{\xib}\!\cdot\! \partial^2 e^a_\mu\xi^\mu - \xi\!\cdot\!\partial 
\bar\phi^a_\mu\xi^\mu+{\xib}\!\cdot\!\partial \phi^a_\mu \xi^\mu 
+f^a_\mu \xi^\mu-\xi{\xib}\!\cdot\! \partial^2\lambda^a+ \xi\!\cdot\!\partial \bar D^a 
-{\xib}\!\cdot\!\partial D^a\label{vierbeinchib4}\\
&& -\xi\!\cdot\!\partial e^a\!\cdot\! h+ \phi^a\!\cdot\! h - h\!\cdot\!\partial  
\lambda^a+h\!\cdot\! \partial e^a_\mu \xi^\mu=   \delta_\xi \bar D^a= -\delta_\xib  D^a. \0
\ee

\subsection{The inverse vielbein $\widehat E^\mu_a$}

Here we introduce the inverse vielbein $\widehat E^\mu_a$. Let us write it as
\be
E_\mu^a (X) =e^a_\lambda (x) \left( \delta^\lambda_\mu + \vt \bar \phi^\lambda_\mu(x)+\vtb \phi^\lambda_\mu(x) +\vt\vtb f^\lambda_\mu(x)\right) = e^a_\lambda (1+ X)^\lambda_\mu
\label{E^mu}, \label{EmuaX}
\ee
where $\phi^\lambda_\mu= e_a^\lambda \phi^a_\mu,\bar \phi^\lambda_\mu= e_a^\lambda \bar\phi^a_\mu$ and $f^\lambda_\mu= e_a^\lambda f_\lambda^a$, and $e_a^\lambda (x) e_\lambda^b(x)= \delta_a ^b$. Then we define
\be
\widehat E_a ^\mu= (1-X+X^2)^\mu_\lambda \, e^\lambda_a. \label{Emuinverse}
\ee
The following is evident
\be
\widehat E_a ^\mu \, E_\mu^b = \delta_a^b .\label{E-1E}
\ee
In terms of components we have 
\be
\hat \phi^\mu_a &=&- \hat e_b^\mu \phi_\lambda ^b \hat e^\lambda_a= -   \hat e_b^\mu \delta_\xi e^b_\lambda \hat e^\lambda_a= \delta_\xi \hat e_a^\mu,\label{hatphimuxi}\\
\hat{\bar\phi}^\mu_a &=&- \hat e_b^\mu \bar\phi_\lambda ^b \hat e^\lambda_a= -   \hat e_b^\mu \delta_\xib e^b_\lambda \hat e^\lambda_a= \delta_\xib \hat e_a^\mu,\label{hatphimuxib}
\ee
and
\be
\hat f_a^\mu =\hat e_b^\mu\left(-f_\lambda^b -\bar \phi_\rho^b e^\rho_c \, \phi^c_\lambda + \phi_\rho^b \, e_c ^\rho\, \bar \phi_\lambda^c\right) \hat e _a^\lambda= \delta_\xi\delta_\xib \hat e_a^\mu.
\ee
A simple way to prove the last step is as follows
\be
\delta_\xi\delta_\xib \hat e_a^\mu& =& \delta_\xi \left(- \hat e_b^\mu \delta_\xib e_\lambda^b \hat e_a^\lambda\right)=- \delta_\xi \hat e_b^\mu \delta_\xib e_\lambda^b \hat e_a^\lambda- \hat e_b^\mu\delta_\xi \delta_\xib e_\lambda^b \hat e_a^\lambda+ \hat e_b^\mu \delta_\xib e_\lambda^b\delta_\xib \hat e_a^\lambda\0\\
&=& \hat e_b^\mu\left(-f_\lambda^b -\bar \phi_\rho^b e^\rho_c \, \phi^c_\lambda + \phi_\rho^b \, e_c ^\rho\, \bar \phi_\lambda^c\right) \hat e _a^\lambda= \hat f_a^\mu. \label{detaxixibeamu}
 \ee

\subsection{The inverse supervielbein $\widehat E^M_a$}

In this subsection, as we did for the supermetric,  we try to define the inverse of the supervielbein.
In analogy with what we did for the metric we define
\be
\widehat {\mathbb E}_a = \widehat E_a^M(X) \frac \partial{\partial X^M},\label{mathbbE}
\ee
where
\be
\widehat E^\mu _a (X) &=&\hat e_a^\mu(x) +\vt \hat{\bar \phi}_a^\mu(x)+\vtb \hat\phi_a^\mu(x) +\vt\vtb \hat f_a^\mu(x),\0\\
\widehat E^\vt_a (X) &=& \hat\chi_a (x) + \vt\hat {\bar C}_a(x) +\vtb \hat C_a(x) + \vt\vtb \hat\psi_a(x), \0\\
\widehat E^\vtb _a (X) &=& \hat\lambda_a (x) + \vt \hat{\bar D}_a(x) +\vtb \hat D_a(x) + \vt\vtb \hat\rho_a(x),\label{Einverse(X)}
\ee
and impose the horizontality condition
\be
 \widetilde{\widehat E}_a^M (X)\frac { \partial}{\partial \widetilde X^M}= \hat e_a^\mu \frac \partial{\partial x^\mu}. \label{horinversevierbein}
\ee
The explicit form of the LHS of this equation can be found in Appendix B, see eq.\eqref{horinverse}.
In the latter the coefficient of $\frac {\partial}{ \partial  \vt} $ leads to
\be
\hat\chi_a=\widehat {\overline C}_a= \widehat C_a = \hat\psi_a=0,\label{horinverse1}
\ee
and the quation proportional to  $\frac {\partial}{ \partial  \vtb} $ to
\be
\hat\lambda_a=\widehat {\overline D}_a= \widehat D_a = \hat\rho_a=0.\label{horinverse2}
\ee
Therefore only the components of $\widehat E_a^\mu(X)$ are nonvanishing. The remaining equations give
\be
\hat \phi_a^\mu &=& \xi \!\cdot\! \hat e_a^\mu - \partial_\lambda \xi^\mu \hat e_a^\lambda= \delta_\xi\hat e_a^\mu, \0\\
\hat {\bar \phi}_a^\mu &=& \xib \!\cdot\! \hat e_a^\mu - \partial_\lambda \xib^\mu \hat e_a^\lambda= \delta_\xib \hat e_a^\mu, \0\\
\hat f_a^\mu &=& -\xi\xib \!\cdot\! \partial^2 \hat e_a^\mu -h\!\cdot\! \partial e_a^\mu + \partial_\lambda h^\mu e_a^\lambda +\xi\!\cdot\! \partial \hat {\bar \phi}_a^\mu +  \partial_\lambda \xi^\mu \hat {\bar \phi}_a^\lambda-\xib\!\cdot\! \partial \hat { \phi}_a^\mu -  \partial_\lambda \xib^\mu \hat { \phi}_a^\lambda,\0\\
&&+ \hat e_a^\lambda \partial_\lambda \xib \!\cdot\!\partial \xi^\mu+
\xi \!\cdot\! \partial \hat e_a^\lambda \partial_\lambda \xib^\mu -\hat e_a^\lambda \partial_\lambda \xi \!\cdot\!\partial \xib^\mu-
\xib \!\cdot\! \partial \hat e_a^\lambda \partial_\lambda \xi^\mu =
\delta_\xi \delta_\xib \hat e_a^\mu.\label{inversevierbeincomp}
\ee
If $\hat e_a^\mu $ is the inverse of $e_\mu ^a$, these formulas coincide with those of the previous subsection.

The results of this section confirm what were found in the previous subsection. In the superspace it makes sense to consider only horizontal tensors, i.e. tensors whose components in the anticommuting directions vanish. We will, therefore, continue to define a frame geometry with this charactistic.

\subsection{The spin superconnection}

The spin superconnection is defined as follows:
\be
{\mathbf \Omega}^{ab}_\mu &=& \frac 12\Big [ \widehat E^{a\nu} \left( \partial_\mu E_\nu^b - \partial_\nu E_\mu^b\right) - \widehat E^{b\nu} \left( \partial_\mu E_\nu^a - \partial_\nu E_\mu^a\right) - \widehat E^{a\nu} \widehat E^{b\lambda} \left(\partial_\nu E_\lambda^c - \partial_\lambda E_\nu^c\right)E_{c\mu}\Big]\label{Omegaab}\\
&=& \omega_\mu^{ab} + \vt \overline P_\mu^{ab} + \vtb   P_\mu^{ab}+\vt\vtb Q_\mu^{ab},\0
\ee
where $  \omega_\mu^{ab} $ is the  usual spin connection and
\be
P_\mu^{ab}\!\!\!&=&\!\!\!\frac 12 \Big[ \hat e^{a\nu} \partial_\mu \phi _\nu ^b +\hat \phi ^{a\nu} \partial_\mu e^b_\nu-\hat e^{b\nu} \partial_\mu \phi _\nu ^a -\hat \phi ^{b\nu} \partial_\mu e^a_\nu- \left( \hat e^{a\nu} \hat \phi^{b\lambda}+\hat  \phi^{a\nu} \hat e^{b\lambda}\right) \left( \partial_\nu e^c_\lambda- \partial_\lambda e^c_\nu\right) e_{c\mu}\0\\
&&\quad- \hat e^{a\nu} e^{b\lambda} \left(\partial_\nu \phi^c_\lambda- \partial_\lambda \phi^c_\nu \right) e_{c\mu}-  \hat e^{a\nu} e^{b\lambda} \left(\partial_\nu e^c_\lambda- \partial_\lambda e^c_\nu \right) \phi_{c\mu}\Big]\0\\
&=&\!\!\! \frac 12 \Big[ \hat e^{a\nu} \partial_\mu\delta_\xi e_\nu ^b +\delta_\xi\hat e ^{a\nu} \partial_\mu e^b_\nu-\hat e^{b\nu} \partial_\mu \delta_\xi e _\nu ^a -\delta_\xi\hat e^{b\nu} \partial_\mu e^a_\nu- \left( \hat e^{a\nu}\delta_\xi \hat e^{b\lambda}+\delta_\xi\hat  e^{a\nu} \hat e^{b\lambda}\right) \left( \partial_\nu e^c_\lambda- \partial_\lambda e^c_\nu\right) e_{c\mu}\0\\
&&\quad- \hat e^{a\nu} e^{b\lambda} \left(\partial_\nu \delta_\xi e^c_\lambda- \partial_\lambda  \delta_\xi e^c_\nu \right) e_{c\mu}-  \hat e^{a\nu} e^{b\lambda} \left(\partial_\nu e^c_\lambda- \partial_\lambda e^c_\nu \right) \delta_\xi e_{c\mu}\Big]= \delta_\xi \omega_\mu^{ab} . \label{Pmuab1}
\ee
where $e_\mu^a, \phi_\mu ^a, \hat e^{a\mu}, \hat \phi^{a\mu}$  have been explained earlier
in subsections 5.1 and 5.2.
Similarly, we have the following
\be
\overline P_\mu^{ab} &=&\delta_\xib  \omega_\mu^{ab},\label{ Pmuab2}\\
\overline Q_\mu^{ab} &=& \delta_\xi\delta_\xib \omega_\mu^{ab}.\label{Pmuab3}
\ee
Thus we have recovered the complete set of (anti)BRST transformations for the spin superconnection.

\subsection{The curvature}

The 2-form supercurvature is
\be
{\mathbb R}^{ab} = {\mathbb R}_{\mu\nu}^{ab} dx^\mu \wedge dx^\nu,\label{spincurv}
\ee
where
\be
{\mathbb R}_{\mu\nu}^{ab} &=& \partial_\mu {\mathbf \Omega}_\nu ^{ab} - \partial_\nu {\mathbf \Omega}_\mu ^{ab}+ {\mathbf \Omega}^a_{\mu c} {\mathbf \Omega}_\nu^{cb} -
 {\mathbf \Omega}^a_{\nu c} {\mathbf \Omega}_\mu^{cb}\label{Rmunuab}\\
&=& R_{\mu\nu}^{ab} +\vt \overline\Sigma_{\mu\nu}^{ab}+\vtb \Sigma_{\mu\nu}^{ab}+\vt\vtb S_{\mu\nu}^{ab}.\0
\ee
$ R_{\mu\nu}^{ab}$ is the usual spin connection curvature. Next, we have the following
\be
\Sigma_{\mu\nu}^{ab}&=& \partial_\mu P_\nu^{ab} -\partial_\nu P_\mu^{ab} + P_\mu^{ac}\,\omega_{\nu c}{}^b+ \omega_\mu^{ac} \,P_{\nu c}{}^b -   P_\nu^{ac}\,\omega_{\mu c}{}^b- \omega_\nu^{ac} \,P_{\mu c}{}^b \0\\
&=&  \partial_\mu \delta_\xi \omega_\nu^{ab} -\partial_\nu \delta_\xi \omega_\mu^{ab} +\delta_\xi \omega_\mu^{ac}\,\omega_{\nu c}{}^b+ \omega_\mu^{ac} \,\delta_\xi \omega_{\nu c}{}^b -   \delta_\xi \omega_\nu^{ac}\,\omega_{\mu c}{}^b- \omega_\nu^{ac} \,\delta_\xi \omega_{\mu c}{}^b\0\\
&= &\delta_\xi  R_{\mu\nu}^{ab}. \label{Sigmamunuab1}
\ee
At the same time we have the following identifications:
\be
\overline \Sigma_{\mu\nu}^{ab}&=& \delta_\xib  R_{\mu\nu}^{ab}, \label{Sigmamunuab2}\\
S_{\mu\nu}^{ab}&=&\delta_\xi \delta_\xib  R_{\mu\nu}^{ab}. \label{Sigmamunuab3}
\ee

\subsection{Fermions}

Fermion fields, under diffeomorphisms, behave like scalars. A Dirac fermion superfield has the following expansion
\be
{\mathbf\Psi}(X)= \psi(x) +\vt \overline F(x) +\vtb F(x) +\vt\vtb \Theta(x,\label{Diracfermion}
\ee
$\psi, F, \overline F$ and $\Theta$ are four-components complex column vector fields.
The horizontality condition is
\be
\widetilde {\mathbf \Psi}(\widetilde X) =  \psi(x). \label{horfermion}
\ee
Repeating the analysis of the scalar superfield we get
\be
F(x)=\delta_\xi \psi(x),\quad\quad \overline F(x)= \delta_{\xib} \psi(x), \quad\quad \Theta(x)= \delta_\xi \delta_\xib \psi(x).\label{fermiontransf}
\ee
The covariant derivative of a vector superfield is
\be
{\bf D}_\mu{\mathbf\Psi} = \left(\partial_\mu +\frac 12 {\mathbf \Omega}_\mu \right){\mathbf\Psi},\0
\ee
where ${\mathbf \Omega}_\mu= {\mathbf \Omega}_\mu^{ab} \Sigma_{ab}$, and $\Sigma_{ab} =\frac 14 [\gamma_a,\gamma_b]$  are the Lorentz generators.

The Lagrangian density for a Dirac superfield is
\be
L = \sqrt{g}\, i \,\overline {\mathbf\Psi} \hat \gamma^\mu {\bf D}_\mu{\mathbf\Psi},\label{LDirac}
\ee 
with $\hat \gamma^\mu = \widehat E^\mu_a \gamma_a$. The Lagrangian density $L$ is invariant under Lorentz transformations and, up to total derivatives, under diffeomorphisms.

\subsection{The super-Lorentz transformations}

The fermion superfield transforms under local Lorentz transformations as
\be
\delta_\Lambda {\mathbf\Psi}=\frac 12{\mathbf \Lambda \Psi}, \quad\quad {\mathbf\Lambda} = {\mathbf\Lambda}^{ab}(X) \Sigma_{ab},\label{Lorentztransf}
\ee
where $\Lambda^{ab}(X)$ is an infinitesimal  antisymmetric supermatrix with arbitrary entries: 
\be
{\mathbf\Lambda}^{ab}(X) =\lambda^{ab}(x) + \vt \bar h^{ab}(x) +\vtb h^{ab}(x) +\vt\vtb \Lambda^{ab}(x) .\label{Lambdaab}
\ee
Local Lorentz transformations act on the super-vielbein as follows:
\be
\delta_\Lambda E^a_\mu =E_\mu^b  {\mathbf\Lambda}_b{}^a, \quad\quad \delta_\Lambda \widehat E_a^\mu =\widehat E^\mu_b  {\mathbf\Lambda}^b{}_a.\label{deltaLambdaE}
\ee
Using the definition \eqref{Omegaab} one finds
\be
\delta_\Lambda {\mathbf \Omega}_\mu =\partial_\mu{\mathbf\Lambda} +\frac12 [{\mathbf \Omega}_\mu, {\mathbf\Lambda}], \label{deltaLOmega}
\ee
or
\be
\delta_\Lambda {\mathbf \Omega}_\mu^{ab} = \partial_\mu {\mathbf\Lambda}^{ab} +  {\mathbf \Omega}_\mu^{cb} {\mathbf\Lambda}_c{}^a +  {\mathbf \Omega}_\mu^{ac} {\mathbf\Lambda}_c{}^b. \label{deltaLOmegaab}
\ee
With this and
\be
[\hat \gamma^\mu,{\mathbf \Lambda}]= 2E_a^\mu {\mathbf\Lambda}^{ab} \gamma_b,\label{gammaLambda}
\ee
one can easily prove that \eqref{LDirac} is invariant under local Lorentz transformations.

\section{Superfield formalism and consistent anomalies}

The superfield formalism nicely applies to the description of consistent anomalies. In this section we first summarize the definitions and properties of gauge anomalies and then we apply to them the superfield description. { Basic material  for the following algebraic approach to anomalies can be found in \cite{Stora1,Stora2,Manes,bonora,BCRS} and \cite{Bert}.}

Here formulas refer to a $\sf d$-dimensional spacetime $\sf M$ without boundary.

\subsection{BRST, descent equations and consistent gauge anomalies}

The BRST operation $\mfs$ in \eqref{BRST1} is nilpotent. We represent with the same
symbol $\mfs$ the corresponding functional operator, i.e.
\be
\mfs= \int d^dx \left(\mfs A_\mu^a(x) \frac {\partial}{\partial  A_\mu^a(x)}+ \mfs c^a(x) 
\frac {\partial}{\partial c^a(x)}\right).\label{S}
\ee

To construct the descent equations we start from a symmetric polynomial
in the Lie algebra of degree $n$, i.e. $P_n (T^{a_1}, . . . , T^{a_n} )$, invariant under the adjoint
transformations:
\be
P_n ([X,T^{a_1}], . . . , T^{a_n} )+\ldots +P_n (T^{a_1}, . . . ,[X, T^{a_n}] )=0,\label{PnX}
\ee
for any element $X$ of the Lie algebra ${\mathfrak g}$. In many cases these polynomials are symmetric
traces of the generators in the corresponding representation
\be
P_n (T^{a_1}, . . . , T^{a_n} )= Str(T^{a_1} . . . T^{a_n} ),\label{Symtrace}
\ee
($Str$ denotes the symmetric trace). With this one can construct the 2n-form
\be
\Delta_{2n}(A)= P_n(F,F,\ldots F),\label{PnFFF}
\ee
where $F = dA +\frac 12 [A,A]$. It is easy to prove that
\be
P_n(F,F,\ldots F)= d \left( n \int_0^1 dt\, P_n(A, F_t,\ldots, F_t)\right)
\equiv d \Delta_{2n-1}^{(0)}(A),\label{PnAFF}
\ee
where we have introduced the symbols $A_t = tA$ and its curvature $F_t =
dA_t +\frac 12  [A_t , A_t ]$, where $0 \leq t \leq 1$. 
In the above expressions, the product of
forms is understood to be the exterior product. It is important to recall that in order
to prove eq.(\ref{PnAFF}) one uses in an essential way the symmetry of
$P_n$ and the graded commutativity of the exterior product of forms.

$ \Delta_{2n-1}^{(0)}(A)$ is often denoted also $TP_n(A)$ and
\be
TP_n(A)= n \int_0^1 dt\, P_n(A, F_t,\ldots, F_t),\label{transgression}
\ee
is known as the {\it transgression formula}.

Eq.(\ref{PnAFF}) is the first of a sequence of equations that can be proven
\be
&&\Delta_{2n}(A)- d \Delta_{2n-1}^{(0)}(A) =0,\label{descent1}\\
&&\mfs \Delta_{2n-1}^{(0)}(A) +d \Delta_{2n-2}^{(1)}(A,c)=0,\label{descent2}\\
&& \mfs \Delta_{2n-2}^{(1)}(A,c)+ d \Delta_{2n-3}^{(2)}(A,c)=0,\label{descent3}\\
&&\dots\dots\0\\
&& \mfs \Delta_{0}^{(2n-1)}(c)=0.  \label{descentn}
\ee
All the expressions $\Delta_k^{(p)} (A, c)$ are polynomials {of and $A, c, dA,dc $ and their commutators. }
The lower index
$k$ is the form degree and the upper one $p$ is the ghost number, i.e. the number
of $c$ factors. The last polynomial $ \Delta_{0}^{(2n-1)}(c)$ is a 0-form and clearly a function
only of $c$. All these polynomials have explicit compact form. For instance,
the next interesting case after eq.(\ref{descent1}) is:
\be
\mfs\, \Delta_{2n-1}(A) = -d\left( n(n-1) \int_0^1 dt (1-t) 
P_n( dc ,A, F_t,\ldots F_t)\right).\label{descent2'}
\ee
This means, in particular, that integrating $\Delta_{2n-1}(A)$ over spacetime in ${\sf d} =
2n\! -\!1$ dimensions we obtain an invariant local expression. This gives the
gauge CS action in any odd dimension. But what matters here is that the
RHS contains the general expression of the consistent gauge anomaly in ${\sf d} =
2n\!-\!2$ dimension. For, integrating \eqref{descent3} over spacetime, one gets
\be
\mfs\, {\cal A}[c,A]&=&0, \label{boldA}\\
{\cal A}[c,A]&=& \int d^dx \, \Delta_d^{(1)}(c,A), \quad\quad {\rm where}\0\\
\Delta_d^1(c,A)&=& n(n-1) \int_0^1 dt (t-1) 
P_n( dc ,A, F_t,\ldots F_t),\label{chiralanom}
\ee
where $ {\cal A}[c,A] $ identifies the anomaly up to an overall numerical coefficient.  

Thus, the existence of chiral gauge anomalies relies on the existence of the
adjoint-invariant polynomials $P_n$. One can prove that
the so-obtained cocycles are non-trivial. 

Although the above formulas are formally correct, one should remark that, in order to describe a consistent anomaly in a $\sfd=2n\!-\!2$ dimensional spacetime, we need two forms $P_n(F,\ldots,F)$ and $\Delta_{2n-1}^{(0)}(A)$, which are identically vanishing. This is an unsatisfactory aspect of the previous purely algebraic approach. The superfield formalism overcomes this difficulty and gives automatically the anomaly as well as its descendants, \cite{BBL}.

\subsection{Superfield formalism, BRST transformations and anomalies}

For simplicity, we introduce only one anticommuting variable $\vartheta$ and consider the superconnection (where $\tilde d =d +\frac {\partial}{\partial\vt} d\vt$)
\be
{\cal A}= e^{-\vt c} (\tilde d +A )  e^{\vt c}= A +\vt (dc +[A,c]) + \left( c-\vt\frac 12[c,c]\right)d\vt\equiv \phi + \eta\, d\vt.\label{calA}
\ee
The supercurvature is:
\be
{\cal F} = e^{-\vt c} F  e^{\vt c}= F+\vt [F,c]. \label{calF}
\ee
From these formulas it is immediately visible that the 
derivative with respect to $\vt$ corresponds to the BRST transformation:
\be
\frac {\partial}{\partial\vt} \phi &=& dc +[A,c] = D_Ac = \mfs A, \0\\
\frac {\partial}{\partial\vt} \eta &=& -\frac 12[c,c]=\mfs c,\0\\
\frac {\partial}{\partial\vt} {\cal F}&=& [F,c]=\mfs F.\0
\ee
Let us now consider the transgression formula
\be
TP_n({\cal A})= n \int_0^1 dt\, P_n({\cal A}, {\cal F}_t,\ldots,{\cal F}_t),\label{transgressioncal}
\ee
in which we have replaced $A$ everywhere with ${\cal A}$, {and $F_t$} with ${\cal F}_t =t\, \tilde d {\cal A} + \frac {t^2}2 [{\cal A},{\cal A}]$ ($t$ is an auxiliary parameter varying from 0 to 1). 

The claim is that \eqref{transgressioncal} contains all the information about the gauge anomaly, including the explicit form of all its descendants. To see this, it is enough to expand the polyform $TP_n({\cal A}) $ in component forms as follows
\be
TP_n({\cal A}) = \sum_{i=0}^{2n-1} \tilde \Delta^{(i)}_{2n-i-1}(\phi,\eta) \underbrace{ d\vt\wedge\ldots \wedge d\vt}_{i \,\,{\rm factors}},\label{TPncalAdecomp}
\ee
where $2n\!\!-\!\!i\!\!-\!\!1$ is the spacetime form degree.
Notice that the wedge product is commutative for the $d\vt$ factors. Of course both $ \tilde \Delta_{2n-1}^{(0)}(\phi)\vert_{\vt=0}=TP_n(A)$ and 
$\tilde \Delta_{2n-1}^{(0)}(\phi)$ vanish in dimension $\sfd=2n\!\!-\!\!2$. But the remaining forms are nonvanishing. 

Let us extract the term $ \tilde \Delta^{(1)}_{2n-2}(A,c)$:
\be
\tilde \Delta^{(1)}_{2n-2}(\phi,\eta)\!\!\!\!&=&\!\!\!\! n \int_0^1\!\! dt\,\Big( P_n(\eta,\tilde F_t,\ldots,\tilde F_t)+(n-1) P_n(\phi, t(d\eta- \partial_\vt \phi)+ t^2[\phi,\eta],\tilde  F_t,\ldots,\tilde F_t)\Big),\label{Delta12n-2}
\ee
where $\tilde F_t= td\phi+\frac {t^2}2 [\phi,\phi]$. Let us take the $\vt=0$ part of this.
\be
\tilde  \Delta^{(1)}_{2n-2}(A,c)= n\int_0^1 \!\! dt\,\Big( P_n(c, F_t,\ldots , F_t)+(n-1) P_n( A, ( t^2-t)[A,c],  F_t,\ldots,F_t)\Big).\label{Delta12n-2b}
\ee
Using $\int_0^1dt (1-t) \frac d{dt} f(t) = \int_0^1 dt f(t)$ when $f(0)=0$, we can rewrite this as
\be
\tilde  \Delta^{(1)}_{2n-2}(A,c)= n(n-1)\int_0^1 \!\! dt\,(1-t)\Big( P_n(c,\frac {dF_t}{dt},\ldots , F_t)-P_n( A,[tA,c],  F_t,\ldots,F_t)\Big).\label{Delta12n-2b}
\ee
Using $\frac {dF_t}{dt}=D_{tA} A$ and the ad invariance of $P_n$ we get finally
\be
\tilde  \Delta^{(1)}_{2n-2}(A,c)&=&d\left( n(n-1)\int_0^1 \!\! dt\,(1-t)\Big( P_n(c, A,\ldots , F_t)\right) \0\\
&& -n(n-1)\int_0^1 \!\! dt\,(1-t) P_n(dc,A,  F_t,\ldots,F_t).\label{Delta12n-2c}
\ee
Therefore $\tilde  \Delta^{(1)}_{2n-2}(A,c)$ coincide with the opposite of  $  \Delta^{(1)}_{2n-2}(A,c)$ up to a total spacetime derivative, which is irrelevant for the integrated anomaly.

The $\vt$ derivative of $\tilde \Delta^{(1)}_{2n-2}(\phi,\eta)$ is the  BRST transformation of
$\tilde  \Delta^{(1)}_{2n-2}(A,c)$, and turns out to be a total spacetime derivative. This can be checked with an explicit calculation. 

{\bf Remark} The cocycles $\Delta$ and $\tilde\Delta$ may differ. For instance, in the case $\sfd=2, n=2$ we get
\be
\tilde \Delta_2^{(1)}(A,c) = P_2(c,dA),\quad\quad \tilde \Delta_1^{(2)}(A,c) =\frac 12 P_2(A,[c,c]),\quad\quad \tilde \Delta_0^{(3)}(c)= \frac 16 P_2(c,[c,c]),\label{casetilden=2}
\ee
while
\be
\Delta_2^{(1)}(A,c) =P_2(dc,A),\quad\quad \Delta_1^{(2)}(c)= P_2(dc,c),\quad\quad \Delta_0^{(3)}(c)=\frac 16 P_2(c,[c,c]) .\label{casen=2}
\ee
This originates from the difference of a total derivative between $\tilde \Delta_2^{(1)}(A,c) $ and $\Delta_2^{(1)}(A,c) $.

\subsection{Anomalies with background connection}

{The expressions of anomalies introduced so far are generally well defined in a local patch of spacetime, but they may not be globally well defined on the whole spacetime (they may not be basic forms, in the language of fiber bundles, i.e., they may be well defined forms in the total space but not in the base spacetime ).  To obtain globally well defined anomalies} we need to introduce a background connection $A_0$ which is invariant under BRST transformations: $\mfs A_0=0$. The dynamical connection transforms, instead, in the usual way, see \eqref{BRST1}. We call $F$ and $F_0$ the curvatures of $A$ and $A_0$. Since the space of connections is affine, also $\widehat A_t=tA+(1-t)A_0$, with $0\leq t\leq 1$ is a connection. We call $\widehat F_t$ its curvature, which takes the values $F$ and $F_0$ for $t=1$ and $t=0$, respectively.

The relevant connection is now
\be
\widehat{\cal A}_t&=&t e^{-\vt c} (\tilde d +A )  e^{\vt c}+(1-t) A_0= t{\cal A}+(1-t)A_0\0\\
&=&t\left( A +\vt (dc +[A,c]) + \left( c-\vt\frac 12[c,c]\right)d\vt\right)+(1-t) A_0\equiv\hat\phi_t+ \hat\eta_t\, d\vt,\label{calAt}
\ee
where $\hat\phi_t=t\phi+(1-t)A_0$ and $\hat \eta_t=t\eta$.
We call $\widehat{\cal F}_t $ the curvature of $\widehat{\cal A}_t$. Notice that $\widehat{\cal F}_1={\cal F}$ and $\widehat{\cal F}_0=F_0$, which is straightforward to be checked.

We start again from the transgression formula
\be
{\cal T}P_n({\cal A},A_0)= n \int_0^1 dt\, P_n({\cal A}-A_0, \widehat{\cal F}_t,\ldots,\widehat{\cal F}_t).\label{transgressioncalT}
\ee
In the same way as before one can prove that, if we assume the spacetime dimension is $\sfd=2n\!-\!2$, we have
\be
\tilde d\, {\cal T}P_n({\cal A},A_0)= P_n({\cal F},\dots,{\cal F})- P_n(F_0,\ldots,F_0)= P_n({ F},\dots,{F})- P_n(F_0,\ldots,F_0)=0.\label{tildedtransgression}
\ee
As before we decompose
\be
{\cal T}P_n({\cal A},A_0) = \sum_{i=0}^{2n-1} \widehat \Delta^{(i)}_{2n-i-1}(\phi,\eta,A_0) \underbrace{ d\vt\wedge\ldots \wedge d\vt}_{i \,\,{\rm factors}}.\label{calTPncalAdecomp}
\ee
The relevant term for the anomaly is $ \Delta^{(1)}_{2n-2}(\phi,\eta,A_0)$, i.e.
\be
\widehat\Delta^{(1)}_{2n-2}(\phi,\eta,A_0)&=& n \int_0^1 dt\,\left[ P_n\left(\phi, \widehat{ \Phi}_t,\ldots,\widehat{ \Phi}_t\right)+P_n\left({\cal A}-A_0, \left(d_{\hat \phi_t} \hat\eta_t-\frac {\partial}{\partial \vt} \hat \phi_t\right),\widehat{ \Phi}_t,\ldots,\widehat{ \Phi}_t\right)\right.\0\\
&&+\left. \ldots + P_n\left({\cal A}-A_0,\widehat{\Phi}_t,\ldots,\widehat{ \Phi}_t,  \left(d_{\hat \phi_t} \hat\eta_t-\frac {\partial}{\partial \vt} \hat \phi_t\right)\right)\right],\label{Delta12n-2}
\ee
where $d_{\hat \phi_t} =d +[{\hat \phi_t}, \cdot]$, and $\widehat \Phi_t= d\hat\phi_t+\frac 12 [\hat\phi_t,\hat\phi_t]$. We have to select the $\vt=0$ part
\be
\left.\widehat\Delta^{(1)}_{2n-2}(\phi,\eta,A_0)\right\vert_{\vt=0}&=& n \int_0^1 dt\,\left[ P_n\left(c, \widehat{ F}_t,\ldots,\widehat{F}_t\right)-t(1-t)P_n\left({A}\!-\!A_0,[A\!-\!A_0,c],\widehat{F}_t,\ldots,\widehat{F}_t\right)\right.\0\\
&&-\left. \ldots -t(1-t) P_n\left(A\!-\!A_0,\widehat{F}_t,\ldots,\widehat{F}_t, [A\!-\!A_0,c]\right)\right].\label{Delta12n-2theta=0}
\ee
This is the chiral anomaly with background connection. 
It can be written in a more familiar form by rewriting the first term on the RHS using: $\int_0^1dt (1-t) \frac {d}{dt} f(t)+f(0)= \int_0^1dt f(t) $:
\be
\left.\widehat\Delta^{(1)}_{2n-2}(\phi,\eta,A_0)\right\vert_{\vt=0}&=&n P_n(c,F_0,\ldots,F_0)-
n(n-1)\int_0^1 dt\, (1-t)P_n(d_{A_0}c ,A-A_0,\widehat{F}_t,\ldots,\widehat{F}_t)\0\\
&&+d\left(n(n-1)\int_0^1 dt\, (1-t) P_n (c,A-A_0,\widehat{F}_t,\ldots,\widehat{F}_t)\right).\label{widehatanomaly}
\ee
Integrating over the spacetime $\sf M$, the last term drops out. If we set $A_0=0$ we recover the formula \eqref{Delta12n-2c}. We notice that, as expected, the RHS of \eqref{widehatanomaly} is a basic quantity.

\subsection{Wess-Zumino terms in field theories with the superfield method}
\label{ss:WZFT}

In a gauge field theory with connection $A$, valued in a Lie algebra with generators $T^a$ and structure constants $f^{abc}$, an anomaly ${\cal A}^a$ must satisfy the WZ consistency conditions, \cite{WZ},
\be
X^a(x) {\cal A}^b(y) - X^b(y) {\cal A}^a(x) + f^{abc} {\cal A}^c(x) \delta(x-y)=0,\label{XaAb}
\ee
where
\be
X^a(x) = \partial_\mu \frac{\delta}{\delta A^a_\mu(x)} +f^{abc} A_\mu^b (x)  \frac{\delta}{\delta A^c_\mu(x)}.\label{Xax}
\ee
The equations \eqref{XaAb} are integrability conditions. This means that one can find a functional
of the fields ${\cal B}_{WZ}$ such that
\be
X^a(x) {\cal B}_{WZ}= {\cal A}^a(x).\label{BWZ}
\ee

In this section we show how to construct a term $ {\cal B}_{WZ}$ which, upon BRST variation,
 generates the anomaly:
\be
{\cal A}=- \int_\sfM c^a(x){\cal A}^a(x)= n(n-1) \int_\sfM \int_0^1 dt\,(1-t) P_n(dc, A, F_t,\ldots,F_t),\label{calAPn}
\ee
where $\sfM$ is the spacetime of dimension ${\sf d}=2n\!\!-\!\!2$, and $F_t= t dA+\frac {t^2}2[A,A]$.

This is possible provided we enlarge the set of fields of the theory, by adding new fields as follows.
Let us introduce a set of auxiliary fields $\sigma(x) = \sigma^a (x) T^a$, which under a gauge transformations with parameters $\lambda(x) = \lambda^a(x)T^a$, transform as
\be
e^{\sigma(x)} \longrightarrow e^{\sigma'(x)}= e^{-\lambda(x)} e^{\sigma(x)}.\label{exptransf}
\ee
Using the Campbell-Hausdorff formula, this means $ \delta \sigma(x) = - \lambda(x) - \frac 12 
[\lambda(x),\sigma(x)]+\ldots$. Next we pass to the infinitesimal transformations and replace the infinitesimal parameter $\lambda(x)$ with anticommuting fields $c(x)= c^a(x) T^a$. We have the BRST transformations
\be
 \quad\quad \mfs e^{\sigma(x)} = - c (x)e^{\sigma(x)}, \quad \quad {\mathfrak{s}}\sigma(x) = -c(x) +\frac 12 [\sigma(x), c(x)] -\frac 1{12} [\sigma(x),[\sigma(x),c(x)]]+\ldots\label{mfsc}
\ee

Now we use a superspace technique, by adding to the spacetime coordinates $x^\mu$ and anticommuting one $\vartheta$, but, simultaneously, we enlarge the spacetime with the addition of an auxiliary commuting parameter $s$, $0\leq s\leq 1$. So the local coordinates in the superspace are $(x^\mu,s,\vartheta)$. In particular we have the following
\be
e^{-s\left(\sigma+\vartheta \mfs \sigma\right)} = e^{-s\sigma}+\vartheta\, {\mathfrak{s}} e^{-s\sigma}.\label{expssigma}
\ee

On this superspace the superconnection is
\be
\widetilde \EA(x,s,\vartheta)&=& e^{-s\left(\sigma+\vartheta \mfs \sigma\right)}e^{-\vartheta c} \left({\tilde d}+ A \right) e^{\vartheta c} e^{s\left(\sigma+\vartheta \mfs \sigma\right)}\label{EAtilde}\\
&=& e^{-s\left(\sigma+\vartheta \mfs \sigma\right)} \left(\EA + d+\frac {\partial}{\partial s}ds\right)e^{s\left(\sigma+\vartheta \mfs \sigma\right)},\0
\ee
where $\tilde d= d + \frac {\partial}{\partial s}ds+ \frac {\partial}{\partial \vartheta}d\vartheta$, and
$\EA =A + \vartheta \left(dc+ [A,c]\right) + \left(c-\vartheta cc\right)d\vartheta$.
We decompose $\widetilde \EA$ as follows:
\be
\widetilde\EA(x,s,\vartheta)&=& \phi(x,s,\vartheta)+\phi_s (x,s,\vartheta)ds + \phi_\vartheta(x,s,\vartheta)d\vartheta,\label{EAxsvartheta}\\
 \phi(x,s,\vartheta)&=& A_s + \vartheta \left(dc_s +[A_s,c_s]\right),\0\\
\phi_s(x,s,\vartheta)&=& \sigma + \vartheta\mfs \sigma,\0\\
\phi_\vartheta (x,s,\vartheta)&=& c_s -\vartheta c_s c_s ,\0
\ee
where
\be
A_s&=&  e^{-s\sigma} A  \, e^{s\sigma}+  e^{-s\sigma} d   e^{s\sigma},\0\\
c_s&=&  e^{-s\sigma} c \,  e^{s\sigma}+  e^{-s\sigma} \mfs  e^{s\sigma}.\label{Ascs}
\ee
In particular, $A_0=A$ and $c_s$ interpolates between $c$ and $0$. Since the derivative with respect to $\vartheta$ corresponds to the BRST transformation, we deduce
\be
\mfs A_s = dc_s +[A_s,c_s], \quad\quad\quad \mfs c_s= -c_s c_s.\label{sAsscs}
\ee
Moreover, if we denote by $\widetilde\EF, \EF$ and $F$ the curvatures of $\widetilde \EA, \EA$ and $A$, respectively, we have
\be
\widetilde \EF = e^{-s\left(\sigma+\vartheta \mfs \sigma\right)} \EF\, e^{s\left(\sigma+\vartheta \mfs \sigma\right)}=  e^{-s\left(\sigma+\vartheta \mfs \sigma\right)}e^{-\vartheta c}A\,
 e^{s\left(\sigma+\vartheta \mfs \sigma\right)}e^{\vartheta c}.\label{tildeEFEFF}
\ee
Now suppose the spacetime $\sfM$ has dimension $\sf d$ and choose any ad-invariant polynomial
$P_n$ with $n=\frac {\sf d}2-1$. Then the following holds
\be
P_n (\widetilde \EF,\ldots,\widetilde \EF)=P_n(\EF,\ldots,\EF) = P_n(F,\ldots,F) =0, \label{russian}
\ee
where the last equality holds for dimensional reasons. But now we can write
\be
P_n (\widetilde \EF,\ldots,\widetilde \EF)= \tilde d\left( n\int_0^1 dt\, P_n({\widetilde \EA},\widetilde \EF_t,\ldots, \widetilde \EF_t)\right)\equiv \tilde d \,\left( TP_n(\widetilde \EA)\right).\label{PntildeEA}
\ee
For notational simplicity let us set $\widetilde{\cal Q}(\widetilde \EA)\equiv TP_n(\widetilde \EA)$  and decompose it in the various components according to the form degree
\be
\widetilde{\cal Q}= \sum_{\stackrel {k,i,j}{k+i+j=2n-1} }\widetilde{\cal Q}^j_{(k,i)},\quad\quad
\widetilde{\cal Q}^j_{(k,i)}= \left( {Q}^j_{(k,i)}+ \vartheta  \hat {Q}^j_{(k,i)}\right) (d\vartheta)^j (ds)^i,\label{Qkij}
\ee
where $k$ denotes the form-degree in spacetime, $j$  is the ghost number and $i$ is either 0 or 1. Next let us decompose the equation $\tilde d \widetilde {\cal Q}=0$ in components and select, in particular, the component $(2n-2,1,1)$, i.e.
\be
0= d \widetilde {\cal Q}^1_{(2n-3,1)}+ \frac {\partial}{\partial\vartheta} \widetilde {\cal Q}^0_{(2n-2,1)}d\vartheta+\frac {\partial}{\partial s} \widetilde {\cal Q}^1_{(2n-2,0)}ds,\label{dQkij}
\ee
and let us integrate it over $\sfM$ and $s$. We get
\be
0= \int_{\sfM}\int_0^1 ds\, \mfs{ Q}^0_{(2n-2,1)}+ \int_{\sfM}\int_0^1 ds\,\frac {\partial}{\partial s}  { Q}^1_{(2n-2,0)}.\label{dQkij1}
\ee
Since ${ Q}^1_{(2n-2,0})$ is linear in $c_s$ and $c_0=c, c_1=0$, we get finally
\be
 \int_{\sfM}{ Q}^1_{(2n-2,0)}= \mfs \int_{\sfM}\int_0^1 ds\,{ Q}^0_{(2n-2,1)}.\label{intQk01}
\ee
Now we remark that $ \int_{\sfM}{ Q}^1_{(2n-2,0})$ is linear in $c$ and coincides precisely with the anomaly. On the other hand, ${ Q}^0_{(2n-2,1)}$ has the same expression as the anomaly with $c$ replaced by $\sigma$ and $A$ by $A_s$, i.e.
\be
{ Q}^0_{(2n-2,1)}(\sigma, A_s)= n(n-1) \int_0^1 dt \,(1-t)\, P_n (d\sigma,A_s, F_{s,t},\ldots,F_{s,t}),\label{Q02n-21}
\ee
where  $F_{s,t}=tdA_s +\frac {t^2}2[A_s,A_s]$.  We call
\be
{\cal B}_{WZ}= \int_{\sfM}\int_0^1 ds\,{ Q}^0_{(2n-2,1)},\label{calBWZ}
\ee
the {\it Wess-Zumino term. }

The existence of ${\cal B}_{WZ}$ for any anomaly seems to contradict the non-triviality of anomalies. This is not so, because the price we have to pay to construct the term \eqref{calBWZ} is the introduction of the new fields $\sigma^a$ which are not present in the initial theory. The proof of non-triviality of anomalies is based on a definite differential space formed by $c,A$ and their exterior derivatives and commutators which constrains the anomaly to be a polynomial in these fields. 
Of course, in principle, it is not forbidden to enlarge the theory by adding new fields plus the WZ term. But the resulting theory is different from the initial one. Moreover, the $\sigma^a$ fields have 0 canonical dimension. This means that, except in $2\sf d$, it is possible to construct new invariant action terms with more than two derivatives, which renders renormalization problematic.

\section{HS-YM models and superfield method}

In this section we apply the superfield method to higher-spin\footnote{{Basic literature on higher spin theories can be found in \cite{HS}}} Yang-Mills (HS-YM) models. These models are characterized by a local gauge symmetry,  the higher spin symmetry, with infinite parameters, which encompass, in particular, both ordinary gauge transformations and diffeomorphisms. In a sense they unify ordinary gauge and gravitational theories. This makes them interesting in themselves, but particularly for the superfield method, to whose bases they seem to perfectly adhere.  These models have only recently been introduced in the literature and they are largely unexplored. For this reason we devote a rather long and hopefully sufficiently detailed introduction .

{HS-YM models {in Minkowski spacetime} are formulated in terms of master fields $h_a(x,p) $, which are local in the phase space $(x,p)$, with $[\hat x^\mu,\hat p_\nu]=i\hbar \delta^\mu_\nu$ ($\hbar$ will be set to the value 1), where $\hat x, \hat p$ are the operators whose classical symbols  are $x, p$ according to the Weyl-Wignar quantization.} The master field can be expanded in powers of $p$,
\be
h_a(x,p) &=& \sum_{n=0}^\infty \frac 1{n!} h_a^{\mu_1\ldots,\mu_n}(x) p_{\mu_1}\ldots p_{\mu_n}\0\\
&=&A_a(x) +\chi_a^\mu(x) p_\mu 
+ \frac 12 b_a^{\mu\nu}(x) p_\mu p_\nu+\frac 16 c_a^{\mu\nu\lambda}(x) p_\mu
p_\nu p_\lambda + \ldots,\label{haunn}
\ee
where $ h_a^{\mu_1\ldots\mu_n}(x)$ are ordinary tensor fields, symmetric in $\mu_1,\ldots,\mu_n$. The indices $\mu_1,\ldots,\mu_n$ are upper (contravariant)
Lorentz indices, $\mu_i=0,\ldots,d-1$. The index $a$ is also a vector index, but it is of a different nature. In fact it will be interpreted as a flat index and $h_a$ will be referred to as a {\it frame-like master field}. Of course, when the background metric {is flat} all indices are on the same footing, but it is preferable to keep them distinct to facilitate the correct interpretation.

The master field $h^a(x,p)$ can undergo the following (HS) gauge transformations, whose infinitesimal parameter $\varepsilon(x,p)$ is itself a master field, 
\be
\delta_\varepsilon h_a(x,p) = \partial_a^x 
\varepsilon(x,p)-i [h_a(x,p) \stackrel{\ast}{,} \varepsilon(x,p)] 
\equiv {\cal D}^{\ast x}_a  \varepsilon(x,p),
\label{deltahxp}
\ee
where we have introduced the covariant derivative
\be
{\cal D}^{\ast x}_a = \partial_a^x- i  [h_a(x,p) \stackrel{\ast}{,}\quad].\label{covderivative}
\ee 
The $\ast$ product is the Moyal product, defined by
\be
f(x,p)\ast g(x,p) = f(x,p) e^{\frac i2 \left( \stackrel{\leftarrow}{\partial}_x \stackrel{\rightarrow}{\partial}_p - \stackrel{\leftarrow}{\partial}_p\stackrel{\rightarrow}{\partial}_x\right)}g(x,p), \label{Moyal}
\ee
between two regular phase-space functions $f(x,p)$ and $g(x,p)$.

Like in ordinary gauge theories, we use the compact notation  ${ d}= \partial_a\, dx^a, { h}= h_a dx^a$  and  write \eqref{deltahxp} as
\be
\delta_\varepsilon {  h}(x,p) ={  d} 
\varepsilon(x,p)-i [{ h} (x,p) \stackrel{\ast}{,} \varepsilon(x,p)] \equiv 
 D \varepsilon (x,p),\label{deltahxpbf}
\ee
 where it is understood that $  [{ h} (x,p) \stackrel{\ast}{,} \varepsilon(x,p)]=  [h_a (x,p)
\stackrel{\ast}{,} \varepsilon(x,p)]dx^a$.

Next, we introduce the curvature notation
\be
{ G} = {  d} {  h}
 -\frac i 2 [ {  h}\stackrel {\ast}{,}{  h}],\label{curv1} 
\ee
with the transformation property $\delta_\varepsilon {  G} = -i [{ G}  \stackrel {\ast}{,}\varepsilon]$.

The action functionals we will consider are  integrated polynomials of $ G$ or of
its components $G_{ab}$. To imitate ordinary non-Abelian gauge theories, 
 we need a `trace property', similar to the trace of polynomials of Lie algebra generators.  In this framework, we have
\be
\langle\!\langle f\ast g\rangle\!\rangle\equiv \int d^dx \int \frac
{d^dp}{(2\pi)^d}
f(x,p)\ast g(x,p) = \int d^dx \int \frac {d^dp}{(2\pi)^d} f(x,p) g(x,p)=
\langle\!\langle g\ast f\rangle\!\rangle.  \label{trace}
\ee
From this, plus associativity, it follows that
\be
&&\langle\!\langle f_1 \ast f_2\ast \ldots \ast f_n\rangle\!\rangle=
\langle\!\langle f_1 \ast (f_2\ast \ldots \ast f_n)\rangle\!\rangle\0\\
&&=(-1)^{\epsilon_1(\epsilon_2+\ldots+\epsilon_n)} \langle\!\langle  (f_2\ast
\ldots \ast f_n)\ast
f_1\rangle\!\rangle=(-1)^{\epsilon_1(\epsilon_2+\ldots+\epsilon_n)} 
\langle\!\langle  f_2\ast \ldots \ast f_n\ast f_1\rangle\!\rangle,\label{cycl}
\ee
where $\epsilon_i$ is the Grassmann degree of $f_i$ ({this is usually referred to as  {\it ciclicity} property}).

This property holds also when the $f_i$ are valued in a (finite dimensional) Lie algebra, provided
the symbol $\langle\!\langle   \quad  \rangle\!\rangle$ includes also the trace over the
Lie algebra generators.

\vskip 0,5cm
{\bf The HS Yang-Mills action.}
The curvature components, see \eqref{curv1}, are
\be
G_{ab}= \partial_a h_b - \partial_b h_a -i [h_a \stackrel{\ast}{,} h_b ],
\label{Gab}
\ee
with transformation rule
\be
\delta_\varepsilon G_{ab}=-i [G_{ab}\stackrel{\ast}{,}
\varepsilon].\label{deltaGab}
\ee
If we consider the functional $\langle\!\langle G^{ab} \ast G_{ab} \rangle\!\rangle$, it follows from the above that
\be 
\delta_\varepsilon \langle\!\langle G^{ab} \ast G_{ab} \rangle\!\rangle = -i
\langle\!\langle 
   G^{ab} \ast G_{ab} \ast\varepsilon -\varepsilon \ast  G^{ab} \ast G_{ab}
\rangle\!\rangle=0. \label{YMinvariance}
\ee
Therefore 
\be
{\cal YM}({h})=- \frac 1{4 g^2}\langle\!\langle G^{ab} \ast G_{ab}
\rangle\!\rangle,\label{YMh}
\ee
is invariant under HS gauge transformations and it is a well defined  functional.
This is the HS-YM-like action.

From \eqref{YMh}  we get the following eom:
\be
\partial_b G^{ab} -i [h_b\stackrel{\ast}{,} G^{ab}]\equiv{\cal D}_b^\ast
G^{ab}=0,\label{YMeom}
\ee
which is covariant by construction under HS gauge transformations
\be
\delta_\varepsilon\left( {\cal D}_b^\ast G^{ab}\right)= -i[ {\cal D}_b^\ast
G^{ab},\varepsilon].\label{covYMeom}
\ee
We recall that ${\cal D}_b^\ast$ is the covariant $\ast$-derivative and $\varepsilon$ the HS gauge parameter.

\vskip 0,5cm
All that has been said so far can be repeated for
non-Abelian models with minor changes. But for simplicity here we limit ourselves to the Abelian case.

\vskip 0,5cm
{\bf Gravitational interpretation.}
The novel property of HS YM-like theories is that, nothwithstanding their evident similarity with ordinary YM theories, they can describe also gravity. To see this, let us expand the gauge master parameter $\varepsilon(x,p)$
\be
\varepsilon(x,p)&=& \epsilon(x) +\xi^\mu(x) p_\mu+\frac 12
\Lambda^{\mu\nu}(x)p_\mu
p_\nu+\frac 1{3!} \Sigma^{\mu\nu\lambda}(x)p_\mu
p_\nu p_\lambda+\ldots\label{epsxu}
\ee
 In Appendix C we show that the parameter $\epsilon(x)$ is the usual $U(1)$ gauge parameter for the field $A_a(x)$, while $ \xi^\mu(x) $ is the parameter for general coordinate transformations and $\chi_a^\mu(x)$ can be interpreted as the fluctuating inverse vielbein field.

\vskip 0.5cm
{\bf Scalar and spinor master fields.} To HS YM-like theories we can couple matter-type fields of any spin. For instance, a complex multi-index boson field 
\be
\Phi(x,p)= \sum_{n=0}^\infty \frac 1{n!}\Phi^{\mu_1\mu_2\ldots \mu_n}(x)
 p_{\mu_1}p_{\mu_2}\ldots p_{\mu_n},\label{Phixu}
\ee
which, under a master gauge transformation \eqref{deltahxp}, transforms like $
\delta_\varepsilon \Phi = i \varepsilon\ast\Phi$.
The covariant derivative is $ \ED^\ast_a \Phi= \partial_a \Phi -i h_a \ast \Phi$ with the property $\delta_\varepsilon \ED^\ast_a \Phi= i\,\varepsilon \ast \ED^\ast_a \Phi$.
With these properties the kinetic action term $
\frac 12 \langle\!\langle(\ED_\ast^a
\Phi)^\dagger\ast\ED^\ast_a \Phi\rangle\!\rangle$ and the potential terms 
$ \langle\!\langle(\Phi^\dagger \ast \Phi)^n_\ast\rangle\!\rangle$ are HS-gauge invariant.

\vskip 0.5cm

In a quite similar manner, we can introduce master spinor fields,
\be
\Psi(x,p) = \sum_{n=0}^\infty\frac 1{n!} \Psi_{(n)}^{\mu_1\ldots \mu_n}(x)
p_{\mu_1} \ldots p_{\mu_n}, \label{Psi}
\ee
where $\Psi_{(0)}$ is a Dirac field. The HS gauge transformations are
$\delta_\varepsilon \Psi = i \varepsilon\ast\Psi $, and the covariant derivative
is $ \ED^\ast_a \Psi= \partial_a \Psi -i h_a \ast \Psi$ with $\delta_\varepsilon (\ED^\ast_a \Psi)=i\varepsilon \ast (\ED^\ast_a \Psi)$.

With these properties the action integral
\be
S(\Psi,h) = \langle\!\langle \overline \Psi i\gamma^a \ED_a \Psi
\rangle\!\rangle
=  \langle\!\langle \overline \Psi \gamma^a\left(i\partial_a+ h_a \ast\right)
\Psi \rangle\!\rangle,
\label{Spsih}
\ee
is invariant.

\vskip 0.5cm
{\bf BRST quantization of HS Yang-Mills.}
Fixing the Lorenz gauge  with parameter $\alpha$ and applying the standard
Faddeev-Popov approach, the quantum action becomes
\be
{\cal Y}{\cal M}(h_a,c,B)=\frac 1{g^2} \langle\!\langle  -\frac 1{4 } G_{ab}
\ast G^{ab} -
h^a\ast \partial_a B-i \partial^a \overline c\ast {\cal D}_a^\ast c +\frac
{\alpha}2 B\ast B  \rangle\!\rangle,\label{YMhquantum}
\ee
where $c, \overline c$ and $B$ are the ghost, antighost and Nakanishi-Lautrup
master fields, respectively. $c, \overline c$ are anticommuting fields, while
$B$ is commuting.

The action \eqref{YMhquantum} is invariant under the BRST transformations
\be
s h_a = {\cal D}_a^\ast c,\quad\quad 
s c= i c\ast c = \frac i2 [c \stackrel{\ast}{,} c], \quad\quad s \overline c= i B,\quad\quad
s B=0,\label{BRSTtr}
\ee
which are nilpotent. In particular $s( {\cal D}_a^\ast c)=0$ and $s (c\ast c)=0.$

\subsection{Anomalies in HS theories}

The effective action of HS-YM theories are functional of the master field $h_a(x,p)$ defined as follows:
\be
\EW[h] &=&\EW[0]\label{EW}\\
&&+\sum_{n=1}^\infty\, \frac {i^{n-1}}{n!}\int \prod_{i=1}^n d^dx_i\,
\frac {d^dp_i}{(2\pi)^d}\, \langle J_{a_1}(x_1,p_1) \ldots  
J_{a_n}(x_n,p_n)\rangle\, h^{a_1}(x_1,p_1) \ldots  h^{a_n}(x_n,p_n),\0
\ee 
where $J_a(x,p)$ are fermion master currents coupled to  $h_a(x,p)$. In a quantum theory it may happen that the HS symmetry is not preserved (the Ward identity is violated)
\be
\delta_\varepsilon \EW[h] = \EA[\varepsilon, h]\neq 0, \label{A1}
\ee
but in this case we have a consistency condition. Since
\be 
\left(\delta_{\varepsilon_2 }\delta_{\varepsilon_1 }- \delta_{\varepsilon_1
}\delta_{\varepsilon_2}\right) h_a(x,p) &=&  i\left( \partial_x
[{\varepsilon_1
}\stackrel{\ast}{,}{\varepsilon_2 }](x,p) -i [h_a(x,p)\stackrel{\ast}{,} 
[{\varepsilon_1 }\stackrel{\ast}{,}{\varepsilon_2 }](x,p) ]] \right)
\0 \\
&=&   i\, {\cal D}^{\ast x}_{a} [{\varepsilon_1}\stackrel{\ast}{,}{\varepsilon_2
}](x,p),\label{e1e2}
\ee
holds, we must have
\be
\delta_{\varepsilon_2}\EA[\varepsilon_1, h]-\delta_{\varepsilon_1}
\EA[\varepsilon_2, h]=
\EA[[\varepsilon_1\stackrel{\ast}{,} \varepsilon_2],h].\label{A2}
\ee
If we exclude the possibility that $\EA[\varepsilon, h]$ is
trivial (i.e. $\EA[\varepsilon, h]= \delta_\varepsilon \EC[h]$, for an  integrated local counterterm
$\EC[h]$), then we are faced with a true anomaly, which breaks the covariance of the effective action.

As illustrated above, in ordinary gauge theories, the form of chiral anomalies (and the CS action) is given by elegant formulas: the descent equations.  It is natural to inquire whether in HS-YM theories there are similar formulas.  We would be tempted to derive the relevant descent equations by mimicking the constructions of the previous sections. For instance, beside \eqref{curv1} we can introduce the standard (ordinary gauge theory) definitions
\be
 { G}_t = {d}
{h}_t
 -\frac i 2 [ {h}_t \stackrel {\ast}{,}{h}_t],\quad\quad {h}_t= t
{ h},\quad\quad h=h_a dx^a,\quad d=\partial_a d x^a. \label{Gt}
\ee

The difference in the HS case is that, unlike in ordinary gauge theories, we
cannot use graded commutativity. There is also another difficulty, the already signaled trace problem. We can use eqs.(\ref{trace},\ref{cycl}), but we have to integrate over the full phase-space. Therefore it is impossible to reproduce the unintegrated descent
equations in the same way as in the ordinary gauge theories. The best we can do is try to reproduce each  equation separately in integrated form. But while it is rather simple, starting from the expression with $n$ $ G$ entries  
\be
\langle\!\langle G \ast G \ast \ldots\ast G \rangle\!\rangle,\label{GGG}
\ee
to derive the Chern-Simons action in $2n\!\!-\!\!1$ dimension
\be
{\cal CS}(h)=n \int_0^1 dt \langle\!\langle h  \ast G_t \ast \ldots
\ast G_t
 \rangle\!\rangle,\label{CSh}
\ee
and prove that $\delta_\varepsilon {\cal CS}(h)=0$,  the remaining derivations are unfortunately lengthy and very cumbersome.  

It is at this point that the superfield formulation of HS-YM theories comes to our rescue and gives us all these relations for free.

\subsection{Superfield formulation of HS YM}

We introduce the master superfield 1-form
\be
{\mathbb H}= {\mathbf h}_a(x,p,\vt,\vtb) dx^a + {\boldsymbol{ \phi}} (x,p,\vt,\vtb) d\vtb +\overline {\boldsymbol {\phi}}(x,p,\vt,\vtb) d\vt ,\label{mathsfH}
\ee
where all the component masterfields ${\mathbf h}_a, {\boldsymbol{\phi}}, \overline {\boldsymbol{\phi}}$ are valued in the Lie algebra with generators $T^\alpha$. The component master fields are
\be
{\mathbf h}_a(x,p,\vt,\vtb)&=& h_a(x,p) +\vt \bar \zeta_a(x,p) + \vtb \zeta_a(x,p) + \vt\vtb t_a(x,p),\label{mathbbh}\\
{\boldsymbol{ \phi}}(x, p, \vt,\vtb)&=& c(x,p)+ \vt \bar B(x,p)+\vtb R(x,p) + \vt\vtb \varsigma(x,p),\label{boldphi}\\
\overline{\boldsymbol{ \phi}}(x, p, \vt,\vtb)&=& \bar c(x,p)+ \vt \bar R(x,p)+\vtb B(x,p) + \vt\vtb \bar\varsigma(x,p).\label{boldbarphi}
\ee

The supercurvature is
\be
{\mathbb G}={\tilde  d}   {\mathbb H}-\frac i2 [  {\mathbb H} \stackrel{\ast}, {\mathbb H}], \label{mathbbG}
\ee
where ${\tilde d}= d+ \frac {d}{d\vt} d{\vt}+  \frac {d}{d\vtb} d{\vtb}$. Notice that the $\ast$ commutator on the RHS persists also in the Abelian case. Setting
\be
 {\bf G}\equiv d {\bf h} -\frac i2 [  {\bf h} \stackrel{\ast}, {\bf h}],\quad\quad 
{\bf h}=  {\mathbf h}_a(x,p,\vt,\vtb) dx^a ,\label{bgG}
\ee
the horizontality condition in this case is
\be
{\mathbb G}= {\bf G} .\label{hormathbfG}
\ee
The procedure is the same as in the ordinary YM case except that ordinary products are replaced by $\ast$ products. From the vanishing of the $dx^\mu\wedge d\vt$ and  $dx^\mu\wedge d\vtb$  component of ${\mathbb G}$, we get
\be
&&\zeta_a(x,p) = D^\ast_a c(x,p), \quad\quad \overline\zeta_a(x,p) = D^\ast_a \bar c(x,p),\label{zetaaxu}\\
&& t_a(x,p) = D^\ast _a B(x,p) +i [ D_a^\ast c(x,p)\stackrel{\ast}, \bar c(x,p)],\label{zetataxu}
\ee
where $D^\ast_a = \partial_a^x -i[h_a(x,p)\stackrel{\ast},\quad]$, see \eqref{covderivative}.
From the $d\vt \vee d\vtb$, $d\vt \vee d\vt$ and $d\vtb \vee d\vtb$ components we get the CF restriction
\be
B(x,p)+\overline B(x,p)-i [c(x,p)\stackrel{\ast}, \bar c(x,p)]=0, \label{CFgauge}
\ee
and 
\be
&&	R(x,p) = \frac i2 [c(x,p) \stackrel{\ast}, c(x,p)], \quad\quad \overline R(x,p) =\frac i2 [\bar c(x,p) \stackrel{\ast},\bar  c(x,p)],\label{RbarR}\\
&& \varsigma (x,p) =- i[\overline B(x,p) \stackrel{\ast}, c(x,p)],\quad\quad
 \bar \varsigma (x,p) = i[B(x,p) \stackrel{\ast}, \bar c(x,p)].\0
\ee
 
Let us recall again that the $\ast$-commutators are present also in the Abelian case we are considering. Expanding, for instance, the $c$ ghost master field in ordinary field components according to eq.\eqref{epsxu}, we get
\be
c(x,p)&=&{c}(x) +{c}^\mu(x) p_\mu+\frac 12
{c}^{\mu\nu}(x)p_\mu
p_\nu+\frac 1{3!} { c}^{\mu\nu\lambda}(x)p_\mu
p_\nu p_\lambda+\ldots, \label{cxu}
\ee
where now all the component fields are anticommuting. ${c}$ is the gauge ghost field,
${c}^\mu$ is the diffeomorphism ghost, and the subsequent are the HS ghost fields.
For instance, the BRST transformations of $A_a$ and $\chi_a^\mu$ are
\be
sA_a&=& \partial_a \, { c} - {c} \!\cdot\! \partial A_a -\partial_\lambda \, { c}_a^\lambda+\ldots,\label{sAa}\\
s\chi_a^\mu &=& \partial_a{ c}^\mu + {c}\!\cdot \! \chi_a^\mu - \partial_\rho {c}^\mu \chi_a^\rho + \partial^\rho A_a { c}_\rho{}^\mu - \partial_\rho {c} \, b_a{}^{\rho\mu}+\ldots \label{schiamu}
\ee

For later use we introduce the notations
\be
\frac {d}{dt} {\mathbb G}_t&=&\tilde d{\mathbb H} -i t[{\mathbb H} \stackrel {\ast}{,} {\mathbb H}]= \tilde d_{t}{\mathbb H}, \quad\quad
\tilde d_{t} = \tilde d -i [{\mathbb H}_t\stackrel {\ast}{,}\quad],\label{dGt}\\
\tilde d {\mathbb G}_t& =& i [{\mathbb H}_t\stackrel {\ast}{,}{\mathbb G}_t],\quad\quad \delta {\mathbb G}_t =
\tilde d \delta {\mathbb H}_t -i [ {\mathbb H}_t \stackrel {\ast}{,} \delta{\mathbb H}_t] = \tilde d_t \delta
{\mathbb H}_t,\label{deltaGt}
\ee
and analogous formulas for $\bfG_t=t\left( \tilde d \, \bfh -i\frac {t}2[\bfh \stackrel{\ast}{,} \bfh]\right)$ and $G_t=t\left( \tilde d \, h -i\frac {t}2[h \stackrel{\ast}{,} h]\right)$.

\subsection{Derivation of HS-YM  anomalies}

In order to limit the size and complication of formulas we limit ourselves to only one anticommuting variable $\vartheta$ and consider the superderivation ($\tilde d =d +\frac {\partial}{\partial\vt} d\vt$)
\be
{\mathbb H}= e^{i\vt c} \ast (i\tilde d+ h \ast)  e^{-i\vt c}= h +\vt (dc -i[h\stackrel {\ast}{,}c]) + \left( c+i\vt\frac 12[c\stackrel{\ast}{,}c]\right)d\vt\equiv {\bf h} + {\boldsymbol \phi}\, d\vt,\label{mathbbH}
\ee
where $c=c(x,p)$.
It follows that the supercurvature is
\be
{\mathbb G} = e^{i\vt c} \ast G \ast  e^{-i\vt c}={\bf G}\equiv  G-i\vt [F\stackrel{\ast}{,}c]. \label{mathbbG}
\ee
From these formulas it is immediately visible that the 
derivative with respect to $\vt$ corresponds to the BRST transformations:
\be
\frac {\partial}{\partial\vt} {\bf h} &=& dc -i[ h\stackrel{\ast}{,}c] = D^\ast_h c = \mfs h, \0\\
\frac {\partial}{\partial\vt} {\boldsymbol \phi} &=& \frac i2[c\stackrel{\ast}{,}c]=\mfs c,\0\\
\frac {\partial}{\partial\vt} {\bf G}&=&-i [G\stackrel{\ast}{,}c]=\mfs G.\0
\ee

 Let us start from the phase space integral with $n$ $ {\mathbb G} $ entries  
\be
\langle\!\langle {\mathbb G} \ast  {\mathbb G} \ast \ldots\ast  {\mathbb G} \rangle\!\rangle,
\ee
where $\langle\!\langle\quad \rangle\!\rangle$ means integration over a phase space of dimension $<4n$. Let us  introduce the notations
\be
\frac {d}{dt} {\mathbb G}_t&=&\tilde d\, {\mathbb H} -i t[{\mathbb H} \stackrel {\ast}{,} {\mathbb H}]= \tilde d_{t}\,{\mathbb H}, \quad\quad
\tilde d_{t} = \tilde d -i [{\mathbb H}_t\stackrel {\ast}{,}\quad],\quad\quad
\tilde d \,{\mathbb G}_t= i [{\mathbb H}_t\stackrel {\ast}{,}{\mathbb G}_t].\label{deltaGt}
\ee
Then, consider the expression with $n\!\!-\!\!1$ ${\mathbb G}_t$ entries
\be
\int_0^1 dt \langle\!\langle {\tilde d} \left({\mathbb H} \ast {\mathbb G}_t \ast \ldots
\ast{\mathbb G}_t \right) \rangle\!\rangle&\!\!=\!\!& \int_0^1 dt \langle\!\langle{\tilde d}\, {\mathbb H} \ast
{\mathbb G}_t \ast \ldots\ast {\mathbb G}_t  \rangle\!\rangle-  \int_0^1 dt \langle\!\langle  {\mathbb H} \ast {\tilde d} \,{\mathbb G}_t \ast
\ldots\ast{\mathbb G}_t  \rangle\!\rangle-\ldots\0\\
\dots- \int_0^1 dt \langle\!\langle {\mathbb H}
\ast{\mathbb G}_t \ast \ldots\ast {\tilde d}\, {\mathbb G}_t  \rangle\!\rangle&\!\!=\!\!&\int_0^1 dt \Big(\langle\!\langle {\tilde d}\,  {\mathbb H} \ast {\mathbb G}_t \ast \ldots\ast
{\mathbb G}_t  \rangle\!\rangle-   i \langle\!\langle  {\mathbb H} \ast
[ {\mathbb H}_t\stackrel{\ast}{,}{\mathbb G}_t ]\ast \ldots\ast{\mathbb G}_t 
\rangle\!\rangle-\ldots\0\\
&& \ldots-  i \langle\!\langle  {\mathbb H} \ast{\mathbb G}_t \ast \ldots\ast 
[ {\mathbb H}_t\stackrel{\ast}{,}{\mathbb G}_t ]\Big)  \rangle\!\rangle,\label{cs1}
\ee
using the last of \eqref{deltaGt}. Then, using the first of \eqref{deltaGt} together
with \eqref{cycl}, this becomes 
\be
&=&
\int_0^1 dt \langle\!\langle({\tilde d}\, {\mathbb H}-i [ {\mathbb H}_t\stackrel{\ast}{,}{\mathbb H}]) \ast
{\mathbb G}_t \ast \ldots\ast {\mathbb G}_t  \rangle\!\rangle
= \int_0^1 dt \langle\!\langle(\frac {d}{dt}{\mathbb G}_t   \ast {\mathbb G}_t \ast
\ldots\ast {\mathbb G}_t  \rangle\!\rangle\label{cs2}\\
&=&\frac 1n \int_0^1dt \, \frac d{dt} \langle\!\langle {\mathbb G}_t   \ast {\mathbb G}_t \ast
\ldots\ast {\mathbb G}_t  \rangle\!\rangle
= \frac 1n \langle\!\langle {\mathbb G} \ast {\mathbb G} \ast \ldots\ast {\mathbb G}
\rangle\!\rangle=\frac 1n \langle\!\langle {\bf G} \ast {\bf G} \ast \ldots\ast {\bf G}
\rangle\!\rangle=0. \0
\ee
Since they are integrated over a spacetime of dimension $\sfd<2n$, these expressions
vanish.However this is the way we identify the primitive functional action for HS CS (if the spacetime dimension is  $\sfd=2n\!\!-\!\!1$). In other words the HS CS action is
\be
{\cal CS}(h)=n \left.\int_0^1 dt \langle\!\langle \bfh  \ast \bfG_t \ast \ldots
\ast\bfG_t
 \rangle\!\rangle\right\vert_{\vt=0}= n \int_0^1 dt \langle\!\langle h  \ast G_t \ast \ldots
\ast G_t
 \rangle\!\rangle,\label{CSh}
\ee
where $\langle\!\langle\quad \rangle\!\rangle$ means now integration over a phase space of dimension $4n\!-\!2$.

Expressions relevant to anomalies appear if the spacetime dimension is  $\sfd=2n\!\!-\!\!2$ and the phase space one is $\sfd=4n\!\!-\!4$. In this case, the (unintegrated) expression
\be
&&\int_0^1 dt\left({\mathbb H} \ast {\mathbb G}_t \ast \ldots\ast{\mathbb G}_t +{\mathbb G}_t \ast{\mathbb H}\ast\dots \ast {\mathbb G}_t+\ldots + {\mathbb G}_t \ast{\mathbb G}_t  \ast \ldots \ast{\mathbb H}\right) \0\\
&&= \sum_{i=0}^{2n-1}Q_{2n-i-1}^{(i)}(\bfh,\boldsymbol{\phi}) \underbrace{d\vt \wedge \ldots \wedge d\vt}_{i\,\,{\rm factors}},\label{polyform}
\ee
is a spacetime polyform of degree $ 2n\!\!-\!\!1, \ldots,1,0$: $i$ represents the ghost number and $2n\!\!-\!\!i\!\!-\!\!1$ is the spacetime form degree. Of course its components of degree $2n$ and $2n\!\!-\!\!1$ vanish for dimensional reasons. Then, excluding vanishing factors and recalling that $\frac {\partial}{\partial\vt}$ corresponds to the BRST transform $\mfs$, the integrand of eq.\eqref{polyform} can be written as follows
\be
&& \sum_{i=2}^{2n-1}d Q_{2n-i-1}^{(i)}(\bfh,\boldsymbol{\phi})\underbrace{d\vt \wedge \ldots \wedge d\vt}_{i\,\,{\rm factors}}+  \sum_{i=1}^{2n-1}\frac {\partial}{\partial\vt} Q_{2n-i-1}^{(i)}(\bfh,\boldsymbol{\phi})\underbrace{d\vt \wedge \ldots \wedge d\vt}_{i+1\,\,{\rm factors}}\0\\
&=& \left(\sum_{i=2}^{2n-1}d Q_{2n-i-1}^{(i)}(\bfh,\boldsymbol{\phi})+ \sum_{i=2}^{2n-2} s Q_{2n-i }^{(i-1)}(\bfh,\boldsymbol{\phi})\right) \underbrace{d\vt \wedge \ldots \wedge d\vt}_{i\,\,{\rm factors}}.\label{decomposition}
\ee
Now this decomposition must be inserted inside the integration symbol $\langle\!\langle\cdot \rangle \!\rangle$. This symbol needs a specification and must be interpreted as follows: any form $Q_{2n-i-1}^{(i)}\sim dx^{\mu_1}\wedge d^{\mu_{2n-i-1}}$ is understood to be multiplied by a trivial factor $dx^{\mu_{2n-i}}\wedge \ldots \wedge dx^{\mu_{2n-2}}$, so that integration over spacetime makes sense. In conclusion, the following equation
\be
\int_0^1 dt \langle\!\langle {\tilde d} \left({\mathbb H} \ast {\mathbb G}_t \ast \ldots
\ast{\mathbb G}_t \right) \rangle\!\rangle=0,\label{cs0}
\ee
means
\be 
s \langle \!\langle Q_{2n-2}^{(1)} (\bfh,\boldsymbol{\phi})\rangle\!\rangle&=&0 ,\label{Q2n-21}\\
s \langle \!\langle Q_{2n-3}^{(2)} (\bfh,\boldsymbol{\phi})\rangle\!\rangle&=&0 ,\label{Q2n-32}\\
\ldots&=& \dots\0\\
s\langle \!\langle Q_{0}^{(2n-1)} (\boldsymbol{\phi})\rangle\!\rangle&=&0. \label{Q02n-1}
\ee
The anomaly is given by  $\left.\langle \!\langle Q_{2n-2}^{(1)} (\bfh,\boldsymbol{\phi})\rangle\!\rangle\right\vert_{\vt=0}$, and
\be
\langle \!\langle Q_{2n-2}^{(1)} (\bfh,\boldsymbol{\phi})\rangle\!\rangle &=&
n\int_0^1dt\, \Big( \langle \!\langle \boldsymbol{\phi}\ast {\bfG}_t\ast\ldots \ast \bfG_t \rangle\!\rangle + \langle \!\langle \bfh \ast \left(d_t \boldsymbol{\phi}-\partial_\vt \bfh\right)\ast  {\bfG}_t\ast\ldots \ast \bfG_t \rangle\!\rangle\0\\
&&\quad\quad\quad+\ldots+  \langle \!\langle \bfh \ast  {\bfG}_t\ast\ldots\ast \left(d_t \boldsymbol{\phi}-\partial_\vt \bfh\right)\Big) ,\label{Q2n-21a}
\ee
where $d_t= d-i t[\bfh,\quad]$.
It follows that
\be
\langle \!\langle Q_{2n-2}^{(1)} (\bfh,\boldsymbol{\phi})\rangle\!\rangle \vert_{\vt=0}&=& 
n\int_0^1dt\Big( \langle \!\langle c \ast {G}_t\ast\ldots \ast G_t \rangle\!\rangle -it (1-t)\langle \!\langle h \ast [h\stackrel{\ast}{,}c]\ast  {G}_t\ast\ldots \ast G_t \rangle\!\rangle\0\\
&&\quad\quad\quad+\ldots-i t (1-t)\langle \!\langle h \ast  {G}_t\ast\ldots\ast  G_t\ast   [ h\stackrel{\ast}{,}c]\Big). \label{Q2n-21a}
\ee
Now, using $\frac {d}{dt} G_t= d_{th} h= dh-it [h\stackrel{\ast}{,} h]$ and $dG_t= it[h,G_t]$ one can easily find the more familiar expression of the anomaly:
\be
\langle \!\langle Q_{2n-2}^{(1)} (\bfh,\boldsymbol{\phi})\rangle\!\rangle \vert_{\vt=0}&\!\!=\!\!& \!\!-
n\int_0^1\!dt\Big( \langle \!\langle dc\ast h \ast {G}_t\ast\ldots \ast G_t \rangle\!\rangle +\ldots+
\langle \!\langle dc \ast {G}_t\ast\ldots \ast G_t\ast h \rangle\!\rangle \Big).\label{Q2n-2final}
\ee

\section{The superfield formalism in supersymmetric gauge theories}

To conclude this review, we will present the ``BRST supergeometry" of an N=1 supersymmetric gauge theory formulated in the superspace. Let us start with a summary of the superspace presentation of this theory.  

\subsection{The supermanifold formulation of SYM}
 
From ch. XIII of \cite{WB}, a supersymmetric gauge theory can be introduced as
follows. One starts from a torsionful (but flat) superspace with
supercoordinates $z^M=(z^m, \theta^\mu,\bar\theta^{\dot\mu})$ and introduces a supervielbein basis
\be 
e^A(z)=dz^M\, e_M{}^A(z)\0
\ee
where $A=(a,\al,\dal)$ are flat indices. The vielbein satisfy
\be
e_A{}^Me_M{}^B= \delta_A{}^B,\quad\quad e_M{}^Ae_A{}^N= \delta_M{}^N,\quad\quad
\delta_M{}^N=\left(\begin{matrix}\delta_m{}^n&0&0\\
0&\delta_\mu{}^\nu &0\\ 0&0&\delta^{\dot\mu}{}_{\dot\nu}\end{matrix}\right).\0
\ee
The vielbein are {\it chosen} to be
\be
e_A{}^M= \left( \begin{matrix} \delta_a{}^m &0&0\\ i\sigma^m_{\al\dal}
\theta^{\dal}& \delta_\al{}^\mu&0\\
i\bar\theta^\al\sigma_{\al\dbeta }^m\epsilon^{\dbeta\dal}&
0&\delta^{\dal}{}_{\dot\mu}\end{matrix}\right),\quad\quad
e_M{}^A= \left( \begin{matrix} \delta_m{}^a& 0&0\\-i\sigma_{\mu\dot\mu}^a \bar
\theta^{\dot\mu}&\delta_\mu{}^\al &0\\
-i\theta^\nu \sigma_{\nu\dot\nu}^a \epsilon^{\dot\nu\dot\mu}&
0&\delta^{\dot\mu}{}_{\dal}\end{matrix}\right).\0
\ee 
In such a type of supergeometry one has
\be 
&&de^A= dz^Mdz^N \frac{\d}{\d z^N}e_M{}^A(z),\quad\quad{\rm i.e.}\0\\
&& de^a=-2i e^\alpha \sigma^a{}_{\al\dal} e^{\dal},\label{gaugeconstr}\\
&& de^{\al}=0,\0\\
&&de^{\dal}=0.\0
\ee
The flat indices derivatives $D_A= e_A{}^M \d_M$ correspond to
\be 
D_a= e_a^m \d_m= \d_a, \quad\quad D_\al= \frac {\d}{\d \theta^\al} +i
\sigma^m{}_{\al\dal} \bar \theta^{\dal} \d_m,\quad\quad \bar D_{\dal}=- \frac {\d}{\d
\bar\theta^{\dal}} -i\theta^\al  \sigma^m{}_{\al\dal}  \d_m ,\0
\ee
because in flat space $ e_a^m =\delta^m_a$. Moreover we have the following
\be 
\{D_\al , \bar D_{\dal}\}= -2i \sigma^m{}_{\al\dal} \d_m\0,\quad\quad \{D_\al ,
D_{\beta}\}=
\{\bar D_{\dal} , \bar D_{\dbeta}\}=0.\label{DaDb}
\ee
The superconnection is defined by
\be 
\phi = e^A \,\phi_A, \quad\quad \phi_A=i T^r\phi_A^r, \quad\quad 
\phi_m^r\Big{\vert}_{\theta=\bar\theta=0}=v_m^r,\label{superconnection}
\ee
where $v_m^r$ is the {\it ordinary} non-Abelian potential and $T^r$ are the Hermitean
generators of the gauge Lie algebra.

The gauge curvature is given by the superform
\be 
F=d\phi-\phi\phi= \frac 12 e^A e^B F_{BA}.\0\
\ee
On the flat basis, this becomes
\be 
F=de^A \phi_A +\frac 12 e^Ae^B\left(D_B\phi_A -(-)^{ab} D_A\phi_B
-\phi_B\phi_A+(-)^{ab}\phi_A\phi_B \right),\label{FAB}
\ee
where the torsion term is the first on the RHS. We have\footnote{{A reviewer of this paper pointed out to us that this result holds in the gauge-real representation of SYM theory. }}
\be 
F_{ab}\big{\vert}_{\theta=\bar\theta=0}=i T^r v_{ab}^r.\0
\ee 
The dynamics is determined by the super-Bianchi identity, $\ED F=d
F-[\phi,F]=0$. They are solved by the following conditions
\be 
F_{\al\beta}=F_{\dal\dbeta}= F_{\al\dbeta}=0,\label{F=0}
\ee
with further restrictions coming from :
\be 
\sigma^a{}_{\al\dot\gamma} F_{a\beta}+ \sigma^{a}{}_{\beta\dot\gamma}
F_{a\al}=0,\quad\quad \sigma^a{}_{\gamma\dbeta} F_{a\dot\al}+ \sigma^{a}{}_{\gamma\dot\al}
F_{a\dbeta}=0. \label{sigmaF=0}
\ee
This allows us to write
\be 
F_{a\al}= -i \sigma_{a\al\dbeta}\bar W^{\dbeta}, \quad\quad \bar W^{\dal}
=-\frac i4 \bar \sigma^{a\dal\al}F_{a\al}.\label{Walphadot}
\ee
Similarly, we have
\be 
F_{a\dal}= -i W^{\beta}\sigma_{a\beta\dal}  , \quad\quad W^{\al} =-\frac i4
F_{a\dal} \bar\sigma^{a\dal\al}.\label{Walpha}
\ee
Moreover the $W$'s must satisfy
\be 
\bar\ED \bar W-\ED W=\bar \ED_{\dal}\bar W^{\dal}-\ED^{\al}W_{\al}=0, \quad\quad
\ED_{\al}\bar W_{\dal}=0,\quad\quad \bar\ED_{\dal} W_{\al}=0,\label{Wid}
\ee
where we have introduced the covariant super-derivative  $\ED_A= D_A -[\phi_A,\,\,\,\,]$.

\subsection{The $\vt,\vtb$ superfield formalism}

Now we will switch on two anticommuting coordinates $\vt$ and $\vtb$ and 
call super-superfield (ss-field) a superfield that is a function {\it also} of these coordinates. In terms of $\tilde Z^{\tilde M} =(x^m,
\theta^\mu,\theta^{\dot\mu},\vartheta,\vtb)=(z^M,\vartheta,\vtb)$ we have
\be
\tilde f(\tilde Z)=\tilde f(z,\vartheta,\vtb)= f(z)+\vartheta\, \bar g(z)+\vtb g(z) +\vt\vtb h(z),\0
\ee
where $f(z),g(z),\bar g(z)$ and $h(z)$ are ordinary supersymmetric superfields. The BRST-anti-BRST interpretation is
\be
g= \mfs  f,\quad\quad \bar g= \bar \mfs f,\quad\quad h= \bar \mfs g=- \mfs\bar g.\label{BRSTinterpr}
\ee
We introduce
also the ss-exterior derivative $\tilde d= d \tilde  Z^{\tilde M} \frac {\d}{\d \tilde Z^{\tilde M}}= d+
d\vartheta\frac {\d}{\d\vartheta}+d\vtb\frac {\d}{\d\vtb} $.
The super-super-connection (ss-connection) is
\be 
\tilde\Phi=\tilde e^{\tilde A} \tilde \Phi_{\tilde A}.\label{SSconn2}
\ee
We choose
\be 
\tilde e^{\tilde A} (\tilde Z)=\left(\begin{matrix} e^A(z)&0&0\\ 0 &
d\vartheta&0\\
0&0&d\vtb\end{matrix}\right).\0
\ee
So we have the following
\be 
\tilde\Phi= \tilde \phi+  d\vartheta\tilde \phi_\vartheta +d\vtb\tilde \phi_{\vtb},\0
\ee
where
\be
&&\tilde \phi = e^A \tilde \phi_A=e^A\left(\phi_A+\vartheta \bar\psi_A+\vtb \psi_A +\vt\vtb \pi_A\right),\0\\
&&\tilde \phi_\vartheta = \varphi_{\vartheta}+\vartheta \bar\psi_{\vartheta}+\vtb \psi_{\vt} + \vt\vtb \varpi_\vartheta,\label{ss-comp2}\\
&&\tilde \phi_{\vtb} = \varphi_{\bar\vartheta}+\vartheta \bar\psi_{\bar\vartheta}+\vtb \psi_{\vtb} + \vt\vtb \varpi_{\bar\vartheta}.\0
\ee
In the above $\phi_A,\psi_A,\dots, \varpi_{\bar\vartheta}$ are ordinary
superfields valued in the gauge Lie algebra with generators $T^r$.

The ss-curvature can be written as
\be 
\tilde {\cal F} &=& \tilde d \tilde \Phi -\tilde \Phi \tilde \Phi \0\\
&=&\tilde F +d\vartheta \left((\d_\vartheta -\tilde \phi_\vartheta)\tilde
\phi-(d-\tilde \phi)\tilde\phi_\vartheta\right)+ d\vtb  \left((\d_{\vtb} -\tilde \phi_{\bar\vartheta})\tilde
\phi-(d-\tilde \phi)\tilde\phi_{\bar\vartheta}\right)\0\\
&&+d\vartheta d\vartheta \left( \d_\vartheta\tilde \phi_\vartheta-  \phi_\vartheta\tilde\phi_\vartheta\right)
+ d\vtb d\vtb \left( \d_{\bar\vartheta}\tilde \phi_{\bar\vartheta}-  \phi_{\bar\vartheta}\tilde\phi_{\bar\vartheta}\right)\0\\
&& + d\vartheta d\vtb\left( \d_\vartheta\tilde \phi_{\bar\vartheta} + \d_{\bar\vartheta}\tilde \phi_{\vartheta}-
\tilde\phi_\vartheta \tilde\phi_{\bar\vartheta}- \tilde\phi_{\bar\vartheta}\tilde\phi_\vartheta\right),\label{supercurvexp}
\ee
where $\tilde F$ has nonzero components only in the $x^\mu,\theta^\mu, \theta^{\dot \mu}$ directions. The horizontality condition is
\be
\tilde d \tilde \Phi -\tilde \Phi \tilde\Phi =\tilde F .\label{horizon2}
\ee
It gives rise to the following set of equations
\be
&&\left.\tilde{\cal  F}\right\vert_{d\vt=0=d\vtb}=\tilde F, \label{constraint1}\\
&&(\d_\vartheta -\tilde \phi_\vartheta)\tilde \phi-(d-\tilde \phi)\tilde\phi_\vartheta=0,\label{constraint2}\\
&&(\d_{\vtb} -\tilde \phi_{\bar\vartheta})\tilde \phi-(d-\tilde \phi)\tilde\phi_{\bar\vartheta} =0,\label{constraint3}\\ 
&& \d_\vartheta\tilde \phi_\vartheta- \phi_\vartheta\phi_\vartheta=0, \label{constraint4}\\ 
&&\d_{\bar\vartheta}\tilde \phi_{\bar\vartheta}-  \phi_{\bar\vartheta}\phi_{\bar\vartheta}=0,\label{constraint5}\\ 
&& \d_\vartheta\tilde \phi_{\bar\vartheta} + \d_{\bar\vartheta}\tilde \phi_{\vartheta}-
\phi_\vartheta \phi_{\bar\vartheta}- \phi_{\bar\vartheta}\phi_\vartheta=0.\label{constraint6}
\ee
Eqs.(\ref{constraint4},\ref{constraint5}) yield the identifications\footnote{In this section the square bracket notation $[\quad,\quad]$ denotes a {\it graded commutator}, with 
grading according to total Grassmannality $\epsilon$ of the two entries
\be 
[A,B]=   AB-(-1)^{\epsilon(A) \epsilon{B}}BA .\label{Grassman}
\ee
The total Grassmannality $\epsilon$ includes both the one related to supersymmetry and to the BRST symmetry.}
\be  
&&\bar \psi_\vartheta= \varphi_\vartheta \varphi_\vartheta, \quad\quad\quad \quad\quad\quad\quad\quad\quad   \psi_{\bar\vartheta}=  \varphi_{\bar\vartheta} 
\varphi_{\bar\vartheta},\label{identif1}\\
&& \varpi_\vartheta =[ \psi_\vartheta,\varphi_\vartheta],\quad\quad\quad \varpi_{\bar\vartheta} =-[\bar \psi_{\bar\vartheta},\varphi_{\bar\vartheta}] .\label{identif2}\\
\ee
The remaining equations
\be 
&&\bar \psi_\vartheta\varphi_\vartheta = \varphi_\vartheta \bar \psi_\vartheta,\quad\quad
\psi_{\bar\vartheta}\varphi_{\bar\vartheta} = \varphi_{\bar\vartheta} \psi_{\bar\vartheta},\0\\
&&\varpi_\vartheta \varphi_\vartheta + \varphi_\vartheta \varpi_\vartheta +\bar\psi_{\vartheta}
\psi_{\vartheta}-\psi_{\vartheta}\bar \psi_{\vartheta}=0,\quad\quad\varpi_{\bar\vartheta} \varphi_{\bar\vartheta} + \varphi_{\bar\vartheta} \varpi_{\bar\vartheta} 
+\bar\psi_{\bar\vartheta}
\psi_{\bar\vartheta}-\psi_{\bar\vartheta}\bar \psi_{\bar\vartheta}=0,\label{verif1}
\ee
are identically satisfied.

The lowest component of $\varphi_\vartheta$ is an anticommuting scalar valued in
the gauge Lie algebra, and is to be identified with the ghost field $c= c^r(x)
\, T^r$. Its BRST transform is $\bar \psi_\vartheta$ where  $\varphi_\vartheta$ is the BRST transform parameter.
The lowest component of $\varphi_{\bar\vartheta}$  is to be identified with the dual ghost field $\bar c=\bar c^r(x)\, T^r$. 
Its anti-BRST transform is $\psi_{\bar\vartheta}$. 
 
Eq.(\ref{constraint6}) gives the relation
\be
\psi_\vartheta+ \bar \psi_{\bar\vartheta}=\varphi_\vartheta  \varphi_{\bar\vartheta} + \varphi_{\bar\vartheta}   \varphi_\vartheta,\label{CF}
\ee
which is to be interpreted as the Curci-Ferrari relation, \cite{CF}, and  $\psi_\vartheta, \bar \psi_{\bar\vartheta}$ are the Nakanishi-Lautrup superfields. Using (\ref{CF}), the remaining relations
\be
&&\varpi_\vartheta=[\varphi_\vartheta ,\bar \psi_{\bar\vartheta}] +  [\varphi_{\bar\vartheta} ,\bar \psi_{\vartheta}],\0\\
&&\varpi_{\bar\vartheta}=-[\varphi_\vartheta , \psi_{\bar\vartheta}] -  [\varphi_{\bar\vartheta} ,\psi_{\vartheta}],\0\\
&&[\bar\psi_\vartheta , \psi_{\bar\vartheta}]+[\bar\psi_{\bar\vartheta} , \psi_{\vartheta}]+[\varpi_\vartheta, \varphi_{\bar\vartheta}]
+[\varpi_{\bar\vartheta}, \varphi_{\vartheta}]=0,\label{verif2}
\ee
turn out to be identically verified.

Let us come next to the constraint (\ref{constraint2}). It implies the definitions
\be
&&\bar \psi_A = D_A \varphi_\vartheta -[\phi_A,\varphi_\vartheta] = \ED_A \varphi_\vartheta,\label{identif3}\\
&& \pi_A= \ED_A \psi_\vartheta -[ \psi_A, \varphi_\vartheta],\label{identif4}
\ee
and the identities
\be
&&\ED_A\bar \psi_\vartheta - [\bar \psi_A, ,\varphi_\vartheta]=0,\0\\
&& \ED_A \varpi_\vartheta-[\pi_A, \varphi_\vartheta]+ [\bar \psi_A, ,\psi_\vartheta]- [ \psi_A, ,\bar\psi_\vartheta]=0, \label{verif3}
\ee
while, from (\ref{constraint3}), we get the definitions
\be
&& \psi_A = D_A \varphi_{\bar\vartheta} -[\phi_A,\varphi_{\bar\vartheta}] = \ED_A \varphi_{\bar\vartheta},\label{identif5}\\
&& \pi_A= -\ED_A \bar\psi_{\bar\vartheta} +[\bar \psi_A, \varphi_{\bar\vartheta}],\label{identif6}
\ee
and the identities
\be
&&\ED_A\psi_{\bar\vartheta} - [ \psi_A, \varphi_{\bar\vartheta}]=0,\0\\
&& \ED_A \varpi_{\bar\vartheta}-[\pi_A, \varphi_{\bar\vartheta}]+ [ \bar\psi_A, ,\psi_{\bar\vartheta}]- 
[ \psi_A, ,\bar\psi_{\bar\vartheta}]=0 .\label{verif4}
\ee
The superfield $\psi_A, \bar \psi_A, \pi_A$ are easily recognized as the (anti)BRST transform.
The equivalence of (\ref{identif4}) and (\ref{identif6}) can be proven by means of the CF condition.
Next let us come to (\ref{constraint1}). In general, using (\ref{FAB}), one can show that
\be
\tilde F_{AB} = F_{AB} -\vartheta [F_{AB}, \varphi_\vartheta] - \bar \vartheta [F_{AB}, \varphi_{\bar\vartheta}] -
\vartheta\bar \vartheta\left(  [F_{AB}, \psi_\vartheta]- [[F_{AB}, \varphi_{\bar\vartheta}],\varphi_\vartheta]\right).\label{FABtransf}
\ee
In proving this, a particular attention must be paid to the $(A,B)=(\alpha, \dot\beta)$ case. The definition (\ref{FAB}) includes in this case also
a contribution from the supertorsion; but this contribution is exactly canceled by an analogous term coming from the first commutator (\ref{DaDb}).

From (\ref{FABtransf}) it is evident that the constraints (\ref{F=0})  can be covariantly 
implemented in the BRST formalism. Next we have to consider the constraints \eqref{sigmaF=0}. 
But instead of solving the ss-Bianchi identity,
we prefer to covariantize the constraints extracted from it in chapter XIII
of \cite{WB}.  In the same way as (\ref{F=0}) we can covariantly 
implement also (\ref{sigmaF=0}) in the BRST formalism. To this end we introduce the BRST covariant definitions of $W^\alpha, \overline W^{\dot \alpha}$
\be 
\widetilde  W^{\al} =-\frac i4
\tilde F_{a\dal} \bar\sigma^{a\dal\al},\quad\quad \widetilde{\overline W}^{\dal}==-\frac i4 \bar \sigma^{a\dal\al}\tilde F_{a\al}.\label{Walphadot}
\ee
Therefore, the ss-field expression for $\widetilde W_{\al}$ is
\be
\widetilde W_{\al} =W_\al   -\vartheta [W_{\al} , \varphi_\vartheta] - \bar \vartheta [ W_{\al}, \varphi_{\bar\vartheta}] -
\vartheta\bar \vartheta\left(  [W_{\al} , \psi_\vartheta]- [[W_{\al} , \varphi_{\bar\vartheta}],\varphi_\vartheta]\right),\label{Wtransf}
\ee
and an analogous one for $\widetilde {\overline W}_{\dal}$. The next issue is now to BRST-covariantize the constraints (\ref{Wid}).

Let us use the compact notation $\underline \alpha$ to denote both $\al$ and $\dal$ and introduce the BRST super-covariant derivative 
\be
\widetilde \ED_{\underline \alpha}\widetilde W_{\underline \beta} = D_{\underline \alpha}  \widetilde  W_{\underline \beta} - 
[\tilde \phi_{\underline \alpha}, \widetilde W_{\underline \beta} ],
\label{redsscovder}
\ee
then it is lengthy but straightforward to prove that
\be
\widetilde \ED_{\underline \alpha}\widetilde W_{\underline \beta}=\ED_{\underline \alpha}    W_{\underline \beta}-
\vartheta [ \ED_{\underline \alpha}    W_{\underline \beta},\varphi_\vartheta]- \bar\vartheta [ \ED_{\underline \alpha}    W_{\underline \beta},\varphi_{\bar\vartheta}]
- \vartheta \bar\vartheta \left(  [ \ED_{\underline \alpha}    W_{\underline \beta},\psi_\vartheta]-
 [[ \ED_{\underline \alpha}    W_{\underline \beta},\varphi_{\bar\vartheta}],\varphi_\vartheta]\right).\label{covWconstr}
\ee
This allows us to write down the  constraints (\ref{Wid}) in a BRST covariant form. They combine perfectly with the BRST superfield formalism.

This section shows that the BRST formalism can be consistently embedded in a supermanifold that 
encompasses also the supersymmetric spinorial directions.

\section{Conclusion and comments}

In this paper we have reviewed (or proposed for the first time) several applications of the superfield method to represent the BRST and anti-BRST algebra in field theories with both gauge and diffeomorphism symmetries\footnote{For recent applications of the superspace/supervariable approach     to (non-)susy 1d and 2d diffeomorphism invariant theories, see \cite{malikdiff}.}. We have shown, in many examples, that it correctly reproduces the transformations and, in more complicated cases, it helps finding them. Beyond that, we have shown that it is instrumental in practical applications, such as in the subject of consistent gauge anomalies and their integration (Wess-Zumino terms). In such instances it can be of invaluable help as an effective algorithmic method, as opposed to laborious alternative trial and error methods.

In Appendix A, we have reported a geometrical description of the BRST and anti-BRST algebra based on the geometry of principal fiber bundles and infinite dimensional groups of gauge transformations. Although elegant and with the appeal of classical geometry, this description can hardly be extended to the full quantum theory, as defined by the perturbative expansion.  The superfield description, which incorporates gauge, ghosts and auxiliary fields in a unique expression,  seems instead to be more apt for this purpose, although a full attempt to exploit this possibility, to our best of knowledge, has never been made. 

Elementary examples, in this sense, are the gauge-fixing action terms. The gauge-fixing is a necessary step of quantization. There is freedom in choosing the gauge-fixing terms, except for a few obvious limitations: they must break completely the relevant gauge symmetry, have zero ghost number, then they must be real and of canonical dimension not larger than 4 (in 4d) in order to guarantee unitarity and renormalizability. But, of course, they must be BRST (and, {if possible,} anti-BRST) invariant. Having at our disposal the superfield method, it is relatively easy to construct such terms. For instance, a well-known case is that of non-Abelian gauge theories. With reference to the notation in section 2.1 one such term is the trace of $\frac {\partial}{\partial \vt} \frac {\partial}{\partial \vtb} \left(\Phi_\mu \Phi^\mu\right)$, which, being the coefficient of $\vt\vtb$, is automatically BRST and anti-BRST invariant. It gives rise to the action term ${\rm tr}(  A_\mu \partial^\mu B -\partial^\mu \bar c D_\mu c)$. Another possibility is the trace of $\left(\frac {\partial\bar \eta}{\partial \vt}\right)^2$ and $\left(\frac {\partial \eta}{\partial \vtb}\right)^2$, which give rise to the action terms ${\rm tr} (\bar B^2)$ and ${\rm tr} (B^2)$, respectively. And so on. 

Against the backdrop of these well known examples, we would like to produce a few analogous terms in the case of gravity.
The Einstein-Hilbert action for gravity in the superfield formalism can be written as follows, \cite{Delbourgo},
\be
S= \kappa \int d^4x d\vtb d\vt \,\vt \vtb\, \sqrt{G}\, {\bf R}= \kappa \int d^4x \, R.\label{EH}
\ee
 On the same footing, we easily add matter fields and their interaction with gravity. Then gauge-fixing terms invariant under BRST and anti-BRST transformations can be easily produced with the superfield formalism of section 4 and 5. For instance, the coefficients of $\vt \vtb$ in any local expression of the superfields of dimension 4.  One such term is
\be
L^{(1)}_{g.f.}=(\partial_\vt \partial_\vtb G_{\mu\nu}(X))\, (\partial_\vt \partial_\vtb \widehat G^{\mu\nu}(X))= V_{\mu\nu}(x) \widehat V^{\mu\nu}(x).\label{VmunuVmunu}
\ee 
The dimensional counting is based on assigning to $X^M$ dimension -1: $[x^\mu]=[{\vt}]=[\vtb]=-1$ and to $\tilde d X^M$ dimension 0. Of course $[G_{MN}]=0$. This fixes the dimensions of all the component fields. For instance $[h]=1, [\xi^\mu]=0, etc.$ Another possible term is
\be
L^{(2)}_{g.f.}=(\partial_\vt \partial_\vtb G_{\vt\vtb}(X) )^4 = G(x)^4,\label{G4}
\ee
which is, however, quartic in $h$.

Another way to get BRST invariant term is the following: consider
\be
\eta^{\mu\nu}(\partial_\vt \partial_\vtb G_{\mu\underline{\vt}}(X))\,(\partial_\vt \partial_\vtb G_{\nu\underline{\vt}}(X)),\0
\ee 
where $\underline{\vt}$ stands either for $\vt$ or $\vtb$. The only nonvanishing term is 
\be
L^{(3)}_{g.f.}= \eta^{\mu\nu} \Gamma_\mu(x)\overline \Gamma_\nu (x). \label{L4}
\ee
Using the supervierbein one can construct
\be
L^{(4)}_{g.f.}=(\partial_\vt \partial_\vtb E^a_{\mu}(X))( \partial_\vt \partial_\vtb \widehat E_a^{\mu}(X))=
f^a_\mu(x)\hat f_a^\mu(x), \label{famufamu}
\ee
and
\be
L^{(5)}_{g.f.}=\eta_{ab}(\partial_\vt \partial_\vtb E^a_{\vt}(X))( \partial_\vt \partial_\vtb E^b_{\vtb}(X))=
\eta_{ab}\psi^a(x)\rho^b(x).\label{rhoarhob}
\ee
{It is remarkable that these gauge-fixing terms are generally more than quadratic in the fields, which implies, in particular, that not only linear gauge fixing terms enjoy both BRST {\it and} anti-BRST symmetry. We see from this that aspects of BRST quantization of gauge and gravity field theories need further investigation and may reserve surprises. We plan to return to them.  }

\vskip 2cm
{\bf Acknowledgements}. We would like to dedicate this paper to the memory of Mario Tonin, to whom we are both indebted. L.B. would like to thank Mauro Bregola, Paolo Cotta-Ramusino, Maro Cvitan, Stefano Giaccari, Paolo Pasti, Predrag Dominis-Prester, Maurizio Rinaldi, Jim Stasheff, Tamara Stemberga  for their precious collaboration over the years.
\vskip 1cm
\appendix

\noindent{\Large {\bf Appendices.}}

\section{Evaluation map and BRST}

The purpose of this Appendix is to present an interpretation of the BRST transformations within the framework of the geometry of a principal fiber bundle, \cite{BCRS}, so that one is in the condition to appreciate the remarkable similarity between this geometry and the superfield formalism.

Let $\sf P(M,G)$ be a principal bundle with a $\sfd$-dimensional manifold $\sf M$ as base, structure group $\sf G$, which we suppose to be compact, and total space $\sf P$. $\pi$ will denote the projection $\pi: \sf  P\to M$. 
An automorphism is a diffeomorphisms of $\sf P$,  $\psi:\sf P\to \sf P$, such that $\psi(p g) = \psi(p)g$, for any $p\in\sf P$ and any $g\in \sf G$. A vertical automorphism does not move the base point: $\pi(\psi(p))=\pi (p)$. Vertical authomorphisms form a group denoted $\sf Aut_v(P) $. The latter is to be identified with the group ${\cal G}$ of gauge transformations. The corresponding Lie algebra will be denoted by  $\sf aut_v(P)\equiv Lie ({\cal G})$; it is a space of vector fields in $\sf P$ generated by one-parameters subgroups of  $\sf Aut_v(P)$.
The reason of this identification is clear from the way a connection transforms under vertical automorphisms. Let $A$ be a connection with curvature $F=dA+\frac 12 [A,A]$. In local coordinates it takes the form $A=A^a_\mu T^a dx^\mu$ where $T^a$ are the generators of $\sf Lie(\sf G)$. Let $\psi$ be a vertical automorphism: we can associate to it a map $\gamma:\,\sf P\to G$ defined by $\psi(p) =p\,\gamma(p)$ satisfying $\gamma(pg) = g^{-1} \gamma(u) g$. Then one can show that the following holds:
\be
\psi^\ast A = \gamma^{-1} A \gamma + \gamma^{-1} d \gamma,\quad\quad \psi^\ast F= \gamma^{-1} F \gamma.\0
\ee
 Next we introduce the evaluation map
\be
ev:\sf P\times {\cal G} \rightarrow \sf P, \quad\quad ev(p,\psi)=\psi(p).\label{evmap}
\ee
We suppose that  $\sf P\times {\cal G}$ is a principal fiber bundle over $M\times {\cal G} $ with group $\sf G$. This means that pulling back a connection $A$ from $\sf P$ we obtain a connection ${\cal A} =ev^\ast A$ in  $P\times {\cal G}$. This connection contains all information about FP-ghosts and BRST transformations. Let us see how this  comes about. 

We evaluate $ev^\ast A $ over a couple $({ X}, {\rm Y})$. Here ${ X}\in T_pP$ and $ {\rm Y}\in T_\psi {\cal G}$, where $T_pP,T_\psi {\cal G}$ denote the tangent space of $\sf P$ at $p$ and of ${\cal G}$ at $\psi$, respectively. Since ${\cal G}$ is a Lie group, there exists a ${\rm Z}\in T_{id}{\cal G}$, such that
$\psi_\ast{\rm  Z}={\rm Y}$. If $\psi_t$ ($t\in {\mathbb R}$) with $\psi_0=\psi$ generates $\rm Y$, i.e. given
$f\in C^{\infty} ({\cal G})$, we have the following
\be
\left. {\rm Y}f = \frac d{dt} f\left(\psi_t\right) \right\vert_{t=0}, \label{Y}
\ee 
then $\mathring{ \psi}_t=\psi^{-1} \psi_t$ generates $\rm Z$:
\be
\left. {\rm Z}f = \frac d{dt} f\left(\mathring{\psi}_t\right) \right\vert_{t=0} = (\psi^{-1}_\ast{\rm Y})f.\label{Z}
\ee 

Now it is useful to introduce two auxiliary maps:
\be
&&ev_p : {\cal G}\rightarrow{ \sf P}, \quad\quad ev_p(\psi) = \psi(p),\0\\
&&ev_\psi :{\sf P}\rightarrow{\sf  P} ,\quad\quad ev_\psi (p)= \psi(p).\0
\ee
For any $h\in C^\infty (\sf P)$, we have 
\be
(ev_{p\ast}{\rm Y} )h = \left. \frac d{dt}  h(ev_p\circ \psi_t)\right\vert_{t=0}= \left. \frac d{dt}  h( \psi_t(p))\right\vert_{t=0}=  Y_{\psi (p)}h.\label{evYh}
\ee
It follows that
\be
(ev^\ast A)_{p,\psi}({X},{\rm Y}) &=&A_{\psi(p)} \left( ev_{\psi\ast} {X}\right)+A_{\psi(p)} \left(ev_{p\ast}{\rm Y} \right)= A_{\psi(p)} \left( \psi_\ast {X}\right)+A_{\psi(p)} \left( {\rm Y} _{\psi(p)}\right)\0\\
&=& \left(\psi^\ast A\right)_{p} \left(  {X}\right)+\left(\psi^\ast A \right)_{p} \left( {\rm Z} _p\right)=\left(\psi^\ast A\right)_{p} \left(  {X}\right)+\left( i_{\psi_\ast^{-1}({\rm Y})} \psi^\ast A\right)_{\psi(p)}\0\\
&=&\left(\psi^\ast A\right)_{p} \left(  {X}\right)+\left( i_{\psi_\ast^{-1}(\cdot)} \psi^\ast A\right)_{\psi(p)}  ({\rm Y}). \label{evstarAXY}
\ee
At $\psi= id$, the identity of ${\cal G}$, this formula can be written in the compact form
\be
ev^\ast A= A+i_{(\cdot)} A,\label{evstarA}
\ee
where $i$ is the interior product and $i_{(\cdot)} A$ denotes the map ${\rm Z}\to i_{\rm Z} A$ that associates to every ${\rm Z}\in \sf Lie({\cal G})$ the map $\xi_{\rm Z}= A({\rm Z}):{\sf  P}\to\sf Lie(\sf G)$. Here $\xi_{\rm Z}$ is an infinitesimal gauge transformation, for let us recall that the action of ${\rm Z}$ over the connection $A$ is given by the Lie derivative $L_{\rm Z}$, which takes the following form:
\be
L_{\rm Z}A &=& \left(di_{\rm Z}+ i_{\rm Z} d\right) A  = d\left(i_{\rm Z} A\right) + i_{\rm Z} \left(dA\right) =  d\left(i_{\rm Z} A\right) + i_{\rm Z} \left(F-\frac 12 [A,A]\right)\0\\
&=& i_{\rm Z} F +   d\left(i_{\rm Z} A\right) + [A , i_{\rm Z} A]= d\xi_{\rm Z} +[A,\xi_{\rm Z}], \label{LZA}
\ee
because $ i_{\rm Z} F =0$, ${\rm Z}$ being a vertical vector, while $F$ is a basic (i.e. horizontal) two-form.

The above means, in particular, that $i_{(\cdot)}A$ behaves like the Maurer-Cartan form on the group $\sf G$.  From now on, our purpose is to justify the fact that $i_{(\cdot)} A$ can play the role of the ghost field $c$.

If $Q(A)$ is a polynomial of $A, dA$ and the exterior product $A\wedge A$, the formula  \eqref{evstarA} is generalized to the following expression 
\be
ev^\ast Q(A) = Q(A) + i_{(\cdot)} Q(A) -  i_{(\cdot)} i_{(\cdot)} Q(A) -\ldots+(-1)^{\frac {k(k-1)}2} \underbrace { i_{(\cdot)}\ldots i_{(\cdot)}}_{k\,\, {\rm terms}}Q(A)+ \dots, \label{evstarQA}
\ee
where the interior products are understood with respect to vectors in $ Lie({\cal G})$. Of course,
we have also
\be
ev^\ast Q(A) = Q(ev^\ast A)= Q( A+i_{(\cdot)} A), \label{evstarQAb}
\ee
because the pull-back ``passes through" exterior product and differential. So the RHS of this equation equals the RHS of \eqref{evstarQA}. This allows us to read off the meaning of the expression $Q( A+i_{(\cdot)} A)$.

Let us consider an example. A remarkable consequence of \eqref{evstarA} is
\be
{\cal F}= ev^\ast F = F,\label{russian}
\ee
because $i_{\rm Z} F=0$ for any ${\rm Z}\in Lie({\cal G})$. Let us write down the explicit form of $ev^\ast F$:
\be
ev^\ast F&=&F +i_{(\cdot)} F + i_{(\cdot)} i_{(\cdot)} F=
F+ i_{(\cdot)} dA + [ i_{(\cdot)} A,A] + i_{(\cdot)} i_{(\cdot)} dA + \frac 12[i_{(\cdot)} A,i_{(\cdot)} A].\label{evstarF}
\ee
On the other hand, if $\hat\delta$ is the exterior differential in ${\cal G}$, we have 
\be
F(ev^\ast A)&=&(d+\hat\delta ){\cal A} + \frac 12 [{\cal A},{\cal A}]\label{FevstarA}\\
&=& F+ \hat\delta A +d i_{(\cdot)}A+\frac 12[A,  i_{(\cdot)}A] +\frac 12  [ i_{(\cdot)}A,A]+\hat\delta  i_{(\cdot)}A+\frac 12 [ i_{(\cdot)}A, i_{(\cdot)}A],\0
\ee
which means, splitting it according to the form degree,
\be
\hat\delta A &=& -d i_{(\cdot)}A-\frac 12[A,  i_{(\cdot)}A] -\frac 12  [ i_{(\cdot)}A,A],\label{hatdeltaA}\\
\hat\delta  i_{(\cdot)}A&=&-\frac 12 [ i_{(\cdot)}A, i_{(\cdot)}A].\label{hatdeltaiA}
\ee
The first equation must correspond with the term $i_{(\cdot)} F $ in eq.\eqref{evstarF}. But
\be
i_{(\cdot)} F = i_{(\cdot)}dA+[ i_{(\cdot)}A,A] = L_{(\cdot)} A -d i_{(\cdot)}A-[A, i_{(\cdot)}A] =0.\label{idotF}
\ee
Therefore, we must understand that $\hat\delta= (-1)^k \mfs$, where $k$ is the order of the form in $P $ it acts on, and $\mfs$ is the ordinary BRST variation. Moreover $[i_{(\cdot)}A,A]= [A,  i_{(\cdot)}A]$, i.e. $A$ and $ i_{(\cdot)}A$ behave like one-forms (remember eq.\eqref{AccA}!). Finally, we have the following
\be
\mfs A &=& d i_{(\cdot)}A+[A,  i_{(\cdot)}A] , \label{truedeltaA}\\
\mfs  i_{(\cdot)}A&=&-\frac 12 [ i_{(\cdot)}A, i_{(\cdot)}A].\label{truedeltaiA}
\ee
On the other hand, eq.\eqref{hatdeltaiA} must correspond to the last two terms in eq.\eqref{evstarF}. To see that this is the case one must recall some basic formulas in differential geometry where, for any one-form $\omega$ and any two vector fields ${\sf X,Y}$, we have
\be
d\omega (\sf{X,Y}) =\frac 12 \bigl({\sf X}\, \omega({\sf Y})- {\sf Y}\, \omega({\sf X})\bigr)
-\omega ([{\sf X,Y}]),\label{domegaXY}
\ee
and
\be
L_{\sf X} \omega({\sf Y})={\sf X}\,\omega({\sf Y})- \omega ([\sf{X,Y}]).\label{LXomega}
\ee
The skew double interior product $ i_{(\cdot)} i_{(\cdot)} dA $ in \eqref{evstarF} must be understood as follows
\be
 i_{{\rm Z}_2} i_{{\rm Z}_1} dA=dA({\rm Z}_1,{\rm Z}_2)= \frac 12 \bigl({\rm Z}_1 A({\rm Z}_2)- {{\rm Z}_2}\, A(Z_1)\bigr)
-A ([{\rm Z}_1,{\rm Z}_2])=L_{{\rm Z}_1}A({\rm Z}_2)- L_{{\rm Z}_2}A({\rm Z}_1).\0
\ee
In other words, $ i_{(\cdot)} i_{(\cdot)} dA$ is to be understood as the Lie derivative of $i_{(\cdot)} A$, or its BRST transform, once we interpret $i_{(\cdot)}A$ as the FP ghost, in agreement with 
\eqref{truedeltaiA}.

From the above formulas we see that the evaluation map provides a geometrical interpretation of the BRST transformations. Let us see now the relation with the superfield formalism.
To start with let us remark that the geometrical formula \eqref{evstarA} corresponds to the superfield expression
\be
{\tilde A}\vert_{\vt=0}=A+c \, d\vartheta,\label{A+thetac}
\ee
where $\vartheta $ is our anticommuting variable (see section 2). The expression $i_{(\cdot)} A$ is the component of ${\cal A}$ in the direction of ${\cal G}$ in the product ${\sf P}\times {\cal G}$. Therefore, we see that $\vt$ represents this direction, rather than the vertical direction in $\sf P$. The $\vt$ partners of $A, F$ and $c$ are nothing but the Lie derivatives in this direction with respect to the vector fields ${\rm Z}\in\sf Lie ({\cal G})$. Therefore the superfield method captures the geometry of the principal fiber bundles: ${\cal A}\rightarrow {\cal A}/{\cal G}$ where ${\cal A}$ is the space of connections.

\section{Auxiliary formulae}

In this Appendix we collect a few cumbersome formulas in expanded form with the aim of clarifying the main text of the paper.

\subsection{Expansion of eq.\eqref{horizonvector}}

We start by expanding the LHS of \eqref{horizonvector}:
\be 
&&\tilde A_M(\tilde X) \tilde d\tilde X^M =\Big{(}A_\mu -\left(\vartheta 
{\xib}+\vtb \xi-\vt\vtb h\right)\!\cdot\! \partial A_\mu 
+\vt \vtb\, \xi{\xib}\!\cdot\! \partial^2 A_\mu + \vt \left( \bar \phi_\mu -\vtb 
\xi \!\cdot \! \partial \bar \phi_\mu\right) \0\\
&&+  \vtb \left( \phi_\mu -\vt {\xib} \!\cdot \! \partial \phi_\mu\right)+ 
\vt\vtb \, B_\mu(x)\Big{)}\0\\
&&\cdot \left( dx^\mu 
-\vt\, \partial_\lambda {\xib}^\mu dx^\lambda - \vtb \,\partial_\lambda \xi^\mu 
dx^\lambda+ \vt\vtb \partial_\lambda h^\mu dx^\lambda
-({\xib}^\mu-\vtb h^\mu) d\vt 
-( {\xi}^\mu +\vt h^\mu) d\vtb\right)\label{vector3}\\
&& +\biggl{(}\chi(x) -\left(\vartheta {\xib}+\vtb \xi-\vt\vtb h\right)\!\cdot\! 
\partial\chi(x) + \vt \vtb \,\xi\xib  \!\cdot\! \partial^2 \chi(x) \0\\
&&+   \vt \left( \bar C(x) -\vtb \xi \!\cdot \! \partial \bar C(x) \right)+  
\vtb \left( C(x) -\vt {\xib} \!\cdot \! \partial C(x) \right)
+\vt \vtb \psi(x) \biggr{)}d\vartheta\0\\
&& + \biggl{(}\lambda(x) -\left(\vartheta {\xib}+\vtb \xi -\vt\vtb 
h\right)\!\cdot\! \partial\lambda(x) + \vt \vtb\,\xi\xib \!\cdot\! \partial^2 
\lambda(x)\0\\
&&+ \vt \left( \bar D(x) -\vtb \xi \!\cdot \! \partial\bar D(x) \right)+  \vtb 
\left( D(x) -\vt {\xib} \!\cdot \! \partial D(x) \right)
+\vt \vtb \rho(x) \biggr{)}d\bar\vartheta= A_\mu (x) dx^\mu, \0
\ee
where $\xi {\xib} \!\cdot\! \partial^2=\xi^\mu {\xib}^\nu \partial_\mu 
\partial_\nu  $.

\subsection{Expansion of eq.\eqref{horizontalmetric}}

The next auxiliary formula is the explicit expression of the LHS of \eqref{horizontalmetric}:
\be 
&&  g_{\mu\nu}(x) dx^\mu \vee dx^\nu=\widetilde G_{MN}(\tilde X) \tilde d\tilde X^M \vee 
\tilde d\tilde X^N \label{hormetric}\\
&=&\Big{(} g_{\mu\nu} - \left(\vartheta {\xib}+\vtb \xi -\vt\vtb 
h\right)\!\cdot\! \partial g_{\mu\nu} +  
\vt \vtb\, \xi{\xib}\!\cdot\! \partial^2 g_{\mu\nu} + \vt \left( \bar 
\Gamma_{\mu\nu} -\vtb \xi  \!\cdot \! \partial \bar \Gamma_{\mu\nu}\right) \0\\
&&
\quad\quad+\vtb \left( \Gamma_{\mu\nu} -\vt \xib \!\cdot \! 
\partial\Gamma_{\mu\nu}\right)+ \vt\vtb \, V_{\mu\nu}(x)
\Big{)}\0\\
&& \left( dx^\mu 
-(\vt\, \partial_\lambda {\xib}^\mu   + \vtb \,\partial_\lambda \xi^\mu  
-\vt\vtb \partial_\lambda h^\mu) dx^\lambda-({\xib}^\mu -\vtb h^\mu)d\vt 
- ({\xi}^\mu+\vt h^\mu) d\vtb\right)\0\\
&&\vee \left( dx^\nu 
-(\vt\, \partial_\rho {\xib}^\nu   + \vtb \,\partial_\rho \xi^\nu - \vt\vtb 
\partial_\rho h^\nu) dx^\rho-({\xib}^\nu -\vtb h^\nu)d\vt 
- ({\xi}^\nu+\vt h^\nu) d\vtb \right)\0\\
&&+2\Big{(} \gamma_\mu - \left(\vartheta {\xib}+\vtb \xi-\vt\vtb 
h\right)\!\cdot\! \partial \gamma_{\mu} +  
\vt \vtb\, \xi{\xib}\!\cdot\! \partial^2 \gamma_{\mu}+ \vt \left( \bar g_{\mu} 
-\vtb \xi \!\cdot \! \partial \bar g_{\mu}\right)\0\\
&&\quad\quad+ \vtb \left( g_{\mu} -\vt \xib \!\cdot \! \partial g_{\mu}\right) + 
\vt\vtb \, \Gamma_{\mu}(x)\Big{)}\0\\
&& 
\left( dx^\mu 
-(\vt\, \partial_\lambda {\xib}^\mu  + \vtb \,\partial_\lambda \xi^\mu -\vt\vtb 
\partial_\lambda h^\mu) dx^\lambda-({\xib}^\mu -\vtb h^\mu)d\vt 
- ({\xi}^\mu+\vt h^\mu) d\vtb \right)\vee d\vt\0\\
&&+2\Big{(} \bar\gamma_\mu - \left(\vartheta {\xib}+\vtb \xi\right)\!\cdot\! 
\partial \bar\gamma_{\mu} +  
\vt \vtb\, \xi{\xib}\!\cdot\! \partial^2 \bar\gamma_{\mu}+ \vt \left( \bar 
f_{\mu} -\vtb \xi \!\cdot \! \partial \bar f_{\mu}\right)\0\\
&&+ \vtb \left( f_{\mu} -\vt \xib \!\cdot \! \partial f_{\mu}\right) + \vt\vtb 
\,\bar \Gamma_{\mu}(x)\Big{)}\0\\
&&\left( dx^\mu 
-(\vt\, \partial_\lambda {\xib}^\mu  +\vtb \,\partial_\lambda \xi^\mu  -\vt\vtb 
\partial_\lambda h^\mu) dx^\lambda-({\xib}^\mu -\vtb h^\mu)d\vt 
- ({\xi}^\mu+\vt h^\mu) d\vtb\right)\vee d\vtb\0\\
&&+\Big{(} g - \left(\vartheta {\xib}+\vtb \xi-\vt\vtb h\right)\!\cdot\! 
\partial g +  
\vt \vtb\, \xi{\xib}\!\cdot\! \partial^2 g+ \vt \left( \bar \gamma -\vtb \xi 
\!\cdot \! \partial \bar \gamma\right)+
\vtb \left( \gamma -\vt \xi \!\cdot \! \gamma \right)+\vt\vtb G \Big{)} d\vt 
\vee d\vtb,\0
\ee
where all the fields on the RHS are function of $x$. 

\subsection{Expansion of eq.\eqref{horhatGhatg}}

Here we expand the horizontality condition for the inverse supermetric, eq.\eqref{horhatGhatg}:
\be
&& \hat g^{\mu\nu}(x) \frac {\partial}{\partial x^\mu}\vee  \frac {\partial}{\partial x^\nu}=
\widetilde{\widehat  G}^{MN}(\tilde X) \frac{\partial}{\tilde \partial \tilde X^M} \vee \frac {\tilde\partial}{\tilde \partial \tilde X^N}\label{horhatGhatg1}\\
&=&\Big{(} \hat g^{\mu\nu} - \left(\vt {\xib}+\vtb \xi -\vt\vtb 
h\right)\!\cdot\! \partial \hat g^{\mu\nu} +  
\vt \vtb\, \xi{\xib}\!\cdot\! \partial^2 \hat g^{\mu\nu} + \vt \left(\widehat {\overline 
\Gamma}^{\mu\nu} -\vtb \xi \!\cdot \! \partial \widehat{\overline \Gamma}^{\mu\nu}\right) \0\\
&&
\quad\quad+\vtb \left( \widehat\Gamma^{\mu\nu} -\vt \xib \!\cdot \! 
\partial\widehat \Gamma^{\mu\nu}\right)+ \vt\vtb \,\widehat V^{\mu\nu}(x)
\Big{)}\0\\
&&\cdot\left(\frac {\partial}{\partial x^\mu}+ \left(\vt \partial_\mu {\xib}^\lambda +\vtb \partial_\mu \xi^\lambda - \vt\vtb \left(\partial_\mu h^\lambda + \partial_\mu\xib ^\sigma \partial_\sigma \xi^\lambda -   \partial_\mu\xi ^\sigma \partial_\sigma \xib^\lambda\right)\right) \frac {\partial}{\partial x^\lambda}\right)\0\\
&&\vee\left(\frac {\partial}{\partial x^\nu}+ \left(\vt \partial_\nu {\xib}^\rho +\vtb \partial_\nu \xi^\rho - \vt\vtb \left(\partial_\nu h^\rho + \partial_\nu\xib ^\tau \partial_\tau \xi^\rho -   \partial_\nu\xi ^\tau \partial_\tau\xib^\rho\right)\right) \frac {\partial}{\partial x^\rho}\right)\0\\
&&+2\Big{(}\hat \gamma^\mu - \left(\vartheta {\xib}+\vtb \xi-\vt\vtb 
h\right)\!\cdot\! \partial\hat \gamma^{\mu} +  
\vt \vtb\, \xi{\xib}\!\cdot\! \partial^2\hat \gamma^{\mu}+ \vt \left( \hat{\bar g}^{\mu} 
-\vtb \xi \!\cdot \! \partial \hat{\bar g}^{\mu}\right)\0\\
&&\quad\quad+ \vtb \left(\hat g^{\mu} -\vt \xi \!\cdot \! \partial\hat g^{\mu}\right) + 
\vt\vtb \, \widehat\Gamma_{\mu}(x)\Big{)}\0\\
&& 
\cdot\left(\frac {\partial}{\partial x^\mu}+ \left(\vt \partial_\mu {\xib}^\lambda +\vtb \partial_\mu \xi^\lambda - \vt\vtb \left(\partial_\mu h^\lambda + \partial_\mu\xib ^\sigma \partial_\sigma \xi^\lambda -   \partial_\mu\xi ^\sigma \partial_\sigma \xib^\lambda\right)\right) \frac {\partial}{\partial x^\lambda}\right)\0\\
&&\vee \left(\frac {\partial}{ \partial  \vt} + \left( -\xib^\rho
+\vt \xib\!\cdot \! \partial\xib^\rho + \vtb \Big( -h^\rho +\xib\!\cdot\!\partial \xi^\rho \right)\right.\0\\
&&\left.\quad\quad\quad+\vt\vtb \left(h\!\cdot\!\partial \xib^\rho -\xib \!\cdot\! \partial h^\rho - \xib \!\cdot\! \partial\xib\!\cdot\! \partial \xi^\rho
+ \xib \!\cdot\! \partial\xi\!\cdot\! \partial \xib^\rho \right)\Big) \frac {\partial}{\partial x^\rho}\right)\0\\
&&+2\Big{(} \hat{\bar\gamma}^\mu - \left(\vartheta {\xib}+\vtb \xi\right)\!\cdot\! 
\partial \hat{\bar\gamma}^{\mu} +  
\vt \vtb\, \xi{\xib}\!\cdot\! \partial^2 \hat{\bar\gamma}^{\mu}+ \vt \left(\hat{ \bar 
f}^{\mu} -\vtb \xi \!\cdot \! \partial\hat{ \bar f}^{\mu}\right)\0\\
&&\quad\quad\quad+ \vtb \left( \hat f^{\mu} -\vt \xi \!\cdot \! \partial \hat f^{\mu}\right) + \vt\vtb 
\,\widehat{\overline \Gamma}^{\mu}(x)\Big{)}\0\\
&&\cdot\left(\frac {\partial}{\partial x^\mu}+ \left(\vt \partial_\mu {\xib}^\lambda+\vtb \partial_\mu \xi^\lambda - \vt\vtb \left(\partial_\mu h^\lambda+ \partial_\mu\xib ^\sigma \partial_\sigma \xi^\lambda -   \partial_\mu\xi ^\sigma \partial_\sigma \xib^\lambda\right)\right) \frac {\partial}{\partial x^\lambda}\right)\0\\
&&\vee\left(\frac {\partial}{ \partial  \vtb} + \Big( -\xi^\rho
+\vtb \xi\!\cdot \! \partial\xi^\rho + \vt \left( h^\rho +\xi\!\cdot\!\partial \xib^\rho \right) \right.\0\\
&&\quad\quad\quad+\left.\vt\vtb \left(h\!\cdot\!\partial \xi^\rho -\xi \!\cdot\! \partial h^\rho - \xi \!\cdot\! \partial\xib\!\cdot\! \partial \xi^\rho 
+ \xi \!\cdot\! \partial\xi\!\cdot\! \partial \xib^\rho \right)\Big) \frac {\partial}{\partial x^\rho}\right)\0\\
&&+\Big{(} \hat g - \left(\vartheta {\xib}+\vtb \xi-\vt\vtb h\right)\!\cdot\! 
\partial \hat g +  
\vt \vtb\, \xi{\xib}\!\cdot\! \partial^2 \hat g+ \vt \left(\hat{ \bar \gamma} -\vtb \xi 
\!\cdot \! \partial\hat{ \bar \gamma}\right)+
\vtb \left(\hat \gamma -\vt \xi \!\cdot \!\hat \gamma \right)+\vt\vtb \widehat G \Big{)}  \0\\
&&\cdot \left(\frac {\partial}{ \partial  \vt} + \left( -\xib^\lambda
+\vt \xib\!\cdot \! \partial\xib^ \lambda+ \vtb \Big( -h^\lambda  +\xib\!\cdot\!\partial \xi^\lambda  \right)\right.\0\\
&&\left.\quad\quad\quad+\vt\vtb \left(h\!\cdot\!\partial \xib^\lambda  -\xib \!\cdot\! \partial h^ \lambda - \xib \!\cdot\! \partial\xib\!\cdot\! \partial \xi^\lambda
+ \xib \!\cdot\! \partial\xi\!\cdot\! \partial \xib^ \lambda\right)\Big) \frac {\partial}{\partial x^\lambda }\right)\0\\
&&\vee\left(\frac {\partial}{ \partial  \vtb} + \Big( -\xi^\rho
+\vtb \xi\!\cdot \! \partial\xi^\rho + \vt \left( h^\rho +\xi\!\cdot\!\partial \xib^\rho \right) \right.\0\\
&&\quad\quad\quad+\left.\vt\vtb \left(h\!\cdot\!\partial \xi^\rho -\xi \!\cdot\! \partial h^\rho - \xi \!\cdot\! \partial\xib\!\cdot\! \partial \xi^\rho 
+ \xi \!\cdot\! \partial\xi\!\cdot\! \partial \xib^\rho \right)\Big) \frac {\partial}{\partial x^\rho}\right).\0
\ee

\subsection{Expansion of eq.\eqref{horinversevierbein}}

Finally, we consider the explicit form of the LHS of \eqref{horinversevierbein} which can be expanded as
\be
&&\hat e_a^\mu \frac \partial{\partial x^\mu} = \Big{(} \hat e_a^\mu - \left(\vt {\xib}+\vtb \xi -\vt\vtb 
h\right)\!\cdot\! \partial \hat e_a^\mu  +  
\vt \vtb\, \xi{\xib}\!\cdot\! \partial^2 \hat   e_a^\mu + \vt \left(\hat {\bar 
\phi}_a^\mu -\vtb \xi \!\cdot \! \partial \hat{\bar \phi}_a^\mu\right) \0\\
&&
\quad\quad+\vtb \left( \hat\phi_a^\mu -\vt \xib \!\cdot \! 
\partial\hat \phi_a^{\mu}\right)+ \vt\vtb \,\hat f^{\mu}_a)
\Big{)}\0\\
&&\cdot\left(\frac {\partial}{\partial x^\mu}+ \left(\vt \partial_\mu {\xib}^\lambda +\vtb \partial_\mu \xi^\lambda - \vt\vtb \left(\partial_\mu h^\lambda + \partial_\mu\xib ^\sigma \partial_\sigma \xi^\lambda -   \partial_\mu\xi ^\sigma \partial_\sigma \xib^\lambda\right)\right) \frac {\partial}{\partial x^\lambda}\right)\0\\
&&+ \Big{(} \hat \chi_a  - \left(\vt {\xib}+\vtb \xi -\vt\vtb 
h\right)\!\cdot\! \partial \hat \chi_a +  
\vt \vtb\, \xi{\xib}\!\cdot\! \partial^2 \hat   \chi_a + \vt \left(\widehat {\overline 
C}_a -\vtb \xi \!\cdot \! \partial \widehat{\overline C}_a \right) \0\\
&&
\quad\quad+\vtb \left( \widehat C_a  -\vt \xib \!\cdot \! 
\partial\widehat C_a \right)+ \vt\vtb \,\hat \psi _a)
\Big{)}\0\\
&&\cdot\left(\frac {\partial}{ \partial  \vt} + \left( -\xib^\lambda
+\vt \xib\!\cdot \! \partial\xib^ \lambda+ \vtb \Big( -h^\lambda  +\xib\!\cdot\!\partial \xi^\lambda  \right)\right.\0\\
&&\left.\quad\quad\quad+\vt\vtb \left(h\!\cdot\!\partial \xib^\lambda  -\xib \!\cdot\! \partial h^ \lambda - \xib \!\cdot\! \partial\xib\!\cdot\! \partial \xi^\lambda
+ \xib \!\cdot\! \partial\xi\!\cdot\! \partial \xib^ \lambda\right)\Big) \frac {\partial}{\partial x^\lambda }\right)\0\\
&&+ \Big{(} \hat \lambda_a  - \left(\vt {\xib}+\vtb \xi -\vt\vtb 
h\right)\!\cdot\! \partial \hat \lambda_a   +  
\vt \vtb\, \xi{\xib}\!\cdot\! \partial^2 \hat   \lambda_a + \vt \left(\widehat {\overline
D}_a  -\vtb \xi \!\cdot \! \partial \widehat{\overline D}_a \right) \0\\
&&
\quad\quad+\vtb \left( \widehat D_a -\vt \xib \!\cdot \! 
\partial\widehat D_a \right)+ \vt\vtb \,\hat \rho _a)
\Big{)}\0\\
&& \cdot\left(\frac {\partial}{ \partial  \vtb} + \Big( -\xi^\rho
+\vtb \xi\!\cdot \! \partial\xi^\rho + \vt \left( h^\rho +\xi\!\cdot\!\partial \xib^\rho \right) \right.\0\\
&&\quad\quad\quad+\left.\vt\vtb \left(h\!\cdot\!\partial \xi^\rho -\xi \!\cdot\! \partial h^\rho - \xi \!\cdot\! \partial\xib\!\cdot\! \partial \xi^\rho 
+ \xi \!\cdot\! \partial\xi\!\cdot\! \partial \xib^\rho \right)\Big) \frac {\partial}{\partial x^\rho}\right).\label{horinverse}
\ee

\subsection{Inverse supermetric}

Here is an additional argument (in 4d) that shows the inverse of $G_{MN}$ does not exist.  
Suppose the inverse $\widehat G^{MN}$ of $G_{MN}$ exists. We have the expansions \eqref{superGhat} which involve 76 component functions. The inversion condition is
\be
\widehat G^{ML}G_{LN} = \delta^M{}_N ,\label{GMLGLN}
\ee
where $\delta^\mu{}_\nu =\delta_\nu{}^\mu=\delta_\nu^\mu$  while $1=\delta^{\vt}{}_{\vtb} = - \delta_{\vtb}{}^{\vt} $. Decomposing the condition \eqref{GMLGLN} in components one realizes that it implies 88 quadratic equations. This is to be compared with the ordinary inverse metric $\hat g^{\mu\nu}$ in 4d which has 10  independent components while the independent inversion conditions are also 10. It is clear that \eqref{GMLGLN} cannot be satisfied without imposing constraints on the supermetric components. For instance, one of the equations is 
\be
\hat \gamma^\mu \gamma_\lambda + \hat{\bar \gamma}^\mu {\bar \gamma}_\lambda=0, \label{gammagammabar}
\ee
which means that either $\hat \gamma^\mu$ and $ \hat{\bar \gamma}^\mu $ vanish or they are constrained to one another (no such constraint exists for $\gamma^\mu$ and $\bar \gamma^\mu$).

\section{Gauge transformations in HS-YM}

In this subsection we examine more in detail the gauge transformation \eqref{deltahxp} and propose an interpretation of the lowest spin fields. Let us expand the master gauge parameter  as in \eqref{epsxu} and consider the first few terms in the transformation law of the lowest spin fields ordered in such a way that component fields and gauge parameters are infinitesimals of the same order. To lowest ($\delta^{(0)}$) order the transformation \eqref{deltahxp} reads
\be
&& \delta^{(0)} A_a= \partial_a \epsilon,\0\\
&&\delta^{(0)} \chi_a^{\nu} = \partial_a \xi^\nu , \0\\
&&\delta^{(0)}b_a{}^{\nu\lambda} = \partial_a\Lambda^{\nu\lambda}.
\label{deltaAhb}
\ee

To first  ($\delta^{(1)}$) order we have
\be
\delta^{(1)} A_a &=& \xi\!\cdot\!\partial A_a - \partial_\rho \epsilon
\,\chi_a^{\rho} ,\label{delta1Ahb}\\
\delta^{(1)} \chi_a^{\nu} &=& \xi\!\cdot\!\partial \chi_a^\nu-\partial_\rho
\xi^\nu \chi_a^\rho 
+ \partial^\rho A_a \Lambda_{\rho}{}^\nu   - \partial_\lambda \epsilon
\,b_a{}^{\lambda\nu},\0\\
\delta^{(1)} b_a^{\nu\lambda} &=& \xi\!\cdot\!\partial b_a{}^{\nu\lambda}
-\partial_\rho \xi^\nu b_a{}^{\rho\lambda}- \partial_\rho \xi^\lambda 
b_a{}^{\rho\nu }+{\partial_\rho \chi_a^{\nu} \Lambda^{\rho\lambda}
+\partial_\rho
\chi_a^{\lambda} \Lambda^{\rho\nu}}
- \chi_a^{\rho} \partial_\rho \Lambda_{\nu\lambda} .\0
\ee
The next orders contain three and higher derivatives.

These transformation properties allow us to associate the first two component fields of $h_a$ to an ordinary  U(1) gauge field  and to a vielbein. To see this,
let us denote by {$\tilde A_a$ and $\tilde E_a^\mu = \delta_a^\mu -\tilde \chi_a^\mu$}
the standard gauge and vielbein fields. The standard gauge and diffeomorphism
transformations, are
{ 
\be
\delta \tilde A_a&\equiv& \delta \left(\tilde E_a^\mu \tilde A_\mu\right)\equiv
\delta  \left((\delta_a^\mu -\tilde \chi_a^\mu) \tilde
A_\mu\right)\label{standardtransf}\\
&=&\left(-\partial_a \xi^\mu -\xi\!\cdot\!\partial \tilde \chi_a^\mu +\partial_\lambda \xi^\mu
\tilde \chi_a^\lambda\right) \tilde A_\mu+(\delta_a^\mu -\tilde
\chi_a^\mu)\left(\partial_\mu\epsilon + \xi\!\cdot\! \partial \tilde A_\mu + \tilde A_\lambda \partial_\mu \xi^\lambda\right) \0\\
&= &
\partial_a\epsilon + \xi\!\cdot\! \partial \tilde A_a- \tilde \chi_a^\mu
\partial_\mu\epsilon,\0 
\ee
}
and
{
\be
\delta \tilde E_a^\mu \equiv  \delta  (\delta_a^\mu -\tilde \chi_a^\mu) =
\xi\!\cdot\! \partial\tilde e_a^\mu -\partial_\lambda \xi^\mu \tilde e_a^\lambda = -
\xi\!\cdot\!\partial \tilde \chi_a^\mu -\partial_a \xi^\mu +\partial_\lambda \xi^\mu
\tilde \chi_a^\lambda,\label{deltaeamu}
\ee
}
so that
\be
\delta \tilde \chi_a^\mu= \xi\!\cdot\! \partial\tilde \chi_a^\mu +\partial_a \xi^\mu
-\partial_\lambda \xi^\mu \tilde \chi_a^\lambda,\label{deltaeamu1}
\ee
where we have retained only the terms at most linear in the fields.

Now it is important to understand the derivative $\partial_a$ in eqs.\eqref{deltahxp} and \eqref{deltaAhb} in the appropriate way:  the
derivative $\partial_a$ means $\partial_a = \delta_a^\mu \partial_\mu,$ not {
$ \partial_a = E_a^\mu \partial_\mu= \left(\delta_a^\mu
-\chi_a^\mu\right)\partial_\mu$}.
In fact the linear correction $ -\chi_a^\mu\partial_\mu$ is
contained in the term $ -i [h_a(x,p) \stackrel{\ast}{,} \varepsilon(x,p)]$, see,
for instance, the second term on the RHS of the first equation
\eqref{delta1Ahb}. 

From the above it is now immediate to make the identifications
\be
A_a= \tilde A_a, \quad\quad \chi_a^\mu = \tilde \chi_a^\mu.\label{identAAee}
\ee

The transformations \eqref{deltaAhb}, \eqref{delta1Ahb}
allow us to interpret  $\chi_a^\mu$ as the fluctuation of the inverse 
vielbein, therefore, the HS-YM action may 
accommodate gravity. However, a gravitational interpretation requires also that the frame field transforms under local Lorentz transformations. Therefore, we expect that the master field $h_a$ transforms and the action be invariant under local Lorentz transformations.  In \cite{HSYM}, it was shown that this symmetry can actually be implemented.


\end{document}